\documentclass[12pt]{article}
\usepackage[a4paper]{geometry}
\usepackage{authblk} 
\usepackage{color,bm}
\usepackage{refcount}
\usepackage{graphicx}
\usepackage{slashed}
\usepackage{braket,amsmath,amssymb,mnsymbol} 
\usepackage[export]{adjustbox}
\usepackage{subcaption}
\usepackage{graphics,psfrag,empheq,float}
\usepackage[colorlinks=true,linkcolor=black]{hyperref}
\usepackage{cite}
\usepackage{wasysym}
\usepackage[normalem]{ulem} 
\usepackage{axodraw2}

\def\XXint#1#2#3{{\setbox0=\hbox{$#1{#2#3}{\int}$}
     \vcenter{\hbox{$#2#3$}}\kern-.5\wd0}}

\newcommand{\cN}{{\cal N}}

\newcommand{\hvp}{{\hat{\boldsymbol{p}}}}
\newcommand{\hvz}{{\hat{\boldsymbol{z}}}}
\newcommand{\vn}{{\boldsymbol{n}}}
\newcommand{\vp}{{\boldsymbol{p}}}
\newcommand{\vq}{{\boldsymbol{q}}}
\newcommand{\vx}{{\boldsymbol{x}}}
\newcommand{\vy}{{\boldsymbol{y}}}

\newcommand{\vP}{{\boldsymbol{P}}}

\newcommand{\vr}{\bm{r}}

\newcommand{\Ima}{{\rm Im}}
\newcommand{\Rea}{{\rm Re}}

\newcommand{\bg}{\begin{align}}
\newcommand{\eeg}{\end{align}}
\newcommand{\be}{\begin{equation}}
\newcommand{\ee}{\end{equation}}
\newcommand{\ba}{\begin{eqnarray}}
\newcommand{\ea}{\end{eqnarray}}

\newcommand{\nn}{\nonumber}

\newcommand{\ve}{\varepsilon}

\newcommand{\la}{\langle}
\newcommand{\ra}{\rangle}
\newcommand{\si}{\sigma}
\newcommand{\sig}{\sigma}

\newcommand{\ep}{\epsilon}
\newcommand{\fa}{\alpha}
\newcommand{\lam}{\lambda}

\newcommand{\vf}{\boldsymbol{\phi}}
\newcommand{\De}{\Delta}

\newcommand{\vce}{\boldsymbol{0}}

\begin{document}

\title{Lectures on scattering theory in partial-wave amplitudes}

\author[a]{J. A. Oller\thanks{oller@um.es}}
\affil[a]{\it Departamento de F\'{\i}sica, Universidad de Murcia, E-30071 Murcia,  Spain}

\maketitle

\begin{abstract}

These lectures treat  scattering theory from a non-perturbative point of view. The course begins with a review of formal aspects in scattering theory, discussing the in/out states and the $S$ matrix that connects them. Unitarity relations, phase space, and the Lippmann-Schwinger equation for the collision operator $T$ are discussed. The calculation of cross sections, the optical theorem and Boltzmann $H$-theorem from unitarity of the $S$ matrix are also explained. Special emphasis is given in these lectures to expansions in partial-wave amplitudes of two-body scattering amplitudes, both for massive and massless particles, and to unitarity in partial-wave amplitudes. In this way, partial-wave expansions in terms of  $\ell SJ$ states, and using helicity states with definite total angular momentum $J$ are discussed. Crossing symmetry is also explained, and connected to crossed channels, analyticity and crossed-channel cuts in partial-wave amplitudes. Different non-perturbative techniques and general parameterizations are developed for partial-wave amplitudes that satisfy unitarity and are consistent with the analyticity properties that they must have. Then, the $N/D$ method and additional solutions  generated by adding CDD poles are covered in detail. The change to different non-physical Riemann sheets and the search for resonant poles is discussed as well. A differentiation is made between dynamically generated resonances by the degrees of freedom explicitly accounted for versus pre-existing resonances derived from short-distance dynamics. We  exemplify it with the cases of the $\sigma/f_0(500)$ and the $\rho(770)$ resonances in $\pi\pi$ scattering. The reader is also introduced to final-state interactions, Watson's theorem, and general parameterizations to take them into account. It ends with an appendix dedicated to the Sugawara-Kanazawa theorem regarding the number of subtractions in dispersion relations.

\end{abstract}
\newpage

\tableofcontents

\bigskip
\bigskip
\bigskip

\noindent
{\Large {\bf Extended table of contents}}
\bigskip
\bigskip

\noindent
{\large {\bf 1~~~ $S$ and $T$ matrices. Unitarity}}

\bigskip

\noindent
Free particle states ~$\square$~  Scattering states, in and out states, and the M\"oller operators ~$\square$~ $S$-matrix elements, and the $S$-matrix operator  ~$\square$~ Expression in perturbation theory ~$\square$~  $T$-matrix operator, unitarity relations, and phase space ~$\square$~ Hermitian unitarity, the Lippmann-Schwinger equation, and spherical waves in the scattering states ~$\square$~ Cross section, and the optical theorem ~$\square$~  Boltzmann $H$-theorem 

\bigskip
\bigskip

\noindent
{\large {\bf 2~~~ Two-body scattering. Partial-wave amplitude expansions}}
\bigskip

\noindent
Two-body scattering, and Mandelstam variables ~$\square$~ Normalization, element of phase space, cross section and unitarity ~$\square$~   Rotational invariance

\bigskip

\noindent
{ {\bf 2.1~~~ Partial-wave amplitudes in the $\ell SJ$ basis}}

\bigskip

\noindent
States, standard Lorentz transformation, and transformation under rotations ~$\square$~ States  $|p \ell m,\sig_1\sig_2\ra$ with definite orbital angular momentum, and transformation under rotations ~$\square$~ $\ell S J$ states with definite total angular momentum ~$\square$~ Partial-wave amplitudes, and calculation ~$\square$~ Identical particles, unitary normalization, and including isospin ~$\square$~ Time-reversal invariance, and PWAs are symmetric ~$\square$~ Unitarity in PWAs ~$\square$~ Lippmann-Schwinger equation in PWAs, phase shifts and inelasticity     

\bigskip

\noindent
{ {\bf 2.2~~~ Partial-wave amplitudes in the helicity basis}}

\bigskip

\noindent
Helicity, monoparticle states, and representation of the rotation group ~$\square$~ Two-particle states, common $R(\hvp)$, and normalization ~$\square$~ Partial-wave states,  normalization, and Bose-Einstein symmetry ~$\square$~  Partial-wave amplitudes with definite helicities  

\bigskip
\bigskip

\noindent
{\large {\bf 3~~~ Crossing symmetry}}

\bigskip

\noindent
Crossing symmetry, physical regions, and analyticity ~$\square$~ Example of $\pi\pi$ scattering, and isospin amplitudes ~$\square$~ Crossed channel cuts, and the Lippmann-Schwinger equation ~$\square$~ Kinematical singularities, and Lorentz invariant amplitudes $~\square~$ Cuts in the $s$ plane for PWAs due to poles and cuts in crossed channels $~\square~$ Crossed cuts in  PWAs for nonrelativistic scattering

\bigskip
\bigskip

\noindent
{\large {\bf 4~~~ The $N/D$ method and CDD poles}}

\bigskip
\noindent
Set up for the $N/D$ method $~\square~$ CDD poles, and additional solutions $~\square~$ General formula for PWAs without LHC $~\square~$ Generalization to coupled channels

\bigskip
\bigskip

\noindent
{\large {\bf 5~~~ General parameterization for a PWA and reaching  unphysical Riemann sheets}}

\bigskip

\noindent
General formula for a PWA, the matrices  $\hat{\cal N}(s)$ and $\hat{G}(s)$ $~\square~$ Change to the second Riemann sheet of $g(s)$, and different Riemann sheets of $T(s)$

\bigskip

\noindent
{ {\bf 5.1~~ Dynamically generated resonances \& pre-existing ones}}

\bigskip

\noindent
Low lying $\sigma$ pole, radius of convergence, and calculation of the $I=\ell=0$ $\pi\pi$ PWA  $~\square~$ Natural-value estimate for $a(\mu)$, pole of the $\sigma$, and dynamically generated resonances  $~\square~$ Calculation of the $I=\ell=1$ $\pi\pi$ PWA, and fine tuning of $a(\mu)$ to a large negative value  $~\square~$ Second CDD pole, and the $\rho(770)$ as pre-existing or elementary resonance

\bigskip
\bigskip

\noindent
{\large {\bf 6~~~ Final(Initial)-state interactions. Watson theorem}}

\bigskip

\noindent
Feeble probes, unitarity, PWAs, and Watson's theorem $~\square~$ Generalization to coupled channels, using the $N/D$ method, and an Omn\'es function

\bigskip
\bigskip

\noindent
{\large {\bf A~~~ The Sugawara-Kanazawa theorem and number of subtractions in dispersion relations}}

\bigskip

\noindent
The Sugawara-Kanazawa theorem  $~\square~$ Corollaries

\newpage
\section*{Brief introduction}

We cover some basic topics on scattering theory, emphasizing unitarity and analyticity of the $S$ matrix while keeping an eye on phenomenology. These lectures could be useful to those interested in scattering processes that cannot be calculated by using only perturbative techniques  in Quantum Mechanics or in Quantum Field Theory. The different concepts here are typically introduced from a nonperturbative point of view. The most formal part of the lectures concern the first two chapters, which are the longest ones. The first one is dedicated to formal topics in the theory of the $S$-matrix, while the second chapter develops in detail the expansion in partial-wave amplitudes of scattering amplitudes and related aspects, like unitarity in partial waves. The third chapter is an introduction to crossing symmetry and its applications. The other chapters are typically of narrower scope, typically  more specific, and  treat several interesting topics in partial-wave amplitudes: The $N/D$ method, CDD poles, unphysical Riemann sheets and resonance poles there, their nature, and final-state interactions. There is an added appendix dedicated to the relatively unknown Sugawara-Kanazawa theorem, which is applicable to many situations of interest in phenomenology. There is a bunch of exercises along the manuscript typically asking the reader to complete some derivations or deduce new results. 

These lectures are rather self-contained and could be used with only limited background knowledge on the subject. Books like \cite{martin.200705.1,Barton:1965dr,Weinberg:1995mt,Schweber:1961zz,Weinberg:2013qm,olive.181102.1} have been often consulted by the author along his research trajectory and they have imprinted these lectures. In preparing them I have also used my previous book \cite{Oller:2019rej}. Most of this material was presented in a course given at the Instituto de Física de la Facultad de Ingeniería de la Universidad de la República Uruguay in August and September 2024. 

\section{$S$ and $T$ matrices. Unitarity}
\def\theequation{\arabic{section}.\arabic{equation}}
\setcounter{equation}{0} 

\bigskip
$\bullet$ \underline{\it Free particle states:}
\bigskip

\begin{figure}
  \begin{center}
 \includegraphics[width=.4\textwidth, valign=c]{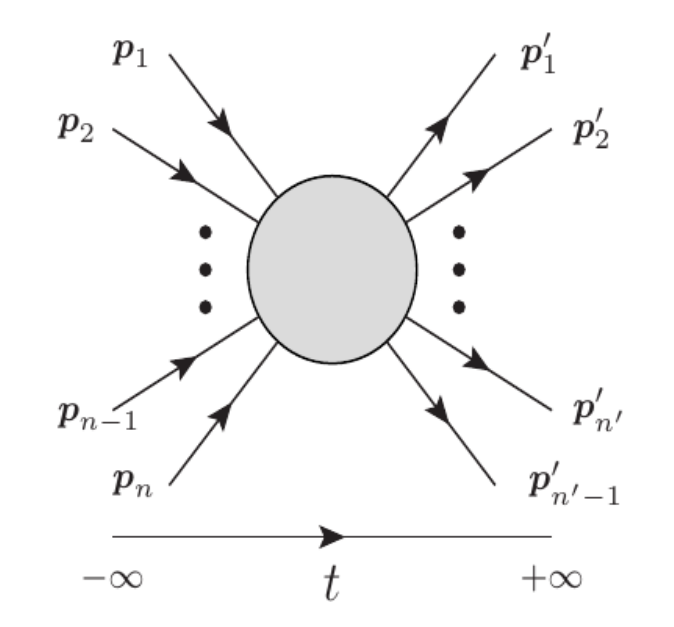}
 \caption{Scattering process for an initial state of $n$ particles giving rise to a final state of $n'$ particles. Time runs from left to right. \label{fig.240809.1}}
    \end{center}
  \end{figure}

A scattering process for a finite-range interaction can be visualized, Fig.~\ref{fig.240809.1}, as a set of $n$ free particles  (behaving individually like plane waves) for asymptotically early times, {\it i.e.} $t\to -\infty$, that interact and then become a final set of $n'$ free particles for asymptotic late times $t\to +\infty$. Each monoparticle state is denoted by $|\vp\sigma\lam\ra$, where $\vp$ is the momentum, $\sigma$ is the third component of spin or helicity (more details will be given below in Sec.~\ref{sec.240811.1}), and $\lam$ comprises the rest of necessary quantum numbers like its mass $m$, its spin $s$, and charges. These monoparticle states have the Lorentz invariant normalization
  \begin{align}
    \label{240808.1}
    \la \vp'\sigma'\lam'| \vp \sigma \lam \ra&=2w(p)(2\pi)^3 \delta(\vp'-\vp)\delta_{\sigma'\sigma}\delta_{\lam'\lam}~,
  \end{align}
  with $w(p)=\sqrt{m^2+\vp^2}$. 

  \bigskip
{\bf Exercise 1:} For particles without spin demonstrate that the normalization in the right-hand side of Eq.~\eqref{240808.1} is Lorentz invariant.
      \bigskip

      A multiparticle state is made by the direct product of several monoparticle states,
\begin{align}
\label{240808.2}
|\vp_1\sigma_1\lam_1,\vp_2\sigma_2\lam_2,\ldots,\vp_n\sigma_n\lam_b\ra&=
|\vp_1\sigma_1\lam_1\ra\otimes|\vp_2\sigma_2\lam_2\ra\otimes\cdots\otimes|\vp_n\sigma_n\lam_n\ra~.
\end{align}

\bigskip
$\bullet$ \underline{\it Scattering states, in and out states, and the M\"oller operators:}
\bigskip

Formally speaking, let $H$ be the Hamiltonian of the system split as $H=H_0+H'$, where $H_0$ is the free part (kinetic energy) and $H'$ is the interaction Hamiltonian. 
The spectrum of the theory comprises the discrete spectrum $\{\Psi_n,E_n\}$,
\begin{align}
  \label{240803.2}
  \text{Discrete spectrum:~} H|\Psi_n\ra=E_n|\Psi_n\ra~,
\end{align}
and the continuum one that consists of the scattering states. For a given energy there are two types of scattering states: The in states,  $|\Psi_\alpha^-\ra$, and the out ones, $|\Psi_\alpha^+\ra$, where the index $\alpha$ is continuum: 
\begin{align}
  \label{240803.3}
  \text{Continuum spectrum:~} H|\Psi_\alpha^\pm\ra=E_\alpha|\Psi_\alpha^\pm\ra~,
\end{align}
There are two types of scattering states because the in states have an outgoing spherical wave component, while for the out ones the spherical wave component is  incoming, as we show below in more detail, cf. Eqs.~\eqref{240811.1} and \eqref{240811.2},  and the discussions around them. The scattering states are in one-to-one correspondence with the free multiparticle states  made out by plane waves (already introduced), and generically denoted by $|\Phi_\alpha\ra$. The relation between the in/out states with the plane-wave states  is established by the M\"oller operators, $\Omega_\pm$, as 
\begin{align}
  \label{240803.4}
|\Psi_\alpha^\pm\ra&=\Omega_\pm|\Phi_\alpha\ra~.
\end{align}
The index $\alpha$ is then the set of quantum numbers used to characterize the free multiparticle states, cf. Eq.~\eqref{240808.2}. Being more specific,  let us introduce the interaction or Dirac picture and obtain a formal expression for the M\"oller operators. Indicating by the subscripts 
$S$ and $I$ the Schr\"odinger the interaction pictures, respectively, we have
\begin{align}
  \label{240808.5}
 |\Psi_I,t\ra&=e^{i H_0 t}|\Psi_S,t\ra=e^{iH_0 t}e^{-iH(t-t_0)}|\Psi_S,t_0\ra=e^{iH_0 t}e^{-iH(t-t_0)}e^{-iH_0t_0}|\Psi_I,t_0\ra~.
\end{align}
From here, it follows that  the evolution operator in the interaction picture is
\begin{align}
  \label{240808.6}
  U_I(t,t_0)&=e^{iH_0t}e^{-iH(t-t_0)}e^{-iH_0 t_0}~.
\end{align}
Now, advocating the adiabatic approximation, with $H_I'(t)\to e^{-\ep |t|}H'_I(t)$ and $\ep\to 0^+$, we then have that
\begin{align}
  \label{240808.7}
  |\Psi_\alpha^-\ra&=\lim_{t_0\to-\infty}U_I(0,t_0)|\Phi_\alpha\ra=\lim_{t_0\to-\infty}e^{iHt_0}e^{-iH_0t_0}|\Phi_\alpha\ra~,\\
  H |\Psi_\alpha^-\ra&=\lim_{t_0\to-\infty}e^{iHt_0}He^{-iH_0t_0}|\Phi_\alpha\ra=\lim_{t_0\to-\infty}e^{iHt_0}e^{-iH_0t_0}(H_0+e^{-\ep |t_0|}H'_I(t_0))|\Phi_\alpha\ra=E_\alpha |\Psi_\alpha^-\ra\,,\nn
\end{align}
where we have used that $H_0|\Phi_\alpha\ra=E_\alpha|\Phi_\alpha\ra$\,. By an analogous reasoning,
\begin{align}
    \label{240808.8}
    |\Psi_\alpha^+\ra&=\lim_{t_0\to+\infty}U_I(0,t_0)|\Phi_\alpha\ra=\lim_{t_0\to+\infty}e^{iHt_0}e^{-iH_0t_0}|\Phi_\alpha\ra~,\\
    H|\Psi_\alpha^+\ra&=E_\alpha|\Psi_\alpha^+\ra\,.\nn
\end{align}

\bigskip
{\bf Exercise 2:} Fill the steps to get the result in the second line of Eq.~\eqref{240808.8}. 
\bigskip

We introduce the operator $\Omega(t_0)$ as
\begin{align}
  \Omega(t_0)=e^{iH t_0}e^{-iH_0 t_0}~.
  \end{align}
From Eqs.~\eqref{240808.7} and \eqref{240808.8} it follows that the M\"oller operators correspond to 
\begin{align}
  \label{240809.1}
  \Omega_\pm=\Omega(\pm \infty)~.
  \end{align}

\bigskip
$\bullet$ \underline{\it $S$-matrix elements, and the $S$-matrix operator:}
\bigskip

The $S$-matrix element $S_{\beta \alpha}$ is
\begin{align}
  \label{240809.2}
S_{\beta \alpha}&=\la \Psi_\beta^+|\Psi_\alpha^-\ra~,  
\end{align}
the probability amplitude between the considered in and  out states. In terms of the M\"oller operators, Eq.~\eqref{240803.4}, and the evolution operator in the interaction picture, Eqs.~\eqref{240808.6} and \eqref{240809.1}, it reads
\begin{align}
  \label{240809.3}
  S_{\beta \alpha}&=\la \Phi_\beta|\Omega_+^\dagger \Omega_-|\Phi_\alpha\ra~,\\
  \label{240809.4}
  S_{\beta\alpha}&=\lim_{{\scriptsize \begin{array}{l}t_2\to +\infty\\t_1\to-\infty\end{array}}}\la \Phi_\beta|e^{iH_0 t_2}e^{-iHt_2}e^{iHt_1}e^{-iH_0t_1}|\Phi_\alpha\ra=\la \Phi_\beta|U_I(+\infty,-\infty)|\Phi_\alpha\ra~.
\end{align}
Therefore, the $S$-matrix operator acting on plane-wave states is
\begin{align}
  \label{240809.5}
  S&=\Omega_+^\dagger\Omega_-=U_I(+\infty,-\infty)\,,
\end{align}
and it is unitarity 
\begin{align}
    \label{240809.6}
  S^\dagger S&=S S^\dagger=I\,.
\end{align}

\bigskip
$\bullet$ \underline{\it Expression in perturbation theory:}
\bigskip

The evolution operator $U_I(t,t_0)$, Eq.~\eqref{240808.6}, satisfies the differential equation
\begin{align}
  \label{240809.7}
  i\frac{\partial }{\partial t}U_I(t,t_0)=H'_I(t)U_I(t,t_0)
  \end{align}
Its solution in the Neumann series is
\begin{align}
  \label{240809.8}
  U_I(t,t_0)&=I+\sum_{n=1}^\infty\frac{(-i)^n}{n!}\int_{t_0}^t dt_1\int_{t_0}^t dt_2\cdots\int_{t_0}^t dt_n T\left[H'_I(t_1)H'_I(t_2)\cdots H_I'(t_n)\right]\\
  &=T \exp\left(-i\int_{t_0}^t dt'H'_I(t')\right)~,\nn
\end{align}
where $T$ indicates  time ordering  the operators $H_I'(t')$ in each term of the series expansion. Then, for the $S$-matrix element in Eq.~\eqref{240809.4} we can also write
\begin{align}
  \label{240809.9}
S_{\beta\alpha}&=\la \Phi_\beta|T \exp\left(-i\int_{-\infty}^{+\infty} dt'H'_I(t')\right)|\Phi_\alpha\ra~.
\end{align}
Denoting by ${\cal H}'(x)$ the Hamiltonian density for the interacting part, with $x=(x^0,\vx)$, we then have 
\begin{align}
\label{240808.10}
S_{\beta\alpha}&=\la \Phi_\beta|T \exp\left(-i\int d^4 x {\cal H}'_I(x)\right)|\Phi_\alpha\ra~.
\end{align}
This expression is actually covariant because it is assumed that $[{\cal H}'_I(x),{\cal H}'_I(y)]=0$ for $(x-y)^2<0$ \cite{Weinberg:1995mt}. This is also referred as local commutativity of the interacting Hamiltonian densities. Therefore, a change of frame does not modify the $T$ product.

\bigskip
$\bullet$ \underline{\it $T$-matrix operator, unitarity relations, and phase space:}
\bigskip

The $T$-matrix operator is defined as
\begin{align}
  \label{240809.11}
S=I+iT~.
\end{align}
Since $S$ is an unitary operator this implies two unitarity relations for $T$:
\begin{align} 
\label{240809.12}
S^\dagger S&=I~ \rightarrow T-T^\dagger=iT^\dagger T ~,\\
\label{240809.13}
S S^\dagger & =I~\rightarrow T-T^\dagger=iT T^\dagger ~.
\end{align} 
Due to space-time translation symmetry the $S$- and $T$-matrix elements contain a global four Dirac delta function of conversation of energy and momentum. In the following, we factorize it out and keep in mind that the initial and final momenta must be the same. Namely,
\begin{align}
\label{240809.14}
S_{\beta\alpha}&\to  (2\pi)^4\delta(P_\beta-P_\alpha)S_{\beta\alpha}~,\\ 
T_{\beta\alpha}&\to  (2\pi)^4\delta(P_\beta-P_\alpha)T_{\beta\alpha}~.\nn 
\end{align}
In the unitarity relation of Eq.~\eqref{240809.12} we introduce in the right-hand side a decomposition of the identity by summing over the $m_{th}$ intermediate states (or channels) containing $n_m$ free particles. Then, it becomes
\begin{align}
\label{240809.15}
T_{\beta\alpha}-T^\dagger_{\beta\alpha}&=i\sum_m\int \left(\prod_{i=1}^{n_m}\frac{d^3p_i}{(2\pi)^32w(\vp_i)}\sum_{\sigma_i,\lam_i}\right)(2\pi)^4\delta(P_T-P_m)\la\Phi_\beta|T^\dagger|\vp_1\sigma_1\lam_1,\ldots,\vp_{n_m}\sigma_{n_m}\lam_{n_m}\ra \\
\times&\la\vp_1\sigma_1\lam_1,\ldots,\vp_{n_m}\sigma_{n_m}\lam_{n_m}|T|\Phi_i\ra~, \nn
\end{align}
where $P_T=P_\beta=P_i$, with $P_m=\sum_{i=1}^m p_i$. Notice that in order to have a nonvanishing contribution from the right-hand side in the previous equation the total energy $E=P_T^0>s_{\rm th,1}$, where $s_{\rm th,1}$ is the smallest threshold for the intermediate channels (we can think that they have been ordered with increasing thresholds).

The $3n_m-4$ infinitesimal element of phase space is defined as
\begin{align}
\label{240809.16}
dQ_m&=\int \left(\prod_{i=1}^{n_m}\frac{d^3p_i}{(2\pi)^32w(\vp_i)}\sum_{\sigma_i,\lam_i}\right)(2\pi)^4\delta(P_T-P_m)~,
\end{align}
with four integrations done to remove the delta of conservation of total energy-momentum.

\bigskip
{\bf Exercise 3:} Why has one to divide by $(2\pi)^32w(\vp_i)$ each $d^3\vp_i$ in Eq.~\eqref{240809.16}?
\bigskip

 In the following, the sum in Eq.~\eqref{240809.15} over the type of intermediate states and the integration over phase space is denoted by the symbol $\displaystyle{\sumint}$. In terms of phase space the unitarity relations in \eqref{240809.15} reads
\begin{align}
\label{240809.17}
T_{\beta\alpha}-T^\dagger_{\beta\alpha}&=i\sumint dQ_m \la \vp_1\sigma_1\lam_1,\ldots,\vp_{n_m}\sigma_{n_m}\lam_{n_m}|T|\Phi_\beta\ra^* \la\vp_1\sigma_1\lam_1,\ldots,\vp_{n_m}\sigma_{n_m}\lam_{n_m}|T|\Phi_\alpha\ra~. 
\end{align}

\bigskip
$\bullet$ \underline{\it Hermitian unitarity, the Lippmann-Schwinger equation, and spherical waves in the scattering states:}
\bigskip

The point of Hermitian unitarity is to show that $T^\dagger(E)=T(E^*)$, with $E$ the total energy. Before proceeding with its demonstration,  when this fact is taken into Eq.~\eqref{240809.17} it implies that (we now explicitly show the argument $E$ with a slight imaginary part $\pm i\epsilon$)
\begin{align}
\label{240809.17b}
T_{\beta\alpha}(E+i\ep)-T_{\beta\alpha}(E-i\ep)&=i\sumint dQ_m \la \vp_1\sigma_1\lam_1,\ldots,\vp_{n_m}\sigma_{n_m}\lam_{n_m}|T(E+i\ep)|\Phi_\beta\ra^*\\
&\times \la\vp_1\sigma_1\lam_1,\ldots,\vp_{n_m}\sigma_{n_m}\lam_{n_m}|T(E+i\ep)|\Phi_i\ra~. \nn
\end{align}

 Then, $T_{\beta\alpha}(E)$ has a right hand cut for $E>s_{\rm th,1}$, with its discontinuity given by the right-hand side of Eq.~\eqref{240809.17b}. This is a crucial point in $S$-matrix theory in which one takes the singularities in the $T$-matrix elements as due to intermediate physical states. There could be other so-called kinematical singularities for particles with spin, associated to the {\it known} properties of these particle states with spin under Lorentz transformations.

 To proceed with its demonstration we first introduce the Lippmann-Schwinger equation for the scattering states $|\Psi^\pm_\alpha\ra$. With $V\equiv H'$, we rewrite Eq.~\eqref{240803.3} as
 \begin{align}
\label{240810.1}
(H_0-E_\alpha)|\Psi_\alpha^\pm\ra&=-V|\Psi_\alpha^\pm\ra~,\\
|\Psi_\alpha^\pm\ra&=|\Phi_\alpha\ra-(H_0-E_\alpha)^{-1}V|\Psi_\alpha^\pm\ra~.
 \end{align}
 This equation is not really meaningful because $H_0-E_\alpha$ has no inverse for real positive $E_\alpha>s_{\rm th,1}$, being zero when acting over $|\Phi_\alpha\ra$. To make it meaningful we add a vanishing imaginary part $\pm i\ep$ to $E_\alpha$ such that
 \begin{align}
\label{240810.2}
|\Psi_\alpha^\pm\ra&=|\Phi_\alpha\ra+(E_\alpha\mp i\ep-H_0)^{-1}V|\Psi_\alpha^\pm\ra~.
 \end{align}
 To check this equation let us show that a wave packet of states $|\Psi_\alpha^\pm\ra$ behaves for $t\to \pm \infty$ as a wave packet of free states, respectively \cite{Weinberg:1995mt}. Let $g(\alpha)$ be a smooth function over some finite range $\Delta E$, and introduce a resolution of the identity in terms of plane waves $|\Phi_\beta\ra$ in between $(E_\alpha\mp i\ep-H_0)$ and $V$ in Eq.~\eqref{240810.2}. Then,
 \begin{align}
\label{240810.3}
\int d\alpha g(\alpha)e^{-iE_\alpha t}|\Psi^\pm_\alpha\ra \! &= \!\int d\alpha g(\alpha)e^{-iE_\alpha t}|\Phi_\alpha\ra+ \int  \!\! d\alpha \int \!\!d\beta  g(\alpha) e^{-iE_\alpha t} (E_\alpha\mp i\ep-E_\beta)^{-1}\langle\Phi_\beta|V|\Psi^\pm_\alpha\ra|\Phi_\beta\ra\,.
 \end{align}
   
 The rapid oscillation of the exponential $e^{-iE_\alpha t}$ for $t\to \pm \infty$ in the second term on the right-hand side of the previous equations kills all contributions to this integral except for $E_\alpha$ near $E_\beta$. This contribution can be evaluated by employing the Cauchy theorem for complex integration.  For $t\to\rightarrow \pm\infty$ the exponential factor $e^{-iE_\alpha t}$ allows to close the integration contour in the complex $E_\alpha$ plane with a semicircle at infinity running over its lower/upper half complex plane, respectively. For definiteness, consider $|\Psi_\alpha^+\ra$ with $t\to +\infty$. Then,  the pole contribution at $E_\alpha=E_\beta+i\ep$ lies outside the integration contour and the integral vanishes. Similarly, for $|\Psi_\alpha^-\ra$ we have the pole contribution at $E_\alpha=E_\beta-i\epsilon$, which again lies outside the integration contour and it gives no contribution.

 The $T$-matrix operators can also be introduced by the relation
 \begin{align}
\label{240810.4}
-V|\Psi_\alpha^-\ra&=-V\Omega_-|\Phi_\alpha\ra=T|\Phi_\alpha\ra~.
 \end{align}

 To connect this definition with Eq.~\eqref{240809.11}, let us consider what an in state looks  at later times $t\to+\infty$. Again, we pick up only the contribution from $E_\alpha\approx E_\beta$ in the double second integral term on the right-hand side of Eq.~\eqref{240810.3}. But now, the pole at $E_\alpha=E_\beta-i\ep$ for the $|\Psi_\alpha^-\ra$ case gives contribution because it lies within the integration contour closed by a semicircle at infinite along the lower half of the complex $E_\alpha$ plane. The result of the integral  is $-2\pi i$ (the integration is closed in the clockwise sense) times the rest of the integrand (once excluded $(E_\alpha-E_\beta-i\ep)^{-1}$):  
 \begin{align}
\label{240810.5}
\int d\alpha g(\alpha)e^{-iE_\alpha t}|\Psi^-_\alpha\ra&=\int d\alpha g(\alpha)e^{-iE_\alpha t}|\Phi_\alpha\ra-2\pi i \int d\alpha \int d\beta \delta(E_\alpha-E_\beta)  g(\alpha) e^{-iE_\alpha t} \langle\Phi_\beta|V|\Psi^-_\alpha\ra|\Phi_\beta\ra \nn \\
&=\int d\alpha g(\alpha)e^{-iE_\alpha t}|\Phi_\alpha\ra+2\pi i \int d\alpha \int d\beta \delta(E_\alpha-E_\beta)  g(\alpha) e^{-iE_\alpha t} \langle\Phi_\beta|T|\Phi_\alpha\ra|\Phi_\beta\ra~.
 \end{align}
  Now, according to Eq.~\eqref{240809.2}, we can expand $|\Psi_\alpha^-\ra$ in terms of out  states $|\Psi_\beta^+\ra$ as (in Eq.~\eqref{240809.2} the total energy-momentum Dirac delta function was not factorized yet, and so we do along this discussion) 
 \begin{align}
\label{240810.6}
|\Psi_\alpha^-\ra&=\int d\beta S_{\beta\alpha}|\Psi_\beta^+\ra~.
 \end{align}
 Upon multiplying by $g(\alpha)e^{-iE_\alpha t}$ and integrating,
 \begin{align}
   \label{240810.7}
   \int d\alpha g(\alpha)e^{-iE_\alpha t }|\Psi_\alpha^-\ra&=\int d\alpha \int d\beta g(\alpha)e^{-iE_\alpha t}S_{\beta\alpha}|\Psi_\beta^+\ra\,,
 \end{align}
 Next, we take into account that $S_{\beta\alpha}$ is nonzero only for $E_\alpha=E_\beta$  and make $t\to +\infty$, so that we can rewrite the  term on the right-hand side as
 \begin{align}
   \label{240810.8}
\int d\beta e^{-iE_\beta t}|\Psi_\beta^+\ra\int d\alpha  g(\alpha)S_{\beta\alpha}
&\xrightarrow[t\to+\infty]{} \int d\alpha \int d\beta g(\alpha)e^{-iE_\beta t}S_{\beta\alpha}|\Phi_\beta\ra\,,\nn\\
   \int d\alpha g(\alpha)e^{-iE_\alpha t }|\Psi_\alpha^-\ra&\xrightarrow[t\to+\infty]{}\int d\alpha \int d\beta g(\alpha)e^{-iE_\alpha t}S_{\beta\alpha}|\Phi_\beta\ra\,.
 \end{align}
 Comparing this result with the right-hand side of Eq.~\eqref{240810.5} we then conclude that
 \begin{align}
   \label{240810.9}
S_{\beta\alpha}&=\delta(\beta-\alpha)+2\pi i \delta(E_\beta-E_\alpha)T_{\beta\alpha}~.
 \end{align}
 By homogeneity of space the matrix element $T_{\beta\alpha}$ is also proportional to a linear momentum Dirac delta function $\delta(\bm{P}_\beta-\bm{P}_\alpha)$. Once this is factorized out, we end with Eq.~\eqref{240809.11}.\footnote{The $T$-matrix operator introduced as in Eq.~\eqref{240810.4} has only a factorizing linear-momentum Dirac delta function. It is a time-independent formalism.\label{foot.240813.1}} 

 From the Lippmann-Schwinger equation for the scattering states, and Eq.~\eqref{240810.4}, we can deduce the Lippmann-Schwinger equation for the $T$-matrix operator. For that, let us  apply  Eq.~\eqref{240810.2} to the in states, and multiply it by $-V$ to the left, 
 \begin{align}
   \label{240810.10}
 -V|\Psi_\alpha^-\ra&=-V|\Phi_\alpha\ra-V(E_\alpha+ i\ep-H_0)^{-1}V|\Psi_\alpha^-\ra\,,
 \end{align}
  and take into account Eq.~\eqref{240810.4}, 
 \begin{align}
 \label{240810.11}
 T(E+i\ep)&=-V+V(E+i\ep-H_0)^{-1}T(E+i\ep)\,.
 \end{align}
 This is the Lippmann-Schwinger equation for the $T$-matrix operator. This equation can also be rewritten as
 \begin{align}
 \label{240826.1}
 T(E+i\ep)&=-V+T(E+i\ep)(E+i\ep-H_0)^{-1}V\,.
 \end{align}
 By iterating the previous integral equation  in the Neumann series one can easily check its equivalence with Eq.~\eqref{240810.11}. 
 Since $V^\dagger=V$ and $H_0^\dagger=H_0$, after taking the Hermitian conjugate of Eq.~\eqref{240810.11} we end with the requirement of Hermitian unitarity,
 \begin{align}
   \label{240810.12}
   T(E)^\dagger=T(E^*)\,.
 \end{align}
 To arrive to this result take also into account the form of the Lippmann-Schwinger equation in Eq.~\eqref{240826.1}.  It is worth stressing that Eq.~\eqref{240810.11} can be extended to the whole complex $E$ plane if $V$ is an analytic function of its arguments.

Let us see the relation between the Green function $(H_0-E_\alpha\pm i\ep)^{-1}$ in Eq.~\eqref{240810.2} and the behavior at asymptotic times $t\to\pm \infty$ of the in/out states by employing elementary nonrelativistic Quantum Mechanics. In this way, and within this narrower context, we reach the conclusions obtained above from Eqs.~\eqref{240810.3} and \eqref{240810.8} in a more straightforward and intuitive way. In configuration space, by taking the inverse Fourier transform of the free-theory Green function, one has
 \begin{align}
   \label{240811.1}
   \int d^3 p \frac{e^{i\vp(\vx-\vy)}}{\frac{\vp^2}{2\mu}-E \mp i \ep}&=-
   \frac{i4\mu\pi}{|\vx-\vy|} \int_0^{+\infty} p dp \frac{e^{ip|\vx-\vy|}-e^{-ip|\vx-\vy|}}{p^2-2\mu E  \mp i \ep}=-\frac{i2\mu\pi}{|\vx-\vy|} \int_{-\infty}^{+\infty} p dp \frac{e^{ip|\vx-\vy|}-e^{-ip|\vx-\vy|}}{p^2-2\mu E\mp i \ep}\,.
 \end{align}
 For  $\exp(\pm ip|\vx-\vy|)$  we can close the integration contour with an infinite semicircle along the upper/lower half $p$ plane, respectively. Let us take $E+i\ep$,  corresponding to the in states. There are two poles at $p_a=\sqrt{2\mu E}+i\ep$ and $p_b=-p_a$, which imply a contribution
 \begin{align}
\label{240811.2}
\frac{\mu(2\pi)^2}{|\vx-\vy|} e^{ip_a|\vx-\vy|}~.
 \end{align}
 For the  time evolution we have to multiply the previous factor by $e^{-iE t}$, similarly as we did before in Eq.~\eqref{240810.3}, but omitting for simplicity the integration associated to the wave packet. Therefore, for $ t\to -\infty$ there is no cancellation between the exponents combined in the form $\sqrt{2\mu E}|\vx-\vy|-Et$, the exponential oscillates wildly and there is no contribution within a wave packet, as it should. Then, the in state behaves as a plane wave from the independent term in Eq.~\eqref{240810.2}. However, for $t\to +\infty$ there is contribution with $|\vx-\vy|/t\approx \sqrt{E/2\mu}$, in the form of outgoing spherical waves. The stationary point in the exponent happens for $\displaystyle{d(\sqrt{2\mu E}|\vx-\vy|-E t)/dE=0}$. This gives $\displaystyle{|\vx-\vy|/t=\sqrt{E/\mu}}$, corresponding to the classical velocity. Therefore,  this elementary analysis illustrates that in wave packets with time dependence incorporated, the in states behave as plane waves for $t\to -\infty$, and  have an outgoing spherical-wave component stemming from the scattering process for $t\to +\infty$. For the out states a similar analysis leads to conclude that they behave like plane waves for $t\to +\infty$, and they also have an incoming spherical wave component in the asymptotic past $t\to-\infty$.

 \bigskip
     {\bf Exercise 4:} Complete the needed steps in the previous discussion to characterize the out states for $t\to \pm \infty$.
       \bigskip

$\bullet$ \underline{\it Cross section, and the optical theorem:}
 \bigskip

The cross section $\sigma_{\beta\alpha}$ is defined as the number of transition events  per unite time from the initial two-particle state $|\vp_1\sigma_1\lam_1,\vp_2\sigma_2\lam_2\ra$ to the final multiparticle state $|\Phi_\beta\ra$. This rate is further normalized by the incident flux $\Phi$ times the number of targets. 

The number of transitions  $\alpha\to\beta$  is 
\begin{align}
   \label{240810.13}
 \int \prod_{i=1}^{n_\beta}\left(\frac{d^3p_i}{(2\pi)^32w(\vp_i)}\sum_{\sigma_i,\lam_i}\right) |T_{\beta\alpha}|^2\left[(2\pi)^4\delta(P_\beta-P_\alpha)\right]^2\,.
\end{align}
There is a four-Dirac delta function squared, which requires a proper interpretation. For that we take its Fourier representation
\begin{align}
\label{240810.14}
(2\pi)^4\delta(P_\beta-P_\alpha)&=\int d^4x e^{-ix(P_\alpha-P_\beta)}\xrightarrow[P_\beta=P_\alpha]{}{\cal V}{\cal T}~,
\end{align}
with ${\cal V}$ and ${\cal T}$ the total volume and time, respectively, to be sent to infinity. The infinite factor ${\cal T}$ cancel because we want the number of transitions per unite time.

Regarding the incident flux and  number of targets we have to take into account the normalization of the monoparticle states, Eq.~\eqref{240808.1}, for $\vp'=\vp$: $\parallel |\vp\sigma\lam\ra \parallel^2={\cal V}2w(p)$. Thus, the number of targets is $2w(p_2){\cal V}$ and the flux of projectiles $\Phi=(\text{relative velocity})\times (\text{density of projectiles})$ is 
\begin{align}
\label{240810.15}
\Phi&=v_{\rm rel}2w(p_1)=\left|\frac{\vp_1}{w(p_1)}-\frac{\vp_2}{w(p_2)}\right|2w(p_1)~,
\end{align}
with $w(p_i)=\sqrt{m_i^2+\vp_i^2}$. 
As a result, the differential cross section is
\begin{align}
\label{240810.16}
d\sigma_{\beta\alpha}&=\frac{ \left(\int \prod_{i=1}^{n_\beta}\frac{d^3p_i}{(2\pi)^32w(\vp_i)}\sum_{\sigma_i,\lam_i}\right)(2\pi)^4\delta(P_\beta-P_\alpha)|T_{\beta\alpha}|^2}{v_{\rm rel}4w(p_1)w(p_2)}
=dQ_\beta \frac{|T_{\beta\alpha}|^2}{4w(p_1)w(p_2)v_{\rm rel}}~.
\end{align}
In the center of mass (CM) frame, $\vp=\vp_1=-\vp_2$ and the relative velocity is
\begin{align}
\label{240810.17}
v_{\rm rel}&=\frac{|\vp|(w(p_1)+w(p_2))}{w(p_1)w(p_2)}~.
\end{align}
In this frame Eq.~\eqref{240810.16} becomes 
\begin{align}
\label{240810.18}
d\sigma_{\beta\alpha}&=\frac{ \int \prod_{i=1}^{n_\beta}\frac{d^3p_i}{(2\pi)^32w(\vp_i)}(2\pi)^4\delta(P_\beta-P_\alpha)|T_{\beta\alpha}|^2}{v_{\rm rel}4w(p_1)w(p_2)}
=dQ_\beta \frac{|T_{\beta\alpha}|^2}{4(w(p_1)+w(p_2))|\vp|}~.
\end{align}
The total cross sections is the full integration of the differential cross section
\begin{align}
  \label{240810.19}
  \sigma_\alpha&=\sumint dQ_\beta \frac{d\sigma_{\beta\alpha}}{dQ_\beta}
  \end{align}
where the sum is over all the possible types of final states or outgoing channels. 

\bigskip
{\bf Exercise 5:} Calculate the cross section at the laboratory frame, where the second particle (target) of the in state is at rest, and show that it is the same as the cross section in the CM frame.
\bigskip

We now apply the unitarity relation Eq.~\eqref{240809.15} to forward scattering, with $\beta=\alpha$ referring to the same two-body state. Since $T_{\alpha\alpha}-T^\dagger_{\alpha\alpha}=2i \Ima T_{\alpha\alpha}$, then it  follows that
\begin{align}
  \label{240810.20}
  \Ima T_{\alpha\alpha}&=\frac{1}{2}\sumint dQ_\beta |T_{\beta\alpha}|^2~.
\end{align}
Comparing the right-hand side of this equation with Eqs~\eqref{240810.18} and \eqref{240810.19} for the total cross section in the CM frame, it follows that
\begin{align}
\label{240810.21}
\Ima T_{\alpha\alpha}&=2|\vp|(w(p_1)+w(p_2))\sigma_{\alpha}~.
\end{align}
This result is known as the optical theorem.

\bigskip
$\bullet$ \underline{\it Boltzmann $H$-theorem:}
\bigskip

Let us consider the two equivalent unitarity relations of Eqs.~\eqref{240809.12} and \eqref{240809.13}. Indeed, since the left-hand sides of these equations are the same it follows that $T^\dagger T=TT^\dagger$. Applying this result for $\beta=\alpha$, and using again the label $\beta$ for the intermediate states, we have that
\begin{align}
  \label{240810.22}
  \sumint dQ_\beta |T_{\beta\alpha}|^2=\sumint dQ_\beta |T_{\alpha\beta}|^2\,.
\end{align}
Let us now define the transition rate from $|\Phi_\alpha\ra$ to a set of states $|\Phi_\beta\ra$. In general terms we have to be careful with the normalization of the monoparticle  states established not to one but to ${\cal V}2w(p)$, as follows from Eq.~\eqref{240808.1}. Because of this,  the transition rate  (probability per unite time) for the initial state to end within a range $d\beta$  of final states is
\begin{align}
\label{240810.23}
  d\Gamma(\alpha\to\beta)&=\frac{dP(\alpha\to\beta)}{{\cal T}}=\frac{dQ_\beta|T(\alpha\to\beta)|^2{\cal V}}{\prod_{i=1}^{n_\alpha}2w(p_i){\cal V}}\equiv \frac{{\cal V}dQ_\beta|T(\alpha\to\beta)|^2}{c_\alpha}\,,\\
\label{240810.24}
  c_\alpha&=(2 {\cal V})^{n_\alpha}w(p_1)\cdots w(p_{n_\alpha})~,
\end{align}
 with the factor ${\cal V}$ due to the four-dimensional Dirac delta squared, cf.~Eq.~\eqref{240810.14}, divided by ${\cal T}$ since it is a rate of change. Notice that it is not necessary to introduce any factor $c_\beta$ in Eq.~\eqref{240810.23} because this is accounted for by the definition of $dQ_\beta$, Eq.~\eqref{240809.16}. Therefore, in terms of the transitions rates we can rewrite Eq.~\eqref{240810.22} as
\begin{align}
\label{240810.25}
\sumint dQ_\beta c_\alpha\frac{d\Gamma(\alpha\to\beta)}{dQ_\beta}&=
\sumint dQ_\beta c_\beta \frac{d\Gamma(\beta\to\alpha)}{dQ_\alpha}~.
 \end{align}

Let $P_\alpha d\alpha$ be the probability that the system is in a volume $d\alpha$ of the space of multiparticle states $|\Phi_\alpha\ra$. We can actually write that
\begin{align}
\label{240810.26}
d\alpha=\int \frac{d^4P_\alpha}{(2\pi)^4}dQ_\alpha=\int\left(\prod_{i=1}^{n_\alpha}\frac{d^3p_i}{(2\pi)^32w(p_i)}\sum_{\sigma_i, \lam_i}\right)\,.
\end{align}
\footnote{Equation~\eqref{240810.26} can also be interpreted in terms of periodic conditions for a particle of momentum $\vp$ in a box of volume $V=L^3$, with $L$ the length of every side. Then, $d^3p=(2\pi)^3/V$. Therefore, $\int d^3p/[(2\pi)^3 2w(p)]$ becomes $\sum 1/[V 2w(p)]$, with the norm squared of the plane-wave state in the denominator to correct for its non-equal to 1 normalization.}
The time derivative of the probability density $P_\alpha$ has two sources. On the one hand, it increases per unite  time  due to transitions from all the other states and, 
on the other hand, it decreases due to transitions to all other states. 
Therefore,
\begin{align}
\label{240810.27}
\frac{dP_\alpha}{dt}&=\sumint d\beta  P_{\beta} \frac{d\Gamma(\beta\to\alpha)}{dQ_\alpha}(2\pi)^4\delta(P_\beta-P_\alpha)-P_\alpha\sumint dQ_\beta \frac{d\Gamma(\alpha\to\beta)}{dQ_\beta}\\
&=\sumint dQ_\beta  P_{\beta} \frac{d\Gamma(\beta\to\alpha)}{dQ_\alpha}-P_\alpha\sumint dQ_\beta \frac{d\Gamma(\alpha\to\beta)}{dQ_\beta} \,.\nn
\end{align}
Let us show that $\sumint d\alpha d P_\alpha/dt=0$. Integrating over $\alpha$ in Eq.~\eqref{240810.27}, and rewriting in the first term on the right-hand side  $\int d\alpha(2\pi)^4\delta(P_\beta-P_\alpha)=\int dQ_\alpha$,  one has 
\begin{align}
  \label{240810.28}
  \sumint d\alpha \frac{dP_\alpha}{dt}&=
  \sumint dQ_\alpha \sumint d\beta  P_{\beta} \frac{d\Gamma(\beta\to\alpha)}{dQ_\alpha}
  -\sumint d\alpha P_\alpha\sumint dQ_\beta \frac{d\Gamma(\alpha\to\beta)}{dQ_\beta}=0\,,
\end{align}
by simply exchanging of the integration variables in one of the two terms on the right-hand side of this equation. 

Rate of change of the entropy $S=-\sumint d\alpha P_\alpha \ln(P_\alpha/c_\alpha)$, with $c_\alpha$ given in Eq.~\eqref{240810.24}: Inside the $\ln$ the probability density $P_\alpha$ is divided by $c_\alpha$ in order to remove the enhancement factor in $P_\alpha$ due to having monoparticle states with a norm squared equal to $2{\cal V}w(p)$, as discussed above. This is not done for the factor $P_\alpha$ outside the $\ln$ because it is multiplied by $d\alpha$, which automatically compensates such normalization factor for the states. Proceeding with the time derivative of the entropy, and taking into account Eq.~\eqref{240810.28}, we have
\begin{align}
  \label{240810.27b}
  \frac{dS}{dt}=-\sumint d\alpha \ln\left(\frac{P_\alpha}{c_\alpha}\right) \frac{dP_\alpha}{dt}
=-\sumint d\alpha \ln\left(\frac{P_\alpha}{c_\alpha}\right)\left(\sumint dQ_\beta P_\beta \frac{d\Gamma(\beta\to\alpha)}{dQ_\alpha}-P_\alpha\sumint dQ_\beta\frac{d\Gamma(\alpha\to\beta)}{dQ_\beta}\right)
\end{align}
Since total four-momentum is conserved we can replace $d\alpha\to dQ_\alpha$ and $dQ_\beta\to d\beta$ in the first term on the right-hand side of the last term, and rewrite Eq.~\eqref{240810.27b} as
\begin{align}
  \label{240810.28b}
  \frac{dS}{dt}=-\sumint dQ_\alpha \ln\left(\frac{P_\alpha}{c_\alpha}\right)\sumint d\beta P_\beta \frac{d\Gamma(\beta\to\alpha)}{dQ_\alpha}+\sumint d\alpha  P_\alpha\ln\left(\frac{P_\alpha}{c_\alpha}\right)\sumint dQ_\beta\frac{d\Gamma(\alpha\to\beta)}{dQ_\beta}~.
\end{align}
Next, exchange the integration variables in the second term on the right-hand side and it results
\begin{align}
\label{240810.29}
\frac{dS}{dt}&=-\sumint dQ_\alpha \ln\left(\frac{P_\alpha}{c_\alpha}\right)\sumint d\beta P_\beta \frac{d\Gamma(\beta\to\alpha)}{dQ_\alpha}
+\sumint d\beta P_\beta\ln\left(\frac{P_\beta}{c_\beta}\right) \sumint dQ_\alpha\frac{d\Gamma(\beta\to\alpha)}{dQ_\alpha}\\
&=\sumint d\beta P_\beta\sumint dQ_\alpha \frac{\Gamma(\beta\to\alpha)}{dQ_\alpha} \ln\left(\frac{P_\beta c_\alpha}{P_\alpha c_\beta}\right)~.\nn
\end{align}
Now, for any positive quantities $x$ and $y$, $y\ln(y/x)\geq y-x$.\footnote{For $y\geq x$ the inequality is clear. For $x=y$ one has the equality to zero. For $x>y$ the derivative with respect to $x$ of the left-hand side minus the right-hand one is $-y/x+1>0$ and the inequality is fulfilled.} Taking this result into Eq.~\eqref{240810.29},
\begin{align}
  \label{240810.30}
\frac{dS}{dt}\geq \sumint d\beta \sumint dQ_\alpha \frac{\Gamma(\beta\to\alpha)}{dQ_\alpha} c_\beta\left(\frac{P_\beta}{c_\beta}-\frac{P_\alpha}{c_\alpha}\right)
\end{align}
Exchanging the integration variables in the second term,
\begin{align}
\label{240810.31}
\frac{dS}{dt}\geq \sumint d\beta \frac{P_\beta}{c_\beta}\sumint dQ_\alpha  \left(c_\beta \frac{\Gamma(\beta\to\alpha)}{dQ_\alpha}-c_\alpha\frac{\Gamma(\alpha\to\beta)}{dQ_\beta}\right)=0~,
\end{align}
because of Eq.~\eqref{240810.25}. This is the Boltzmann $H$-theorem, for the increase in entropy of a closed system. This theorem is usually deduced either by employing the Born approximation for reversible dynamics, in which $|T_{\beta\alpha}|^2$ is symmetric in $\alpha$ and $\beta$ and then $c_\beta d\Gamma(\beta\to\alpha)/dQ_\alpha=c_\alpha d\Gamma(\alpha\to\beta)/dQ_\beta$, or by assuming time-reversal invariance. This implies that $|T_{\beta\alpha}|^2$ is unchanged if $\alpha$ and $\beta$ are interchanged and also if momenta and spins are reversed (we discuss time reversal below in Sec.~\ref{sec.240811.1}). However, neither the Born approximation nor time-reversal invariance are exact, so that it is reassuring that one can deduce the $H$-theorem by only relying on unitarity. 

The total entropy stops growing when the probability $P_\alpha/c_\alpha$ becomes a function of conserved magnitudes, such as total energy and charge.  In this the conservation laws require that $d\Gamma(\beta\to\alpha)/dQ_\alpha$ vanishes unless $P_\alpha/c_\alpha=P_\beta/c_\beta$. Substituting then $P_\beta=P_\alpha c_\beta/c_\alpha$  into the second line of Eq.~\eqref{240810.27}, and using Eq.~\eqref{240810.25}, it follows that $dP_\alpha(t)/dt$=0. Again this is a consequence of the general principle of unitarity, and not of the Born approximation or time-reversal invariance.

\section{Two-body scattering. Partial-wave amplitude expansions}
\def\theequation{\arabic{section}.\arabic{equation}}
\setcounter{equation}{0}

\bigskip
$\bullet$ \underline{\it Two-body scattering, and Mandelstam variables:}
\bigskip

In two-body scattering, two particles form the initial and final states:
\begin{align}
  \label{240811.2b}
  A(p_1)+B(p_2)\to C(p_3)+D(p_4)\,,
\end{align}

In a two-body scattering process one has 4 momenta which implies 12 variables. Four of them can be fixed in terms of the others by the conservation of total four momentum, then we are left with 8. Additionally, a Lorentz transformation involves 6 parameters, so that from these remnant 8 variables we are left only with 2 Lorentz invariants.

The Mandelstam variables for the scattering in Eq.~\eqref{240811.2b} are defined as
\begin{align}
  \label{240811.2c}
  s&=(p_1+p_2)^2~,\\
  t&=(p_1-p_3)^2~,\nn\\
  u&=(p_1-p_4)^2~.\nn
\end{align}
It is straightforward to show that  $s+t+u=m_1^2+m_2^2+m_3^2+m_4^2$\,, and only two of the Mandelstam variables are independent.

In the CM frame $\vp_1+\vp_2=0$ so that
\begin{align}
  \label{240811.2d}
  s=(w(p_1)+w(p_2))^2=E^2~.
\end{align}
In terms of the scattering angle $\theta$, with $\hat{\vp}_1\cdot \hat{\vp}_3=\cos\theta$, we have in the equal mass case ($m_i=m$, $i=1,\ldots,4$)
\begin{align}
  \label{240811.2e}
  s&=4(m^2+\vp^2)~,\\
  t&=-2\vp^2(1-\cos\theta)=-4\vp^2\sin^2\frac{\theta}{2}~,\nn\\
  u&=-2\vp^2(1+\cos\theta)=-4\vp^2\cos^2\frac{\theta}{2}~.\nn
\end{align}

\bigskip
$\bullet$ \underline{\it Normalization, element of phase space, cross section and unitarity:}
\bigskip

 In the normalization of the two-body states we extract the Dirac delta function of conservation of total four-momentum $(2\pi)^4\delta(P'-P)$. These sates are built as the direct product of two monoparticle states with normalization given in Eq.~\eqref{240808.1}. Then, 
\begin{align}
  \label{240811.3}
  \la \vp'_1\si'_1\lam'_1,\vp'_2\si'_2\lam'_2|\vp_1\si_1\lam_1,\vp_2\si_2\lam_2\ra
  &=(2\pi)^4(2\pi)^24w(p_1)w(p_2)\delta(\vp'_1-\vp_1)\delta(\vp'_2-\vp_2)\delta_{\si'_1\si_1}\delta_{\si'_2\si_2}\delta_{\lam'_1\lam_1}\delta_{\lam'_2\lam_2}\,.
\end{align}
The  change of variables from $\{\vp_1,\vp_2\}$ to the set of variables made by the total momentum $\vP=\vp_1+\vp_2$ and the relative momentum $\vp=(\vp_1-\vp_2)/2$ has a Jacobian equal to one. Therefore,
\begin{align}
  \label{240811.4}
  \la \vp'_1\si'_1\lam'_1,\vp'_2\si'_2\lam'_2|\vp_1\si_1\lam_1,\vp_2\si_2\lam_2\ra
  &=(2\pi)^4 \delta(\vP'-\vP)(2\pi)^24w(p_1)w(p_2)\delta(\vp'-\vp)\delta_{\si'_1\si_1}\delta_{\si'_2\si_2}\delta_{\lam'_1\lam_1}\delta_{\lam'_2\lam_2}~.
\end{align}
Now, writing $\delta(\vp'-\vp)=|\vp|^{-2}\delta(|\vp'|-|\vp|)\delta(\hat{\vp}'-\hat{\vp})$, we have in the CM frame
\begin{align}
\label{240811.5a}
\delta(|\vp'|-|\vp|)=\frac{dE}{d|\vp|}\delta(E'-E)=\frac{\delta(E'-E)E|\vp|}{w(p_1)w(p_2)}\,.
\end{align}
Thus,
\begin{align}
\label{240811.6b}
\la \vp'_1\si'_1\lam'_1,\vp'_2\si'_2\lam'_2|\vp_1\si_1\lam_1,\vp_2\si_2\lam_2\ra
 &=(2\pi)^4 \delta(P'-P)\,\frac{16\pi^2\sqrt{s}}{|\vp|}\delta(\hat{\vp}'-\hat{\vp})\delta_{\si'_1\si_1}\delta_{\si'_2\si_2}\delta_{\lam'_1\lam_1}\delta_{\lam'_2\lam_2}\,.
\end{align}

It should be clear that the phase-space element is just the inverse of Eq.~\eqref{240811.6b}, after removing  $(2\pi)^4\delta(P'-P)$: 
\begin{align}
  \label{240811.5}
dQ&=\int\frac{d^3p_1}{(2\pi)^32w(p_1)}\frac{d^3p_2}{(2\pi)^32w(p_2)}(2\pi)^4\delta(P'-P)=\frac{|\vp|d\hat{\vp}}{16\pi^2\sqrt{s}}\,,
\end{align}
with the result expressed in terms of $\vp$ in the CM frame. 

\bigskip
{\bf Exercise 6:} Show explicitly that the phase space element  is given by Eq.~\eqref{240811.5}.
\bigskip

In the following, the two-body states are indicated by their CM momentum, together with the other discrete quantum numbers.

Particularizing Eq.~\eqref{240810.18} to the present case of two-body scattering with $dQ$ given in Eq.~\eqref{240811.5}, one has for the differential cross section:
\begin{align}
  \label{240811.6}
\frac{d\sigma}{d\Omega}&=\frac{\left|\la \vp',\si'_1\lam'_1,\si'_2\lam'_2|T|\vp,\si_1\lam_1,\si_2\lam_2\ra\right|^2}{64\pi^2 s}\frac{|\vp'|}{|\vp|}\,,
\end{align}
with $d\Omega\equiv d\hat{\vp}$. The unitarity relation Eq.~\eqref{240809.17} for two-body scattering particularizes into
\begin{align}
  \label{240811.7}
&  \la \vp_j',\si'_{j1}\lam'_{j1},\si'_{j2}\lam'_{j2}|T|\vp,\si_{i1}\lam_{i1},\si_{i2}\lam_{i2}\ra-
  \la \vp_j',\si'_{j1}\lam'_{j1},\si'_{j2}\lam'_{j2}|T^\dagger|\vp,\si_{i1}\lam_{i1},\si_{i2}\lam_{i2}\ra\\
&  =\sum_n\!\!\sum_{\scriptsize{\begin{array}{l}\si''_{n1},\si''_{n2}\\\lam''_{n1},\lam''_{n2}\end{array}}}\!\!\int\frac{|\vp_n''|\,d\hat{\vp}_n''}{16\pi^2\sqrt{s}}
  \la \vp_{j}',\si'_{j1}\lam'_{j1},\si'_{j2}\lam'_{j2}|T^\dagger|\vp_n'',\si''_{n1}\lam''_{n1},\si''_{n2}\lam''_{n2}\ra \la\vp_n'',\si''_{n1}\lam''_{n1},\si''_{n2}\lam''_{n2}|T|\vp_i,\si_{i1}\lam_{i1},\si_{i2}\lam_{i2}\ra\,,\nn
\end{align}
where the sum over different types of intermediate states (channels) is indicated by the subscript $n$. To make clear that the final and initial states can correspond to different channels, we have also labelled them with the subscripts $j$ and $i$, respectively. This  sum  is restricted to those channels that are open for the total energy $E$.\footnote{A channel $c$ at a given total energy $E$ is open if $E>m_{n1}+m_{n2}$.}  One can also write the previous unitarity relation without the appearance of $T^\dagger$ as
\begin{align}
  \label{240811.8}
&  \la \vp_j',\si'_{j1}\lam'_{j1},\si'_{j2}\lam'_{j2}|T|\vp,\si_{i1}\lam_{i1},\si_{i2}\lam_{i2}\ra -
  \la \vp,\si_{i1}\lam_{i1},\si_{i2}\lam_{i2} |T| \vp_j',\si'_{j1}\lam'_{j1},\si'_{j2}\lam'_{j2}\ra^*\\
  &=\sum_n\!\!\sum_{\scriptsize{\begin{array}{l}\si''_{n1},\si''_{n2}\\\lam''_{n1},\lam''_{n2}\end{array}}}\!\!\int\frac{|\vp_n''|\,d\hat{\vp}_n''}{16\pi^2\sqrt{s}}
  \la \vp_n'',\si''_{n1}\lam''_{n1},\si''_{n2}\lam''_{n2}|T|\vp_j',\si'_{j1}\lam'_{j1},\si'_{j2}\lam'_{j2}\ra^*\la \vp_n'',\si''_{n1}\lam''_{n1},\si''_{n2}\lam''_{n2}|T|\vp_i,\si_{i1}\lam_{i1},\si_{i2}\lam_{i2}\ra\,.\nn
\end{align}

\bigskip
$\bullet$ \underline{\it Rotational invariance}
\bigskip

According to writing convenience we denote a rotation either by a vector $\vf=\phi \hat{\vn}$, where $\phi$ is the rotation angle around the axis of rotation $\hat{\vn}$, or in terms of the three Euler angles $(\phi,\theta,\gamma)$. For the latter, $\phi$ and $\gamma$ are the angles of rotations around the $z$ axis, and $\theta$ is the angle of a rotation around the $y$ axis. Then, we write $R(\vf)$ or $R(\phi,\theta,\gamma)$.  We always indicate the rotation operator by $R$ independently of the vector space on which it acts. It is well known that rotations correspond to a unitary operator, $R(\vf)^\dagger=R(\vf)^{-1}$. 

Rotational invariance follows if the Hamiltonian is invariant under rotations
\begin{align}
  \label{240811.8}
  R(\vf) V R(\vf)^{-1}&=V\,,
  \end{align}
as it should be the case for a closed system. Then, the $T$-matrix operator is also invariant under rotations. From the Lippmann-Schwinger equation \eqref{240810.11},
\begin{align}
  \label{240811.9}
  RTR^{-1}&=-RVR^{-1}+RVR^{-1} \, R(E+i\ep-H_0)^{-1}R^{-1} \, RTR^{-1}\\
  &=-V+V(E+i\ep-H_0)RTR^{-1}~.\nn
\end{align}
Therefore,
\begin{align}
  \label{240811.10}
  R(\vf)TR(\vf)^{-1}=T\,. 
\end{align}

\subsection{Partial-wave amplitudes in the $\ell SJ$ basis}
\label{sec.240811.1}

\bigskip
$\bullet$ \underline{\it States, standard Lorentz transformation, and transformation under rotations:}
\bigskip

The monoparticle states $|\vp\sigma\ra$ for a particle of mass $m>0$ and spin $s$ are defined by the action of a standard Lorentz transformation $U(\vp)$ acting on the rest-frame states $|\vce\sigma\ra$ as (we only indicate the labels $\vp$ and $\sigma$ since they are the only relevant ones for the discussions that follow)
\begin{align}
  \label{240811.11}
  |\vp\sigma\ra&=U(\vp)|\vce\sigma\ra~,\\
  U(\vp)&=R(\phi,\theta,0)B_z(p)R(\phi,\theta,0)^{-1}~,\nn
\end{align}
with $\sigma=-s,\ldots,s$, the third component of spin of the particle in its rest frame. 
Here, $\phi$ and $\theta$ are, respectively, the azimuthal and polar angles of $\hat{\vp}=(\sin\theta\cos\phi,\sin\theta\cos\phi,\cos\theta)$. To simplify the writing we also denote $R(\phi,\theta,0)=R(\hvp)$. This rotation takes the axis $\hvz$ into $\hvp$
\begin{align}
  \label{240811.12}
  R(\hvp)\hvz=\hvp~.
  \end{align}
In turn, $B_z(p)$ is a Lorentz boost along the $z$ axis that takes the particle from rest to having a momentum $|\vp|\hvz$:

\begin{align}
  \label{240811.13}
  B_z(p)\left(\begin{array}{c}m \\ 0 \\ 0 \\ 0 \end{array}
  \right)&=\gamma \left(
  \begin{array}{llll}
    1 & 0 & 0 & v \\
    0 & 1 & 0 & 0 \\
    0 & 0 & 1 & 0 \\
    v & 0 & 0 & 1
    \end{array}
  \right)\left(\begin{array}{c}m \\ 0 \\ 0 \\ 0 \end{array}
  \right)=
  \left(
  \begin{array}{l}
    w(p)\\0\\0\\|\vp|
    \end{array}
  \right)\,,
\end{align}
with $\gamma=1/\sqrt{1-v^2}$ and $v=|\vp|/w(p)$.

Let us determine how $|\vp\sigma\ra$ as defined in Eq.~\eqref{240811.11} transforms under a rotation $R(\vf)$. The transformed momentum $R(\vf)\vp$ is denoted as $\vp'$ and, of course, $|\vp'|=|\vp|$. Then, 
\begin{align}
  \label{240811.14}
  R(\vf)|\vp\sigma\ra&=U(\vp')U(\vp')^{-1}R(\vf)U(\vp)|\vce\sigma\ra\,.
\end{align}
Let us show that $U(\vp')^{-1}RU(\vp)=R$:
\begin{align}
  \label{240811.15}
U(\vp')^{-1}R(\vf)U(\vp)&=R(\hvp')B_z(p)^{-1}R(\hvp')^{-1}R(\vf)R(\hvp)B_z(p)R(\hvp)^{-1}\,.
\end{align}
Acting over $\hvz$
\begin{align}
  \label{240811.16}
R(\hvp')^{-1}R(\vf)R(\hvp)\hvz=R(\hvp')^{-1}R(\vf)\hvp=R(\hvp')^{-1}\hvp'=\hvz\,.
\end{align}
Therefore, $R(\hvp')^{-1}R(\vf)R(\hvp)$ is a rotation around the $z$ axis,
\begin{align}
  \label{240811.16}
  R(\hvp')^{-1}R(\vf)R(\hvp)&=R(\varphi \hvz)~,\\
  R(\hvp')&=R(\vf)R(\hvp)R(\varphi \hvz)^{-1}~.\nn
\end{align}
We take this result to Eq.~\eqref{240811.15}
\begin{align}
  \label{240811.17}
  U(\vp')^{-1}R(\vf)U(\vp)&=R(\hvp')B_z(p)^{-1}R(\varphi\hvz)B_z(p)R(\hvp)^{-1}=R(\vf)~,
\end{align}
where we have taken into account successively the fact that $B_z(p)$ commutes with $R(\varphi\hvz)$ and then the second line  of Eq.~\eqref{240811.16}. Coming back to Eq.~\eqref{240811.14} we finally have 
\begin{align}
  \label{240811.18}
R(\vf)|\vp\sigma\ra=\sum_{\sig'=-s}^{s}D^{(s)}_{\sig'\sig}(R(\vf))|\vp'\sig'\ra~,
\end{align}
where $D^{(s)}_{\sig'\sig}(R(\vf))$ is the rotation matrix corresponding to the representation of spin $s$ of the rotation group for the rotation $R(\vf)$. 

For a two-particle state $|\vp,\sig_1\sig_2\ra=|\vp\sig_1\ra\otimes|-\vp\sig_2\ra$  Eq.~\eqref{240811.18} acts multiplicatively for every monoparticle state
\begin{align}
  \label{240811.19}
R(\vf)|\vp,\sigma_1\sigma_2\ra=\sum_{\sig_1'=-s_1}^{s_1}\sum_{\sig_2'=-s_2}^{s_2}D^{(s_1)}_{\sig'_1\sig_1}(R(\vf))D^{(s_2)}_{\sig'_2\sig_2}(R(\vf))|\vp',\sig'_1\sig'_2\ra~.
  \end{align}

\bigskip
$\bullet$ \underline{\it  States  $|p \ell m,\sig_1\sig_2\ra$ with definite orbital angular momentum, and transformation under rotations:}
\bigskip

Let us define the states $|p\ell m,\sig_1\sig_2\ra$  with definite orbital angular momentum
\begin{align}
  \label{240811.20}
|p\ell m,\sig_1\sig_2\ra&=\frac{1}{\sqrt{4\pi}}\int d\hvp Y_\ell^m(\hvp)|\vp,\sig_1\sig_2\ra\,.
\end{align}

It is important to notice that these states transform as the direct product of the representations $\ell$, $s_1$ and $s_2$ of the rotation group, as we show next:
\begin{align}
  \label{240811.21}
  R(\vf)|p\ell m,\sig_1\sig_2\ra&=\sum_{\sig'_1,\sig'_2} D^{(s_1)}_{\sig'_1\sig_1}(R(\vf))D^{(s_2)}_{\sig'_2\sig_2}(R(\vf)) \frac{1}{\sqrt{4\pi}}\int d\hvp Y_\ell^m(\hvp)|R(\vf)\vp,\sig'_1\sig'_2\ra~.
\end{align}
We now change to the integration variables of $\hvp'=R(\vf)\hvp$ and take into account that the angular measure is invariant under rotations. Then, 
\begin{align}
\label{240811.22}
R(\vf)|p\ell m,\sig_1\sig_2\ra&=\sum_{\sig'_1,\sig'_2} D^{(s_1)}_{\sig'_1\sig_1}(R(\vf))D^{(s_2)}_{\sig'_2\sig_2}(R(\vf))\frac{1}{\sqrt{4\pi}}\int d\hvp' Y_\ell^m(R(\vf)^{-1}\hvp')|\vp',\sig'_1\sig'_2\ra~.
\end{align}
Next, we take into account the property of the spherical harmonics under a rotation of their arguments
\begin{align}
\label{240811.23}
Y_\ell^m(R(\vf)^{-1}\hvp')=\la R(\vf)^{-1}\hvp'|\ell m\ra=\la\hvp'|R(\vf)|\ell m\ra=\sum_{m'}D^{(\ell)}_{m'm}(R(\vf))Y_\ell^{m'}(\hvp')\,.
\end{align}
Using this property in Eq.~\eqref{240811.22} we end with the stated result
\begin{align}
\label{240811.24}
R(\vf)|p\ell m,\sig_1\sig_2\ra&=\sum_{\sig'_1,\sig'_2,m'}D^{(s_1)}_{\sig'_1\sig_1}(R(\vf))D^{(s_2)}_{\sig'_2\sig_2}(R(\vf))D^{(\ell)}_{m'm}(R(\vf))\frac{1}{\sqrt{4\pi}}\int d\hvp Y_\ell^{m'}(\hvp)|\vp,\sig'_1\sig'_2\ra~,\\
&=\sum_{\sig'_1,\sig'_2,m'}D^{(s_1)}_{\sig'_1\sig_1}(R(\vf))D^{(s_2)}_{\sig'_2\sig_2}(R(\vf))D^{(\ell)}_{m'm}(R(\vf))|p\ell m',\sig'_1\sig'_2\ra\,,\nn
\end{align}
where the integration variables $\hvp'$ have been relabelled without the prime in the last term of the right-hand side of the first line.

Equation~\eqref{240811.20} can be inverted by taking into account the completeness relation of spherical harmonics
\begin{align}
  \label{240812.1}
\sum_{\ell=0}^\infty\sum_{m=-\ell}^\ell Y_\ell^m(\hvp')Y_\ell^m(\hvp)^*=\delta(\hvp'-\hvp)~.
  \end{align}
This relation reflects a resolution of the identity for the angular variables $\la \hvp'|\hvp\ra=\sum_{\ell,m}\la \hvp'|\ell m\ra\la\ell m|\hvp\ra$. Then, multiplying Eq.~\eqref{240811.20} by $Y_\ell^m(\hvp')^*$ and summing over $\ell$ and $m$
\begin{align}
  \label{240812.2}
|\vp,\sig_1\sig_2\ra&=\sqrt{4\pi}\sum_{\ell=0}^\infty \sum_{m=-\ell}^\ell Y_\ell^m(\hvp)^*|p\ell m,\sig_1\sig_2\ra~,
  \end{align}
with some exchange for the name of integration variables to remove the prime in the final expression.

\bigskip
$\bullet$ \underline{\it $\ell S J$ states with definite total angular momentum}
\bigskip

Since the states $|p\ell m,\sig_1\sig_2\ra$ transform under rotations as $\boldsymbol{\ell}\otimes \boldsymbol{s_1}\otimes \boldsymbol{s_2}$, cf. Eq.~\eqref{240811.24}, we can combine the spin representations into a total spin one $\boldsymbol{S}$, and the latter together with $\boldsymbol{\ell}$ to give the representation with total angular momentum $\boldsymbol{J}$. Therefore, we define the states
\begin{align}
  \label{240812.3}
|pJ\mu,\ell S \ra&=\sum_{\sig_1,\sig_2,m}(\sig_1\sig_2 \sig|s_1s_2S)(m\sig\mu|\ell S J)|p\ell m,\sig_1\sig_2\ra\,,
  \end{align}
with $(m_1m_2m_3|j_1j_2j_3)$ the Clebsch-Gordan coefficient for the combination of angular momenta $\boldsymbol{j_1}\oplus\boldsymbol{j_2}=\boldsymbol{j_3}$. Inserting Eq.~\eqref{240811.20} for $|p\ell m,\sig_1\sig_2\ra$ in Eq.~\eqref{240812.3}
\begin{align}
  \label{240812.4}
|pJ\mu,\ell S \ra&=\frac{1}{\sqrt{4\pi}}\sum_{\sig_1,\sig_2,m}(\sig_1\sig_2 \sig|s_1s_2S)(m\sig\mu|\ell S J)\int d\hvp Y_\ell^m(\hvp)|\vp,\sig_1\sig_2\ra\,.
\end{align}
Taking into account the orthonormality properties of the Clebsch-Gordan coefficients
\begin{align}
  \label{240812.15}
  \sum_{j_3}\sum_{m_3=-j_3}^{j_3}(m'_1m'_2m_3|j'_1j'_2j_3)(m_1m_2m_3|j_1j_2j_3)=\delta_{j'_1j_1}\delta_{j'_2j_2}\delta_{m'_1m_1}\delta_{m'_2m_2}\,,
\end{align}
let us work out the normalization of the $\ell S J$ states. From Eqs.~\eqref{240812.4}, \eqref{240812.15} and \eqref{240811.6b}:
\begin{align}
  \label{240813.1}
\la p'_j J'_j\mu'_j,\ell'_jS'_j|p_i J_i\mu_i,\ell_iS_i\ra&=\delta_{ij}\delta_{S'_jS_i}\delta_{\ell'_j\ell_i}\delta_{J'_jJ_i}\delta_{\mu'_j\mu_i}\frac{4 \pi \sqrt{s}}{|\vp_i|}\,,
\end{align}
with $i$ and $j$ referring to the type of channel involving different particles. For instance, in $\pi\pi$ scattering coupled to $K\bar{K}$ we have $i,j=1$ for $\pi\pi$ and 2 for $K\bar{K}$.\footnote{We are using here an isospin formalism so that we take as the same particles those belonging to the same isospin multiplet.}  
In Eq.~\eqref{240813.1}  we have used the orthogonality property of the spherical harmonics
\begin{align}
  \label{240812.11}
  \int d\Omega Y_{\ell'}^{m'}(\phi,\theta)Y_\ell^m(\phi,\theta)^*=\delta_{\ell'\ell}\delta_{m'm}\,,
  \end{align}

\bigskip
{\bf Exercise 7:} Complete the steps to deduce Eq.~\eqref{240813.1}.
\bigskip

Using Eq.~\eqref{240812.15} we can easily invert Eq.~\eqref{240812.3}
\begin{align}
  \label{240812.5}
|p\ell m,\sig_1\sig_2\ra&=\sum_{\mu,J,\sig,S}(\sig_1\sig_2\sig|s_1s_2S)(m\sigma\mu|\ell S J)|pJ\mu,\ell S\ra~. 
\end{align}

\bigskip
{\bf Exercise 8:} Check Eq.~\eqref{240812.5} by plugging it into Eq.~\eqref{240812.3}.
\bigskip

Substituting Eq.~\eqref{240812.5} into the right-hand side of Eq.~\eqref{240812.2} we then have
\begin{align}
\label{240812.6}
|\vp,\sig_1\sig_2\ra&=\sqrt{4\pi}\sum_{\ell,m}\sum_{\mu,J,\sig,S}(\sig_1\sig_2\sig|s_1s_2S)(m\sigma\mu|\ell S J)Y_\ell^m(\vp)^*|pJ\mu,\ell S\ra\,.
\end{align}

\bigskip
$\bullet$ \underline{\it Partial-wave amplitudes, and calculation:}
\bigskip

A partial-wave amplitude (PWA) is the matrix element of the $T$-matrix operator between two  $\ell SJ$ states
\begin{align}
  \label{240812.7}
T^{J}_{\ell_j S_j,\ell_i S_i}\equiv \la p_jJ\mu,\ell_j S_j|T|p_i J\mu,\ell_i S_i\ra\,,
\end{align}
where the indices $i$ and $j$ refers to the different possible channels. We notice that this matrix element conserves $J$ due to rotational symmetry, Eq.~\eqref{240811.10}, and is independent of $\mu$ because of the Wigner-Eckart theorem.

To express a PWA in terms of the $T$-matrix elements between two-particle plane waves we make use of rotational invariance. Because of the latter, we first notice that we can always  take the initial state along the $z$ axis, by applying $R(\vp_i)^{-1}$ to $\la \vp_j,\sig_{j1}\sig_{j2}|T|\vp_i,\sig_{i1}\sig_{i2}\ra$ as
\begin{align}
  \label{240812.8}
  &  \la \vp_j,\sig_{j1}\sig_{j2}|T|\vp_i,\sig_{i1}\sig_{i2}\ra=\la \vp_j,\sig_{j1}\sig_{j2}|R(\hvp_i) T R(\hvp_i)^{-1}|\vp_i,\sig_{i1}\sig_{i2}\ra\\
  &=\sum D^{(s_{j1})}_{\sig'_{j1}\sig_{j1}}(R(\hvp_i)^{-1})^*\cdots D^{(s_{i2})}_{\sig'_{i2}\sig_{i2}}(R(\hvp_i)^{-1})
  \la  \vp'_j,\sig'_{j1}\sig'_{j2}|T||\vp_i|\hvz,\sig'_{i1}\sig'_{i2}\ra\,,\nn
\end{align}
with $\vp'_j=R(\hvp_i)^{-1}\vp_j$. 
We can again use rotational invariance and take a rotation around the $z$ axis of angle $-\phi'_j$, $R(\phi'_j\hvz)^{-1}$, in order to take $R(\phi'\hvz)^{-1}\vp'_j$ into the $xz$ plane, while the $\hvz$ direction of the momentum in the initial state is preserved. The angle $\phi'_j$ is the azimuthal angle of $\vp'_j$ that after the rotation becomes $\vp'_{xzj}=|\vp_j|(\sin\theta'_j,0,\cos\theta'_j)$:
\begin{align}
  \label{240812.9}
\la \vp'_j,\sig_{j1}\sig_{j2}|T|\vp_i|\hvz,\sig_{i1}\sig_{i2}\ra&
  =\la \vp'_j,\sig_{j1}\sig_{j2}|R(\phi'_j\hvz)TR(\phi'_j\hvz)^{-1}||\vp_i|\hvz,\sig_{i1}\sig_{i2}\ra\\
  &=e^{-i\sig_{j1}\phi'_j}\cdots e^{i\sig_{i2}\phi'_j} \la \vp'_{xzj},\sig_{j1}\sig_{j2}|T||\vp_i|\hvz,\sig_{i1}\sig_{i2}\ra\,.\nn
\end{align}
From the deduction chain along Eqs.~\eqref{240812.8} and \eqref{240812.9} it is clear that a PWA should be expressible in terms of the matrix elements $\la \vp_{xzj},\sig_{j1}\sig_{j2}|T||\vp_i|\hvz,\sig_{i1}\sig_{i2}\ra$ with the initial momentum along the $z$ axis and the final one contained in the $xz$ plane. We next show it explicitly. We denote the polar angle of $\vp_{xzj}$ by $\theta$, and we also indicate the arguments of a spherical harmonic in terms of the azimuthal and polar angles $Y_\ell(\phi,\theta)$. We recall that $Y_\ell^m(0,\theta)$ is real and that $Y_\ell^m(0,0)=\delta_{m0}\sqrt{(2\ell+1)/4\pi}$. For simplicity in the writing, we remove the subscripts $j$ and $i$ and, for clarity, we explicitly show the spin of each particle. Using Eq.~\eqref{240812.6} we have
\begin{align}
  \label{240812.10}
  \la \vp'_{xz},\sig'_{1}\sig'_{2},s'_1s'_2|T||\vp|\hvz,\sig_{1}\sig_{2},s_1s_2\ra&=\sum \sqrt{4\pi(2\ell+1)} Y_{\ell'}^{m'}(0,\theta)(\sig'_1\sig'_2\sig'|s'_1s'_2S')(m'\sig'\mu|\ell'S'J)(\sig_1\sig_2\sig|s_1s_2S)\nn\\
  &\times (0\sig\mu|\ell S J)\la p'J\mu,\ell'S',s'_1s'_2|T|pJ\mu,\ell S,s_1s_2\ra\,.
\end{align}
Let us notice that in the previous sums only $m'=\sigma-\sigma'$ gives contribution because $\mu=\sigma$. Note also that $\sigma=\sigma_1+\sigma_2$ and $\sig'=\sig'_1+\sig'_2$. 
We now take advantage of the orthogonality properties of the spherical harmonics, Eq.~\eqref{240812.11}, and multiply Eq.~\eqref{240812.10} by $Y_{\ell''}^{\sigma-\sigma'}(0,\theta)$ and integrate
\begin{align}
  \label{240812.12}
 &\int_{-1}^{+1}d\cos\theta Y_{\ell''}^{\sigma-\sigma'}(0,\theta) \la \vp'_{xz},\sig'_{1}\sig'_{2},s'_1s'_2|T||\vp|\hvz,\sig_{1}\sig_{2},s_1s_2\ra=\sum \sqrt{4\pi(2\ell+1)}\int_{-1}^{+1}\!\!\!d\cos\theta Y_{\ell''}^{\sigma-\sigma'}(0,\theta) Y_{\ell'}^{m'}(0,\theta)\nn\\
  &\times (\sig'_1\sig'_2\sig'|s'_1s'_2S')(m'\sig'\mu|\ell'S'J)(\sig_1\sig_2\sig|s_1s_2S)(0\sig\mu|\ell S J) \la p'J\mu,\ell'S',s'_1s'_2|T|pJ\mu,\ell S,s_1s_2\ra\,.
\end{align}

Now the trick is to notice that there is only contribution for $m'=\sigma-\sigma'$, so that we can write
\begin{align}
  \label{240812.12b}
  \int_{-1}^{+1}d\cos\theta Y_{\ell''}^{\sigma-\sigma'}(0,\theta) Y_{\ell'}^{m'}(0,\theta)\delta_{\sigma-\sigma',m'}&=
  \frac{1}{2\pi}\int_0^{2\pi}d\phi\int_{-1}^{+1}d\cos\theta Y_{\ell''}^{\sigma-\sigma'}(\phi,\theta)^* Y_{\ell'}^{m'}(\phi,\theta)\delta_{\sigma-\sigma',m'}\\
  &=\frac{\delta_{\ell''\ell'}\delta_{\sigma-\sigma',m'}}{2\pi}\,,\nn
\end{align}
using also Eq.~\eqref{240812.11}. Taking this result to Eq.~\eqref{240812.12},
\begin{align}
  \label{240812.14}
  &\int_{-1}^{+1}d\cos\theta Y_{\ell''}^{\sigma-\sigma'}(0,\theta) \la \vp'_{xz},\sig'_{1}\sig'_{2},s'_1s'_2|T||\vp|\hvz,\sig_{1}\sig_{2},s_1s_2\ra=\sum \sqrt{\frac{2\ell+1}{\pi}} (\sig'_1\sig'_2\sig'|s'_1s'_2S')(m'\sig'\mu|\ell''S'J)\nn\\
  &\times (\sig_1\sig_2\sig|s_1s_2S)(0\sig\mu|\ell S J) \la p'J\mu,\ell''S',s'_1s'_2|T|pJ\mu,\ell S,s_1s_2\ra\,.
\end{align}
Let us continue by applying the orthonormality properties of the Clebsch-Gordan coefficients, Eq.~\eqref{240812.15}, first for the spin indices 
\begin{align}  \label{240812.16}
 &\int_{-1}^{+1}d\cos\theta  \sum Y_{\ell''}^{\sigma-\sigma'}(0,\theta) (\sig'_1\sig'_2\sig''|s'_1s'_2S'')(\sig_1\sig_2\bar{\sig}|s_1s_2\bar{S})\la \vp'_{xz},\sig'_{1}\sig'_{2},s'_1s'_2|T||\vp|\hvz,\sig_{1}\sig_{2},s_1s_2\ra\\
  &=\sum \sqrt{\frac{2\ell+1}{\pi}} (m'\sig''\mu|\ell''S''J) (0\bar{\sig}\mu|\ell \bar{S} J) \la p'J\mu,\ell''S'',s'_1s'_2|T|pJ\mu,\ell \bar{S},s_1s_2\ra\,.\nn
\end{align}

Next, we note that there is only contribution on the right-hand side when $\mu=\bar{\sig}$, i.e., fixed in terms of values driven by the left-hand side. Then, we multiply Eq.~\eqref{240812.16}  by $(m'\sig''\mu|\ell''S''J')(0\bar{\sig}\mu|\bar{\ell}\bar{S}J')$ and  additionally sum over $\sig''$,  $\bar{\sig}$, $m'$ and $\mu$:
\begin{align} \label{240812.17}
  &\int_{-1}^{+1}d\cos\theta  \sum Y_{\ell''}^{m'}(0,\theta)(\sig'_1\sig'_2\sig''|s'_1s'_2S'')(\sig_1\sig_2\bar{\sig}|s_1s_2\bar{S})
  (m'\sig''\mu|\ell''S''J')(0\bar{\sig}\mu|\bar{\ell}\bar{S}J')\\
  &\times \la \vp'_{xz},\sig'_{1}\sig'_{2},s'_1s'_2|T||\vp|\hvz,\sig_{1}\sig_{2},s_1s_2\ra=\sum \sqrt{\frac{2\ell+1}{\pi}} (m'\sig''\mu|\ell''S''J)   (m'\sig''\mu|\ell''S''J') (0\bar{\sig}\mu|\ell \bar{S} J) (0\bar{\sig}\mu|\bar{\ell}\bar{S}J')\nn\\
  &\times \la p'J\mu,\ell''S'',s'_1s'_2|T|pJ\mu,\ell \bar{S},s_1s_2\ra\,.\nn 
\end{align}
Then, on the right-hand side
\begin{align}
  \label{240812.18}
\sum_{m',\sig''}  (m'\sig''\mu|\ell''S''J)   (m'\sig''\mu|\ell''S''J')=\delta_{JJ'}~.
\end{align}
For the other product of Clebsch-Gordan coefficients on the same side $(J=J')$,
\begin{align}
  \label{240812.19}
\sum_{\bar{\sig},\mu}  (0\bar{\sig}\mu|\ell \bar{S} J') (0\bar{\sig}\mu|\bar{\ell}\bar{S}J')\,,
\end{align}
we are going to make use of the following symmetry property of Clebsch-Gordan coefficients,
\begin{align}
  \label{240812.20}
(m_1m_2m_3|j_1j_2j_3)&=(-1)^{j_2+m_2}\sqrt{\frac{2j_3+1}{2j_1+1}}(-m_2m_3m_1|j_2j_3j_1)\,.
\end{align}
Then, we can rewrite Eq.~\eqref{240812.19} as
\begin{align}
  \label{240812.21}
\frac{2J'+1}{\sqrt{(2\ell+1)(2\bar{\ell}+1)}}\sum_{\bar{\sig},\mu}  (-\bar{\sig}\mu 0| \bar{S} J' \ell) (-\bar{\sig}\mu 0|\bar{S}J'\bar{\ell})=\frac{2J'+1}{2\bar{\ell}+1}\delta_{\ell\bar{\ell}}\,.
\end{align}
After having get rid of the Clebsch-Gordan coefficients on the right-hand side of Eq.~\eqref{240812.17} in virtue of Eqs.~\eqref{240812.18}, \eqref{240812.19} and \eqref{240812.21}, the desired PWA has been isolated, with the result (getting rid of now unnecessary primes and bars):
\begin{align}
  \label{240812.22a}
\la p'J\mu,\ell'S',s'_1s'_2|T|pJ\mu,\ell S,s_1s_2\ra&=\frac{\sqrt{\pi(2\ell+1)}}{2J+1}\sum
(\sig'_1\sig'_2\sig'|s'_1s'_2S')(\sig_1\sig_2\sig|s_1s_2S)(m'\sig'\mu|\ell'S'J)(0\sig\mu|\ell SJ)\nn\\
&\times\int_{-1}^{+1}d\cos\theta Y_{\ell'}^{m'}(0,\theta)
\la \vp'_{xz},\sig'_1\sig'_2,s'_1s'_2|T||\vp|\hvz,\sig_1\sig_2,s_1s_2\ra\,.
\end{align}
The sum is over all the labels that are not on the left-hand side of the equation. Notice that $|\vp|'\neq |\vp|$ for different types of channels involving different particles. 
 An alternative derivation of this result was also given in Ref.~\cite{Oller:2024qkw}, using previous results from Refs.~\cite{Oller:2019rej,Lacour:2009ej}.

\bigskip
$\bullet$ \underline{\it Identical particles, unitary normalization, and including isospin:}
\bigskip

For the case of two bosons/fermions the state must be symmetric/antisymmetric under the exchange of its two constituents. For instance, this is the case for a state made out of two $\pi^0$. However, this situation can be enlarged by introducing an isospin formalism, such that $|\pi\pi\ra$, a two-pion state, must be symmetric and $|NN\ra$, a two-nucleon state, must be antisymmetric, and similarly for other systems. The isospin formalism will be introduced straightforwardly at the end.

A (anti)symmetrized state of two particles with spin $s$ is
\begin{align}
  \label{240812.23}
 |\vp,\sig_1\sig_2\ra_S&=\frac{1}{\sqrt{2}}\left(|\vp,\sig_1\sig_2,s\ra+(-1)^{2s}|-\vp,\sig_2\sig_1,s\ra\right)\,.
\end{align}
Employing Eq.~\eqref{240812.6} for each state on the right-hand side
\begin{align}
  \label{240812.24}
  |\vp,\sig_1\sig_2\ra_S&=\sqrt{2\pi}\sum(\sig_1\sig_2\sig|ssS)(m\sig\mu|\ell SJ)Y_\ell^m(\vp)^*|J\mu,\ell S\ra\\
  &+(-1)^{2s}\sqrt{2\pi}\sum(\sig_2\sig_1\sig|ssS)(m\sig\mu|\ell SJ)Y_\ell^m(-\vp)^*|J\mu,\ell S\ra\,.\nn
\end{align}
Next, we use the parity properties of spherical harmonics and the following symmetry property of the Clebsch-Gordan coefficients
\begin{align}
  \label{240812.25}
  Y_\ell^m(-\hvp)&=(-1)^\ell Y_\ell^m(\hvp)\,,\\
(\sig_2\sig_1\sig|ssS)&=(-1)^{S-2s}(\sig_1\sig_2\sig|ssS)\,.\nn
\end{align}
Equation~\eqref{240812.24} simplifies to
\begin{align}
  \label{240812.26}
  |\vp,\sig_1\sig_2\ra_S&=\sqrt{4\pi}\sum \underbrace{\frac{1+(-1)^{S+\ell}}{\sqrt{2}}}_{\chi(S\ell)}(\sig_1\sig_2\sig|ssS)(m\sig\mu|\ell SJ)Y_\ell^m(\hvp)^*|J \mu,\ell S\ra~,
\end{align}
such that $S+\ell=$even and then $\chi(S\ell)=\sqrt{2}$, otherwise it is zero. For instance, for $\pi^0\pi^0$ with $S=0$ this rule implies that $\ell=\,$even. In turn, for a neutron-neutron state the total spin $S=0\,,1$, and the rule demands that $\ell=\,$even for $S=0$ and $\ell=\,$odd for $S=1$.  

We notice that compared with a state not composed by identical particles, the expression for $|\vp,\sig_1\sig_2\ra_S$ in Eq.~\eqref{240812.26} contains an extra factor of $\sqrt{2}$ from $\chi(S\ell I)$ when it is not zero. This implies that when isolating the  PWA proceeding as in order to arrive to Eq.~\eqref{240812.22a} we have to divide by an extra factor of $\sqrt{2}$ for each state made out of two identical particles. We introduce the switches $\chi_1$ and $\chi_2$ for the initial and final states, respectively, being 1 when the state is made by identical particles and 0 otherwise.  Thus, instead of Eq.~\eqref{240812.22a} we have now
\begin{align}
  \label{240812.22}
\la p'J\mu,\ell'S',s'_1s'_2|T|pJ\mu,\ell S,s_1s_2\ra&=\frac{\sqrt{\pi(2 \ell+1)}}{(2J+1)2^{\frac{\chi_1+\chi_2}{2}}}\sum
(\sig'_1\sig'_2\sig'|s'_1s'_2S')(\sig_1\sig_2\sig|s_1s_2S)(m'\sig'\mu|\ell'S'J)\\
&\times (0\sig\mu|\ell SJ) \int_{-1}^{+1}d\cos\theta Y_{\ell'}^{m'}(0,\theta)\,
{_S\la} \vp'_{xz},\sig'_1\sig'_2,s'_1s'_2|T||\vp|\hvz,\sig_1\sig_2,s_1s_2{\ra_S}\,.\nn
\end{align}
This is the so called unitary normalization \cite{Oller:1997ti}. We notice that in this way we can formally treat the states with identical particles as in the case of states made out of separable particles. However, when calculating ${_S\la} \vp'_{xz},\sig'_1\sig'_2,s'_1s'_2|T||\vp|\hvz,\sig_1\sig_2,s_1s_2{\ra_S}$ in terms of PWAs we have to multiply by $2^{\frac{\chi_1+\chi_2}{2}}$.

Including isospin labels, with  $\alpha_i$ referring to the third component of the isospin  $i$,  instead of Eq.~\eqref{240812.23} we have now 
\begin{align}
  \label{240812.27}
|\vp,\sig_1\sig_2,\alpha_1\alpha_2\ra_S&=
\frac{1}{\sqrt{2}}
\left(
|\vp,\sig_1\sig_2,\alpha_1\alpha_2\ra
+(-1)^{2s}
|-\vp,\sig_2\sig_1,\alpha_2\alpha_1\ra
\right)
\,.
\end{align}
Then, we proceed analogously to Eq.~\eqref{240812.24}, but now introducing the Clebsch-Gordan coefficients for the decomposition of the direct product of two equal isospins $i$ into states with define total isospin $I$ and third component $\alpha$. We also make use of the symmetry relations in Eq.~\eqref{240812.25} applied now the isospin Clebsch-Gordan coefficients. The neat result is
\begin{align}
  \label{240812.28}
|\vp,\sig_1\sig_2,\alpha_1\alpha_2\ra_S&=\sqrt{4\pi}\sum(\sig_1\sig_2\sig|ssS)(m\sig\mu|\ell SJ)(\alpha_1\alpha_2\alpha|iiI)\underbrace{\frac{1+(-1)^{S+\ell+I-2i}}{\sqrt{2}}}_{\chi(S\ell I)}|pJ\mu,\ell S I\ra\,,
\end{align}
and it is required that $S+\ell+I-2i=\,$even. For instance, for $\pi\pi$ with $i=1$ and $S=0$ it is required that $\ell+I=\,$even. For nucleon-nucleon $NN$, $i=s=1/2$ and the requirement is that $S+\ell+I=\,$odd. For the Deuteron bound state $I=0$, $\ell=0$ or 2 and then $S=1$. 

For completeness, we give here the formula for the PWA including isospin
\begin{align}
  \label{240812.29}
\la p'J\mu,\ell'S'I',s'_1s'_2,i_1i_2|T|pJ\mu,\ell S I,s_1s_2,i_1i_2\ra&=\frac{\sqrt{\pi(2 \ell+1)}}{(2J+1)2^\frac{\chi_1+\chi_2}{2}}\sum
  (\sig'_1\sig'_2\sig'|s'_1s'_2S')(\sig_1\sig_2\sig|s_1s_2S)\\
  &\times (m'\sig'\mu|\ell'S'J)(0\sig\mu|\ell SJ)
(\alpha'_1 \alpha'_2\alpha'|i'_1i'_2I')
  (\alpha_1 \alpha_2\alpha|i_1i_2I)\nn\\
  &\times \int_{-1}^{+1}d\cos\theta Y_{\ell'}^{m'}(0,\theta)\,
{_S\la} \vp'_{xz},\sig'_1\sig'_2,\fa'_1\fa'_2|T||\vp|\hvz,\sig_1\sig_2,\fa_1\fa_2{\ra_S}\,,\nn
\end{align}
which can be deduced straightforwardly from Eq.~\eqref{240812.22} since isospin goes straight. 

\bigskip
$\bullet$ \underline{\it Time-reversal invariance, and PWAs are symmetric:}
\bigskip

An important property of PWAs is that they are symmetric if time-reversal invariance is assumed to hold. Let us denote by $\theta$ the time-reversal operator. It is well-known that it is an antiunitary operator. It is simple to give an understanding in Quantum Mechanics why $\theta$ is antiunitary, because $\theta \vr \theta^{-1}=\vr$, and it is necessary that $\theta$ is antiunitary because then $\theta \vp \theta^{-1}=\theta\left(-i\frac{\partial}{\partial \vr}\right)\theta^{-1}=-\vp$, as required.

We assume time-reversal invariance, $\theta V\theta^{-1}=V$. From the Lippmann-Schwinger equation \eqref{240810.12} we then have that

\begin{align}
  \label{240812.30}
  \theta T\theta^{-1}&=-\theta V\theta^{-1}+\theta V \theta^{-1}\,\theta(E_\alpha+i\ep-H_0)^{-1}\theta^{-1}\,\theta T\theta^{-1}\\
  &=- V +  V(E_\alpha-i\ep-H_0)^{-1} V\,\theta T\theta^{-1}\,,\nn
\end{align}
where we have used that $\theta H_0\theta^{-1}=H_0$. 
Therefore, we arrive to the result
\begin{align}
  \label{240812.31}
\theta T\theta^{-1}=T^\dagger\,,
\end{align}
where we have also taken into account Eq.~\eqref{240826.1}.

The action of $\theta$ on a state $|jm\ra$ with definite angular momentum can be written as $\theta|jm\ra=(-1)^{j-m}|j-m\ra$ \cite{gott.240812}.  Recalling that if $|\alpha'\ra=\theta|\alpha\ra$ and $|\beta'\ra=\theta|\beta\ra$ then $\la \alpha'|\beta '\ra=\la\alpha|\beta\ra^*$, we then have for the PWAs
\begin{align}
  \label{240812.32}
  \la p'J\mu,\ell'S'|T|pJ\mu,\ell S\ra^*&=
  (\la p'J\mu,\ell' S'|\theta)\theta T|pJ\mu,\ell S\ra=((-1)^{J-\mu})^2\la p'J-\mu,\ell'S'|\theta T\theta^{-1}|pJ-\mu,\ell S\ra\\
  &=\la p'J-\mu,\ell'S'| T^\dagger|pJ-\mu,\ell S\ra
  =\la p' J-\mu,\ell S|T|p J-\mu,\ell'S'\ra^*\,. \nn
\end{align}
Thus,
\begin{align}
  \label{240812.32}
  \la p' J\mu,\ell'S'|T|p J\mu,\ell S\ra&=\la p J-\mu,\ell S|T|p' J-\mu,\ell'S'\ra
\end{align}
and, since the PWA is independent of $\mu$, we conclude that PWAs are symmetric under the exchange between the initial and final states.

\bigskip
$\bullet$ \underline{\it Unitarity in PWAs:}
\bigskip

We use Eq.~\eqref{240809.12} for the unitarity relation of the $T$ matrix and take its matrix elements between states of definite total angular momentum of the $\ell S J$ type:
\begin{align}
  \label{240812.33}
  \la p'J\mu,\ell' S'|T|pJ\mu,\ell S\ra-\la p'J\mu,\ell' S'|T^\dagger|pJ\mu,\ell S\ra&=i\sum \theta(s-s''_{\rm th})\frac{|\vp|''}{4\pi\sqrt{s}} \la p'J\mu,\ell' S'|T^\dagger|p''J\mu,\ell'' S''\ra \\
  &\times \la p''J\mu,\ell'' S''| T|pJ\mu,\ell S\ra\,,\nn
\end{align}
with $s''_{\rm th}=(m''_1+m''_2)^2$, the threshold of the intermediate state.  
In the sum over the intermediate states we have divided by their normalization Eq.~\eqref{240813.1} in order to sum over states normalized to 1. Expressing Eq.~\eqref{240812.33} only in terms of matrix elements of $T$ 
\begin{align}
  \label{240813.2}
  \la p'J\mu,\ell' S'|T|pJ\mu,\ell S\ra-\la pJ\mu,\ell S|T|p'J\mu,\ell' S'\ra^*&=i\sum \theta(s-s''_{\rm th}) \frac{|\vp''|}{4\pi\sqrt{s}} \la p''J\mu,\ell'' S'' |T|p'J\mu,\ell' S'\ra^* \\
  &\times \la p''J\mu,\ell'' S''| T|pJ\mu,\ell S\ra\,. \nn
\end{align}

Finally, assuming time-reversal invariance and, then,  taking into account the symmetry of the PWAs, 
we have
\begin{align}
  \label{240813.3}
&  \la p'J\mu,\ell' S'|T|pJ\mu,\ell S\ra-\la pJ\mu,\ell S|T|p'J\mu,\ell' S'\ra^*=2i \Ima \la p'J\mu,\ell' S'|T|pJ\mu,\ell S\ra\\
  &=i\sum \theta(s-s''_{\rm th}) \frac{|\vp''|}{4\pi\sqrt{s}} \la p''J\mu,\ell'' S'' |T| p'J\mu,\ell' S' \ra^* \la p''J\mu,\ell'' S''| T|pJ\mu,\ell S\ra\,. \nn
\end{align}
Therefore,
\begin{align}
  \label{240813.3}
  \Ima \la p'J\mu,\ell' S'|T|pJ\mu,\ell S\ra&=\sum \rho''(s) \la p''J\mu,\ell'' S''|T|p'J\mu,\ell' S' \ra^* \la p''J\mu,\ell'' S''| T|pJ\mu,\ell S\ra\,,\\
  \label{240813.4}
  \rho''(s)&=\theta(s-s''_{\rm th}) \frac{|\vp''|}{8\pi\sqrt{s}}\,.
  \end{align}
   
For only one PWA the previous expression simplifies as 
\begin{align}
  \label{240813.3bb}
  \Ima \la pJ\mu,\ell S|T|pJ\mu,\ell S\ra&= \rho(s) \left|\la pJ\mu,\ell S|T|pJ\mu,\ell S \ra\right|^2 \,,
\end{align}
or in other words
\begin{align}
  \label{240813.3b}
  \Ima\frac{1}{\la pJ\mu,\ell S|T|pJ\mu,\ell S\ra}&=-\rho(s)\,,
\end{align}
which is obtained by dividing both sides of Eq.~\eqref{240813.3bb} by the modulus squared of the PWA.

For several coupled PWAs, with or without involving different types of scattered states, we can construct the matrix of PWAs $\hat{T}(s)$, whose matrix elements are $\la p'J\mu,\ell' S'|T|pJ\mu,\ell S\ra$. We also introduce the diagonal matrix of phase space $\hat{\rho}(s)$, of diagonal elements $\rho''(s)$. Since the PWAs are symmetric then $\hat{T}$ is a symmetric matrix. Then, Eq.~\eqref{240813.3} in matrix notation reads
\begin{align}
  \label{240813.3c}
\Ima \hat{T}&=\hat{T}^\dagger\hat{\rho}\hat{T}~. 
\end{align}
Now, by writing $\Ima \hat{T}=(\hat{T}-\hat{T}^\dagger)/2i$ and then multiplying to the left and right by $\hat{T}^{\dagger -1}$ and $\hat{T}^{-1}$, respectively, we then find
\begin{align}
  \label{240813.3d}
\frac{1}{2i}\left(\hat{T}^{\dagger -1}-\hat{T}^{-1}\right)&=\hat{\rho}
\end{align}
that is,
\begin{align}
  \label{240813.3e}
\Ima \hat{T}(s)^{-1}&=-\hat{\rho}(s)\,.
\end{align}

\bigskip
$\bullet$ \underline{\it Lippmann-Schwinger equation in PWAs, phase shifts and inelasticity:}
\bigskip

We introduce a set of intermediate two-body $\ell S J$ states in the second term on the right-hand side of Eq.~\eqref{240810.11}. Recall  that the operators $V$ and $T$ in the Lippmann-Schwinger equation factorize only the Dirac delta function for the conserved total linear momentum, as indicated in the footnote \ref{foot.240813.1}. Then, in the normalization of the $\ell SJ$ in Eq.~\eqref{240813.1} we have to reinsert $2\pi \delta(E'-E)=2\pi \delta(|\vp'|-|\vp|)w(p_1)w(p_2)/(|\vp|\sqrt{s})$, according to Eq.~\eqref{240811.5a}. Thus, 
\begin{align}
\label{240813.5}
\la p'_j J'_j\mu'_j,\ell'_jS'_j|p_i J_i\mu_i,\ell_iS_i\ra&=\delta_{ij}\delta_{S_jS'_i}\delta_{\ell'_j\ell'_i}\delta_{J'_jJ'_i}\delta_{\mu'_j\mu'_i}\frac{8 \pi^2 w(p_{i1})w(p_{i2})}{|\vp_i|^2}\delta(|\vp_j|-|\vp_i|)\,.
\end{align}
Then,
\begin{align}
\label{240813.6}
T^{(J)}_{\ell'S',\ell S}(p',p)&=-V^{(J)}_{\ell'S',\ell S}(p',p)+\sumint_0^{+\infty}\frac{d|\vp''||\vp''|^2}{8\pi^2 w_1(p'')w_2(p'')}\frac{V^{(J)}_{\ell'S',\ell'S''}(p',p'')T^{(J)}_{\ell''S'',\ell S}(p'',p)}{E+i\ep-w_1(p'')-w_2(p'')}~,
\end{align}
where $w_i(\vp'')=\sqrt{{m''_i}^2+{\vp''}^2}$, and 
\begin{align}
\label{240813.7}
T^{(J)}_{\ell'S',\ell S}(p',p)&=\la p' J\mu,\ell'S'|T|p J\mu,\ell S\ra~,\\
V^{(J)}_{\ell'S',\ell S}(p',p)&=\la p' J\mu,\ell'S'|V|p J\mu,\ell S\ra~.\nn
\end{align}

The sum in Eq.~\eqref{240813.6} is over all the coupled-channel states considered, not only over the open ones as in the unitarity relation Eq.~\eqref{240813.3}. 

\bigskip
We come back again to our standard normalization of factorizing out the total-momentum conservation Dirac delta function.

We can build the $S$-matrix in partial waves by taking the matrix elements of Eq.~\eqref{240809.11} between partial-wave states. In doing this we divide by the norm of these states, which implies to multiply by $(2\rho_{\ell S}(s))^{1/2}$, so that we have
\begin{align}
  \label{240817.1}
  S^{(J)}_{\ell' S',\ell S}&=\delta_{\ell' S',\ell S}+2i\rho^{1/2}_{\ell' S'}T^{(J)}_{\ell'S',\ell S}\rho^{1/2}_{\ell S}\,.
\end{align}
Notice that $\rho_{\ell S}(s)$ only depends on $s$ and the masses of the particles making the two-body state, not on $\ell S$, that we have included to differentiate between different channels.  
Since $S$ is a unitary operator then $|S^{(J)}_{\ell S,\ell S}|^2\leq 1$ and its customarily written as $\eta^{(J)}_{\ell S} e^{2i\delta^{(J)}_{\ell s}}$, where $\delta^{(J)}_{\ell s}$ is the phase shifts and $\eta_{\ell S}^{(J)}$ the inelasticity factor.

\bigskip
{\bf Exercise 9:} Show that $|S^{(J)}_{\ell'S',\ell S}|\leq 1$. 
\bigskip

\bigskip
{\bf Exercise 10:} Making use of unitarity in PWAs show that $\hat{S}^{J} \hat{S}^{(J)\dagger}=I$, with $\hat{S}^{(J)}$ the matrix whose matrix elements are given in Eq.~\eqref{240817.1}.  
\bigskip


\subsection{Partial-wave amplitudes in the helicity basis}
\label{sec.240813.1}

\bigskip
$\bullet$ \underline{\it Helicity, monoparticle states, and representation of the rotation group:}
\bigskip

Had we chosen the standard Lorentz transformation $U(\vp)$ as
\begin{align}
  \label{240813.8}
U(\vp)&=R(\hvp)B_z(p)~,  
\end{align}
instead of  Eq.~\eqref{240811.11}, the moving particle would have define spin along its momentum. That is, it would have definite helicity $\sigma_i$.  This is clear because in its rest mass the particle has third component of spin $\sig_1$ along the $z$ axis, then $B_z(p)$ boosts  the particle to have  momentum $|\vp|\hvz$, which does not modify the third component of spin,\footnote{A Lorentz transformation along the $z$ does not modify the kinematics along the $xy$ plane, the one responsible for the component of spin along the $z$ axis.} and, finally, $R(\hvp)$ rotates equally the third component of spin and the momentum.

Next,  we consider the case of states made by particles with zero mass. In such a case, there is no rest frame of a massless particle and the labels $\sigma_i$ have the meaning of  helicity \cite{Weinberg:1995mt}. We concentrate here in the necessary developments for calculating the PWAs for two-graviton scattering, giving and extended analysis to that in Refs.~\cite{Blas:2020dyg,Oller:2024neq}. Let us recall that gravitons have helicity $\sig_i=\pm 2$. The same type of steps can be applied to any other zero-mass particle cases. 

 The monoparticle state with momentum $\vp$ is defined by applying the rotation $R(\hvp)=R(\phi,\theta,0)$, where $\phi$ and $\theta$ are the azimuthal and polar angles of $\hvp$, respectively,  to the state moving along the $z$ axis,
\begin{align}
  \label{240813.9}
  |\vp \sig\ra&=R(\hvp)||\vp|\hvz\sig\ra\,,\\
  R(\hvp)\hvz&=\hvp.\nn
\end{align}
Let us see how these monoparticle states transform under rotations:
\begin{align}
  \label{240813.18}
R(\vf)|\vp\sig\ra&=R(\vf)R(\hvp)||\vp|\hvz\sig\ra=R(\hvp')R(\hvp')^{-1}R R(\hvp)||\vp|\hvz\sig\ra\,,
\end{align}
with $\hvp'=R(\vf)\hvp$. Using Eq.~\eqref{240811.16} we then have
\begin{align}
  \label{240813.19}
  R(\vf)|\vp\sig\ra&=R(\hvp') R_z(\gamma)||\vp|\hvz\sig\ra=e^{-i\gamma \sig} R(\hvp')||\vp|\hvz\sig\ra=e^{-i\gamma \sig}|\vp'\sig\ra\,.
\end{align}
This is a representation of the rotation group. Indicating by $\phi_2$ and $\theta_2$ the spherical angular variables for $\hvp''=R(\vf_2)\hvp'$, and by $\phi_1$, $\theta_1$ those for $\hvp'=R(\vf_1)\hvp$, for the compositions of two rotation matrices $R(\vf_2)R(\vf_1)$ we have
\begin{align}
  \label{240813.20}
  R(\vf_2) R(\vf_1)|\vp\sig\ra&=
  R(\phi_2,\theta_2,0)R(-\phi_2,-\theta_2,0)R(\vf_2)R(\vf_1)|\vp\sig\ra\\
&=R(\phi_2,\theta_2,0)
  \underbrace{R(-\phi_2,-\theta_2,0)R(\vf_2)R(\phi_1,\theta_1,0)}_{R_z(\gamma_2)}\underbrace{R(-\phi_1,-\theta_1,0)R(\vf_1)R(\phi,\theta,0)}_{R_z(\gamma_1)}||\vp|\hvz\sig\ra  \nn\\
    &=R_z(\gamma_2+\gamma_1)|\vp''\sig\ra\,.\nn
\end{align}
It is simple to see that $R_z(\gamma(-\vf))$ for $R(\vf)^{-1}$ is $R_z(-\gamma(\vf))$:
\begin{align}
  \label{240813.21}
|\vp\sig\ra=  R(\vf)^{-1}R(\vf)|\vp\sig\ra&=R(\hvp)\underbrace{R(\hvp)^{-1}R(\vf)^{-1}R(\hvp')}_{R_z(\gamma(-\vf))}\underbrace{R(\hvp')^{-1}R(\vf)R(\hvp)}_{R_z(\gamma(\vf)))}||\vp|\sig\ra\,,
\end{align}
as we wanted to show. Notice that in the derivation we have used that $R(\vf)^{-1}\hvp'=\hvp$. Equation~\eqref{240813.21} is also a consequence of the composition law in Eq.~\eqref{240813.20}.

\bigskip
$\bullet$ \underline{\it Two-particle states, common $R(\hvp)$, and normalization:}
\bigskip

A two-graviton state in the CM frame is defined as the direct product
\begin{align}
  \label{240813.10}
  |\vp,\sig_1\sig_2\ra&=|\vp\sig_1\ra\otimes |-\vp\sig_2\ra\,.
\end{align}
Let us show that we can also write
\begin{align}
  \label{240813.10}
  |\vp,\sig_1\sig_2\ra&=R(\hvp)||\vp|\hvz,\sig_1\sig_2\ra\,.
\end{align}

For that we have to analyze,
\begin{align}
  \label{240813.11}
R(\hvp)|-|\vp|\hvz,\sig_2\ra&=R(-\hvp)R(-\hvp)^{-1}R(\hvp)R_y(\pi)||\vp|\hvz,\sig_2\ra
\,,
\end{align}
being $R(-\hvp)=R(\pi+\phi,\pi-\theta,0)$. Now,
\begin{align}
\label{240813.12}
R(-\hvp)^{-1}R(\hvp)R_y(\pi)&=R_y(-\pi+\theta)R_z(-\pi-\phi)R_z(\phi)R_y(\theta)R_y(\pi)=R_y(-\pi+\theta)R_z(-\pi)R_y(\theta)R_y(\pi)\nn\\
&=R_y(-\pi+\theta)\underbrace{R_z(-\pi)R_y(\pi+\theta)R_z(\pi)}_{R_y(-\pi-\theta)}R_z(-\pi).
\end{align}
We notice that $R_z(-\pi)R_y(\pi+\theta)R_z(\pi)$ is a rotation of the same angle $\pi+\theta$ but around the axis of rotation $-\hat{\boldsymbol{y}}$, which is the same as a rotation around $\hat{\boldsymbol{y}}$ in the opposite sense. Thus, 
\begin{align}
\label{240813.13}
 R(-\hvp)^{-1}R(\hvp)R_y(\pi)&=R_y(-2\pi)R_z(-\pi)\,.
\end{align}
Taking this result to Eq.~\eqref{240813.11}
\begin{align}
\label{240813.14}
R(\hvp)|-|\vp|\hvz,\sig_2\ra&=R(-\hvp)R_y(-2\pi)R_z(-\pi)||\vp|\hvz,\sig_2\ra=
e^{i\sig_2\pi}e^{i\sig_2 2\pi}R(-\hvp)||\vp|\hvz,\sig_2\ra=e^{i\sig_2\pi}e^{i\sig_2 2\pi}|-\vp,\sig_2\ra\,.
\end{align}
For a graviton $\sig_2=\pm 2$ and the exponential factor $e^{i\sig_2\pi}e^{i\sig_2 2\pi}=+1$. Then, it follows Eq.~\eqref{240813.10}. 
  
The normalization of these two particle states is given by Eq.~\eqref{240811.6b} but taking into account that the particles are massless. Then, $\sqrt{s}=2|\vp|$ and this equation simplifies as
\begin{align}
\label{240813.15}
\la \vp',\sig'_1\sig'_2|\vp,\sig_1\sig_2\ra&=32\pi^2\delta(\hvp'-\hvp)\delta_{\sig'_1\sig_1}\delta_{\sig'_2\sig_2}\,,
\end{align}
where the total momentum delta function has been removed.

\bigskip
$\bullet$ \underline{\it Partial-wave states,  normalization, and Bose-Einstein symmetry:}
\bigskip

We now proceed with  the decomposition of these plane-wave states into states with total angular momentum well defined, that we call the partial-wave states. For that we notice that a state $||\vp|\hvz,\sig_1\sig_2\ra$ has angular momentum $\sig\equiv \sig_1-\sig_2$  along the $z$ axis, which  can take the values $\pm 4$ or 0.  Therefore, we can write
\begin{align}
\label{240813.16}
||\vp|\hvz,\sig_1\sig_2\ra&=\sum_{J\geq |\sig|} C_{J\sig_1\sig_2}|J\sig,\sig_1\sig_2\ra\,.
\end{align}
Now, acting with $R(\hvp)$,
\begin{align}
\label{240813.17}
|\vp,\sig_1\sig_2\ra&=R(\hvp)||\vp|\hvz,\sig_1\sig_2\ra=\sum_J C_{J\sig_1\sig_2}R(\hvp)|J\sig,\sig_1\sig_2\ra=\sum_J \sum_{M=-J}^J C_{J\sig_1\sig_2} D^{(J)}_{M\sig}(\phi,\theta,0)|JM,\sig_1\sig_2\ra\,.
\end{align}
The coefficients $C_{J\sig_1 \sig_2}$ could depend on $\sig_1$, $\sig_2$, but since these remain invariant under rotations the coefficient $C_{J\sig_1\sig_2}$ is just a constant for every multiplet $J$ (each of them with its invariant values of $\sig_1$, $\sig_2$). Then, the coefficients $C_{J\sig_1\sig_2}$ can be fixed by the normalization of the partial-wave states.

Before reaching this point, let us invert Eq.~\eqref{240813.17}. For that, we are going to use the orthogonality properties of the rotation matrices
\begin{align}
\label{240813.22}
\int_0^{2\pi}d\phi\int_{-1}^{+1}d\cos\theta\int_0^{2\pi}d\varphi D^{(J')}_{M'\sig'}(\phi,\theta,\varphi)
D^{(J)}_{M\sig}(\phi,\theta,\varphi)^*=\frac{8\pi^2}{2J+1}\delta_{JJ'}\delta_{MM'}\delta_{\sig\sig'}\,.
\end{align}
We multiply and integrate  Eq.~\eqref{240813.17} as (we only keep the subscript $J$ in $C_{J\sig_1\sig_2}$)
\begin{align}
\label{240813.23}
\int_0^{2\pi}d\phi\int_{-1}^{+1}d\cos\theta D^{(J')}_{M'\sig}(\phi,\theta,0)^*|\vp,\sig_1\sig_2\ra&=
\sum_{J,M} C_{J}|JM,\sig_1\sig_2\ra\\
&\times
\int_0^{2\pi}d\phi\int_{-1}^{+1}d\cos\theta D^{(J')}_{M'\sig}(\phi,\theta,0)^*D^{(J)}_{M\sig}(\phi,\theta,0)\,.\nn
\end{align}
In order to make use of Eq.~\eqref{240813.22} we multiply the integrand on the right-hand side of Eq.~\eqref{240813.23} by $1=e^{i\sig\varphi}e^{-i\sig\varphi}$, so that the integral is the same as
\begin{align}
\label{240813.24}
\frac{1}{2\pi}\int_0^{2\pi}d\phi\int_{-1}^{+1}d\cos\theta\int_0^{2\pi}d\varphi D^{(J)}_{M\sig}(\phi,\theta,\varphi)^*D^{(J')}_{M'\sig}(\phi,\theta,\varphi)&=\frac{4\pi}{2J+1}\delta_{JJ'}\delta_{MM'}\,.
\end{align}
Employing this result in Eq.~\eqref{240813.23} (removing unnecessary primes to show the final result)
\begin{align}
\label{240813.25}  
|JM,\sig_1\sig_2\ra&=\frac{2J+1}{4\pi C_J}\int_0^{2\pi}d\phi\int_{-1}^{+1}d\cos\theta D^{(J)}_{M\sig}(\phi,\theta,0)^*|\vp,\sig_1\sig_2\ra\,.
\end{align}
To simplify the writing we denote by $\int d\hvp=\int_0^{2\pi}d\phi\int_{-1}^{+1}d \cos\theta$. 
Regarding the normalization of these states
\begin{align}
  \label{240813.26}
&  \la J'M',\sig'_1\sig'_2|JM,\sig_1\sig_2\ra=\frac{(2J+1)(2J'+1)}{(4\pi)^2C_JC_{J'}}\int d\hvp'\int d\hvp D^{(J')}_{M'\sig'}(\phi',\theta',0)D^{(J)}_{M\sig}(\phi,\theta,0)^*
  \underbrace{\la\vp',\sig'_1\sig'_2\ra|\vp,\sig_1\sig_2\ra}_{32\pi^2\delta(\hvp'-\hvp)\delta_{\sig'_1\sig_1}\delta_{\sig'_2\sig_2}}\nn\\
  &=32\pi^2\frac{(2J+1)(2J'+1)}{(4\pi)^2C_JC_{J'}}\frac{\delta_{\sig'_1\sig_1}\delta_{\sig'_2\sig_2}}{2\pi}
  \int_0^{2\pi}d\phi\int_{-1}^{+1}d\cos\theta\int_0^{2\pi}d\varphi D^{(J')}_{M'\sig}(\phi,\theta,\varphi)D^{(J)}_{M\sig}(\phi,\theta,\varphi)^*\\
  &=\frac{8\pi(2J+1)}{C_J^2}\delta_{MM'}\delta_{JJ'}\delta_{\sig'_1\sig_1}\delta_{\sig'_2\sig_2}\,.\nn
  \end{align}
Here, we have applied  the same trick as in Eq.~\eqref{240813.24} when the integration over $\varphi$ was also included. Notice that the fact of having $\delta_{\sig'_1\sig_1}\delta_{\sig'_2\sig_2}$ implies also that $\sig'=\sig$, and this is why we have taken advantage of it to put them equal in the subscripts of the rotation matrices to introduce $1=e^{i\sig\varphi}e^{-i\sig\varphi}$. Finally, we require the normalization
\begin{align}
  \label{240813.27}
  \la J'M',\sig'_1\sig'_2|JM,\sig_1\sig_2\ra&=\frac{2}{\pi}\delta_{MM'}\delta_{JJ'}\delta_{\sig'_1\sig_1}\delta_{\sig'_2\sig_2}\,,
\end{align}
which fixes
\begin{align}
  \label{240813.28}
C_J=2\pi\sqrt{(2J+1)}\,.
\end{align}
Taking this to Eq.~\eqref{240813.25}, the partial-wave states with definite angular momentum are finally 
\begin{align}
  \label{240813.29}
  |JM,\sig_1\sig_2\ra&=\frac{\sqrt{2J+1}}{8\pi^2}\int_0^{2\pi}d\phi\int_{-1}^{+1}d\cos\theta D^{(J)}_{M\sig}(\phi,\theta,0)^*|\vp,\sig_1\sig_2\ra\,,
\end{align}
and from Eq.~\eqref{240813.17} it follows also that
\begin{align}
\label{240813.30}
|\vp,\sig_1\sig_2\ra&=\sum_{J,M} 2\pi \sqrt{2J+1} D^{(J)}_{M\sig}(\phi,\theta,0)|JM,\sig_1\sig_2\ra\,.
\end{align}

Let us now take into account that two-gravitons state must be symmetric under the exchange of their constituents:
\begin{align}
  \label{240813.31}
|\vp,\sig_1\sig_2{\ra}_S&=\frac{1}{\sqrt{2}}\left(|\vp,\sig_1\sig_2\ra+|-\vp,\sig_2\sig_1\ra\right)\,.
\end{align}
Using for each state on the right-hand side the Eq.~\eqref{240813.30}
\begin{align}
\label{240813.31}
|\vp,\sig_1\sig_2{\ra_S}&=\frac{1}{\sqrt{2}}\sum_{J,M} 2\pi\sqrt{2J+1}\left[
D^{(J)}_{M\sig}(\phi,\theta,0)|JM,\sig_1\sig_2\ra
 +D^{(J)}_{M-\sig}(\pi+\phi,\pi-\theta,0)|JM,\sig_2\sig_1\ra
\right]\,.
\end{align}
Now, recalling that
\begin{align}
\label{240813.32}
D^{(J)}_{M\sig}(\phi,\theta,0)=e^{-iM\phi}d^J_{M\sig}(\theta)\,,
\end{align}
with $d^{(J)}_{M\sig}(\theta)$ the  Wigner small $d$ matrices,  we have
\begin{align}
\label{240813.33}
D^{(J)}_{M-\sig}(\pi+\phi,\pi-\theta,0)&=e^{-iM\pi}e^{-iM\phi}d^{J}_{M-\sig}(\pi-\theta)\,.
\end{align}
Next, we use the property
\begin{align}
\label{240813.34}
d^J_{M-\sig}(\pi-\theta)&=(-1)^{J+M}d^J_{M\sig}(\theta)\,.
\end{align}
Therefore, we have that Eq.~\eqref{240813.33} can be rewritten as
\begin{align}
\label{240813.35}
D^{(J)}_{M-\sig}(\pi+\phi,\pi-\theta,0)&=e^{-iM\phi}d^J_{M\sig}(\theta)(-1)^{2M}(-1)^J
=D^{(J)}_{M\sig}(\phi,\theta,0)(-1)^J\,.
\end{align}
We use this result in Eq.~\eqref{240813.31} and find
\begin{align}
\label{240813.36}
|\vp,\sig_1\sig_2{\ra}_S&=2\pi\sum_{J,M}\sqrt{2J+1}D^{(J)}_{M\sig}(\phi,\theta,0)\frac{1}{\sqrt{2}}\left(
|JM,\sig_1\sig_2\ra+(-1)^J|JM,\sig_2\sig_1\ra\right)\,.
\end{align}
We introduce the notation,
\begin{align}
\label{240813.37}
|JM,\sig_1\sig_2\ra_S=\frac{1}{\sqrt{2}}\left(
|JM,\sig_1\sig_2\ra+(-1)^J|JM,\sig_2\sig_1\ra\right)\,.
\end{align}
In the case that $\sig_1=\sig_2$ it follows that only $J=$\,even is allowed, otherwise $|JM,\sig_1\sig_2\ra_S=0$

The final expression for Eq.~\eqref{240813.36} is:
\begin{align}
\label{240813.38}
|\vp,\sig_1\sig_2{\ra}_S&=2\pi\sum_{J,M}\sqrt{2J+1}D^{(J)}_{M\sig}(\phi,\theta,0)|JM,\sig_1\sig_2\ra_S\,.
\end{align}
Its inversion is analogous to Eq.~\eqref{240813.29}:
\begin{align}
\label{240813.39}
|JM,\sig_1\sig_2\ra_S&=\frac{\sqrt{2J+1}}{8\pi^2}\int_0^{2\pi}d\phi\int_{-1}^{+1}d\cos\theta\,
D_{M\sig}^{(J)}(\phi,\theta,0)^*|\vp,\sig_1\sig_2{\ra}_S\,.
\end{align}

Regarding the normalization of the symmetrized states, we have from Eqs.~\eqref{240813.27} and \eqref{240813.37}
\begin{align}
\label{240813.40}
{_S\la}J'M',\sig'_1\sig'_2|JM,\sig_1\sig_2{\ra}_S&=\frac{2}{\pi}\delta_{JJ'}\delta_{MM'}\left(\delta_{\sig'_1\sig_1}\delta_{\sig'_2\sig_2}+(-1)^J\delta_{\sig'_1\sig_2}\delta_{\sig'_2\sig_1}\right)\,,\\
{_S\la}JM,\sig_1\sig_2|JM,\sig_1\sig_2{\ra}_S&=\frac{2}{\pi}\left(1+(-1)^J\delta_{\sig_2\sig_1}\right)\,.
\end{align}
Thus, 
\begin{align}
\label{240813.41}
{_S\la}JM,\sig_1\sig_2|JM,\sig_1\sig_2\ra_S &=\frac{4}{\pi 2^{|\sig|/4}} \,,
\end{align}
with $\sig=\sig_1-\sig_2$, as defined above.

\bigskip
$\bullet$ \underline{\it Partial-wave amplitudes with definite helicities:}
\bigskip

Following the same type of argument based on rotational invariance as in Sec.~\ref{sec.240811.1}, we know that it should be possible to calculate the PWAs in terms of the scattering amplitudes with the initial momentum along the $z$ axis and the final one contained in the $\vp_{xz}$ plane. Let us explicitly work out the resulting formula.

Using Eq.~\eqref{240813.38}
\begin{align}
  \label{240814.1}
        {_S\la}\vp_{xz},\sig'_1\sig'_2|T||\vp|\hvz,\sig_1\sig_2{\ra}_S&=\sum_{J,M}(2\pi)^2(2J+1)d^J_{M\sig'}(\theta)\underbrace{{_S\la}JM,\sig'_1\sig'_2|T|J\sig,\sig_1\sig_2\ra_S}_{\propto \delta_{M\sig}}\\
   &=\sum_{J}(2\pi)^2(2J+1)d^J_{\sig\sig'}(\theta)\,{_S\la}J\sig,\sig'_1\sig'_2|T|J\sig,\sig_1\sig_2\ra_S\,.\nn
\end{align}
The following orthogonality property of the Wigner small $d$ matrices is going to be used
\begin{align}
  \label{240814.2}
  \int_{-1}^{1}d\cos\theta d^{J'}_{\sig'\sig}(\theta)d^{J}_{\sig'\sig}(\theta)=\frac{2\delta_{JJ'}}{2J+1}\,.
  \end{align}
Multiplying Eq.~\eqref{240814.1} by $d^J_{\sig'\sig}(\theta)$ and integrating, it follows from Eq.~\eqref{240814.2}
\begin{align}
  \label{240814.3}
{_S\la}J\sig,\sig'_1\sig'_2|T|J\sig,\sig_1\sig_2\ra_S&=\frac{1}{8\pi^2}\int_{-1}^1d\cos\theta d^J_{\sig'\sig}(\theta){_S\la}\vp_{xz},\sig'_1\sig'_2|T||\vp|\hvz,\sig_1\sig_2\ra_S\,.
  \end{align}


\section{Crossing symmetry}
\def\theequation{\arabic{section}.\arabic{equation}}
\setcounter{equation}{0}   
\label{181024.2}

\bigskip
$\bullet$ \underline{\it Crossing symmetry, physical regions, and analyticity:}
\bigskip

Taking the set up of perturbative Quantum Field Theory, a generic quantum filed $\phi_i(x)$ contains both
 annihilation $a_i(\vp)$ and creation $b_i(\vp)^\dagger$ operators for particles and antiparticles, respectively  \cite{Weinberg:1995mt}. The former term is multiplied by the space-time exponential factor
$\exp(-ipx)$ and the latter by $\exp(ipx)$:
 \begin{align}
   \label{240814.4}
\phi_i(x)&=\int\frac{d^3p}{(2\pi)^32w_i(p)}\left(a(\vp)e^{-ipx}+b(\vp)^\dagger e^{ipx}\right)\,.
 \end{align}

 The simplest case to get the basic idea of crossing involves only zero-spin fields.
 Because of Eq.~\eqref{240814.4}, the same vertices in a given scattering process
 can be associated with an initial 
particle of momentum $p$ or with a final antiparticle with momentum $-p$, and viceversa (when considering $\phi_i^\dagger$). 
As a result, the scattering amplitude for  
\begin{align}
\label{181102.3}
a_1(p_1)+a_2(p_2)+\ldots \to b_1(p'_1)+b_2(p'_2)+\ldots
\end{align}
 governs any other process in which one or several particles
are crossed to their antiparticles  from the initial/final to the final/initial channels and, simultaneously, their momenta change sign. For instance, for the previous reaction we could have many others
related by crossing, in particular
\begin{align}
\label{181102.3b}
a_1(p_1)+a_2(p_2)+\ldots+\bar{b}(-p'_1) \to b_2(p'_2)+\ldots
\end{align}
where the bar corresponds to the antiparticle. 

Let us particularize crossing to the two-body scattering $a+b\to c+d$. We can then
distinguish the following related processes 
\begin{eqnarray}
\label{181102.4}
\text{$s$-channel process:}~& a(p_1)+b(p_2)&\to~~ c(p_3)+d(p_4)~,\\
\label{181102.4b}
\text{$t$-channel process:}~& a(p_1)+\bar{c}(-p_3)&\to~~ \bar{b}(-p_2)+d(p_4)~,\\
\label{181102.4c}
\text{$u$-channel process:}~& a(p_1)+\bar{d}(-p_4)&\to~~ c(p_3)+\bar{b}(-p_2)~.
\end{eqnarray} 
The $s$-channel is also called the direct one, while the $t$- and $u$-channels are also
 denoted as crossed channels.
Apart from the scattering processes in Eqs.~\eqref{181102.4}-\eqref{181102.4c},
there are other three processes in which  $a(p_1)\to \bar{a}(-p_1)$ is exchanged instead of $b(p_2)\to\bar{b}(-p_2)$
from the initial to the final state. These  new scattering processes also stem from applying CPT invariance to those shown in these equations.

By changing signs in the momenta, the $s$, $t$ and $u$ variables for every channel can be related.
Let us denote by the subscript $t$ and $u$ the Mandelstam variables in the $t$- and $u$-channels, respectively. Therefore: 
\begin{eqnarray}
\label{181102.5}
\text{$t$-channel:}&s_t&\!\!\!\!=(p_1-p_3)^2=t~,\\
&t_t&\!\!\!\!=(p_1+p_2)^2=s~,\nn\\
&u_t&\!\!\!\!=(p_1-p_4)^2=u~,\nn
\end{eqnarray}
\begin{eqnarray}
\label{181102.6}
\text{$u$-channel:}&s_u&\!\!\!\!=(p_1-p_4)^2=u~,\\
&t_u&\!\!\!\!=(p_1-p_3)^2=t~,\nn\\
&u_u&\!\!\!\!=(p_1+p_2)^2=s~.\nn
\end{eqnarray}
Notice that the physical regions for the $s$-, $t$- and $u$-channels are disjoint. To simplify matters  let us take that the four
particles have the same mass $m$, e.g. this is the case of $\pi\pi$ scattering. The  Mandelstam variables $s$, $t$ and $u$  in the CM are given in Eq.~\eqref{240811.2e}, from where it is obvious that
\begin{align}
  \label{181102.11}
  s+t+u=4m^2\,.
\end{align}
Therefore, only two of the three variables are independent. 

From Eq.~\eqref{240811.2e}, the physical region for the $s$-channel is
\begin{align}
  \label{181102.8}
s&\geq 4m^2~,~t\leq 0~,~u\leq 0\,.
\end{align}
Similarly,  for the $t$-channel:
\begin{align}
\label{181102.9}
t&=s_t\geq 4m^2~,~s=t_t\leq 0~,~u=u_t\leq 0~\,,
\end{align}
and for the $u$-channel,
\begin{align}
\label{181102.10}
u&=s_u\geq 4m^2~,~s=u_u\leq 0~,~t=t_u\leq 0\,.
\end{align}

Analyticity connects the scattering amplitudes in the three disjoint physical regions for
the $s$-, $t$- and $u$-channels  by the same analytical function $A(s,t,u)=A((p_1+p_2)^2,(p_1-p_3)^2,(p_1-p_4)^2)$ of $s$ and $t$
[$u$ is then given by Eq.~\eqref{181102.11}].
The physical values for the three channels correspond to the boundary values of this analytic function.

\bigskip
$\bullet$ \underline{\it Example of $\pi\pi$ scattering, and isospin amplitudes:}
\bigskip

Let us exemplify crossing symmetry and analyticity with the case of $\pi\pi$ scattering. We take as the $s$ channel the process with the scattering amplitude $A(s,t,u)$:
\begin{eqnarray}
  \label{240814.5}
&\text{$s$-channel}& \pi^0(p_1)\pi^0(p_2)\to \pi^+(p_3)\pi^-(p_4)\,,\\
&\text{$t$-channel}& \pi^0(p_1)\pi^-(-p_3)\to \pi^0(-p_2) \pi^-(p_4)\,,\nn\\
&\text{$u$-channel}& \pi^0(p_1)\pi^+(-p_4)\to \pi^+(p_3)\pi^0(-p_2)\,.\nn
\end{eqnarray}
The scattering amplitude for the $t$-channel is $B(s,t,u)=A(t,s,u)$ because of Eq.~\eqref{181102.5}. Notice that the first argument in $A(s,t,u)$ refers to $(p_1+p_2)^2$ that goes to $(p_1-p_2)^2$ in the $t$ channel and there it is the $t$ variable, and this is why the replacements in the arguments. Similarly, for the $u$ channel the scattering amplitude is $C(s,t,u)=A(u,t,s)$. 

A useful application of the previous relations is the determination of the three different isospin scattering amplitudes in $\pi\pi$ scattering. For that one takes into account that $|\pi^0\ra=|I=1,i_3=0\ra$, $|\pi^-\ra=|1,-1\ra$ and $|\pi^+\ra=-|1,1\ra$, cf. Eq.~\eqref{181103.1}. Then, it results: 
\begin{align}
  |\pi^0\pi^0\ra&=\sqrt{\frac{2}{3}}|20\ra-\frac{1}{\sqrt{3}}|00\ra\,,\\
  |\pi^+\pi^-\ra&=-\frac{1}{\sqrt{6}}|20\ra-\frac{1}{\sqrt{2}}|10\ra-\frac{1}{\sqrt{3}}|00\ra\,,\nn
\end{align}
and similarly for the other two pion states in the crossed channels. 
We denote the isospin amplitudes by $T_I(s,t,u)$. Thus,
\begin{eqnarray}
  &\pi^0\pi^0\to\pi^+\pi^- & A(s,t,u)=-\frac{1}{3}T_2(s,t,u)+\frac{1}{3}T_0(s,t,u)\,,\\
  &\pi^0\pi^-\to\pi^0\pi^- & A(t,s,u)=\frac{1}{2}T_2(s,t,u)-\frac{1}{2}T_1(s,t,u)\,,\nn\\
  &\pi^0\pi^+\to\pi^+\pi^0&  A(u,t,s)=\frac{1}{2}T_2(s,t,u)+\frac{1}{2}T_1(s,t,u)\,.
\end{eqnarray}
This is a linear system of algebraic equations from where one has
\begin{align}
  \label{240814.6}
  T_0(s,t,u)&=3A(s,t,u)+A(t,s,u)+A(u,t,s)\,,\\
  T_1(s,t,u)&=A(u,t,s)-A(t,s,u)\,,\nn\\
  T_2(s,t,u)&=A(u,t,s)+A(t,s,u)\,.\nn
\end{align}

\bigskip
$\bullet$ \underline{\it Crossed channel cuts, and the Lippmann-Schwinger equation:}
\bigskip

We already discussed in relation with Hermitian unitarity, cf. Eq.~\eqref{240809.17b}, that unitarity implies the existence of a cut in the scattering amplitude for $s\geq s_{\rm th,1}$. Following with the simplified kinematics of the equal mass case, this corresponds to $s>4m^2$. This applies also to the crossed channels. 
In particular, 
the unitarity cut  in the $u$-channel with a  constant value of $t$  
gives rise to a new  cut in the complex $s$ plane, apart from the $s$-channel unitarity cut. For the case referred, the cut runs for $u\geq 4m^2$ so that
it correspond to the values
\begin{align}
\label{181102.12}
s&=4m^2-t-u\leq -t~,
\end{align}
in addition to the  unitary cut  for $s\geq 4m^2$. 
This is a simple example of a crossed-channel cut, also referred as an unphysical cut, because it implies
unphysical values of the Mandelstam variables in the $s$-channel. They arise 
 from a branch point singularity attached to a two(multi)-body
threshold.

    The presence of the unitarity cuts either in the direct $s$-channel or in the crossed $t$- and $u$-channels  is a reflection of the Lippmann-Schwinger equation \eqref{240810.11}, considered for each channel separately. Notice that it has a cut for $E_\alpha>\sqrt{s_{\rm th,1}}$ in the $s$-channel but, similarly, this is also the case  for $E_\alpha>\sqrt{t_{\rm th,1}}$ and $E_\alpha>\sqrt{u_{\rm th,1}}$ in the $t$- and $u$-channels, respectively. They just arise in each case from the energy flowing through the scattering process, irrespectively of the value of the other independent variable. 

\bigskip
$\bullet$ \underline{\it Kinematical singularities, and Lorentz invariant amplitudes:}
\bigskip

For particles with spin the analytical continuation of the scattering amplitude in the complex $s$ and $t$
planes is more delicate because of the so-called kinematical singularities, which origin is not
dynamical like for the unitarity cuts in the $s$- or crossed channels.
They arise from the algebraic dependence on the Mandelstam variables of the solutions to the relativistic
equations for the particles with spins \cite{martin.200705.1,olive.181102.1}, e.g. they could introduce square roots, etc. Once these kinematical dependences are separated,  the singularities in the different components of the scattering amplitudes  are due to values of the Mandelstam variables for which intermediate particles are on shell.  

A possible way to deal with the kinematics singularities is to isolate Lorentz invariant functions out of
the scattering amplitudes, which plays the role of the Lorentz invariant amplitudes for zero-spin meson-meson scattering. 
Take for example $\pi^a(q)N(p,\sigma;\alpha)\to \pi^{a'}(q')N(p',\sigma';\alpha')$,
where $a$ and $a'$ denote the isospin Cartesian coordinates of pions. 
The charged pions correspond to the combinations
\begin{align}
\label{181103.1}
\sum_a \frac{\pi^a\tau^a}{\sqrt{2}}=\left(
\begin{matrix}
\frac{\pi^0}{\sqrt{2}} & \pi^+\\
\pi^- & -\frac{\pi^0}{\sqrt{2}}
\end{matrix}
\right)~.
\end{align}
Firstly, it is clear that there are  two  isospin amplitudes corresponding
to $I=1/2$ and 3/2, because the pions and nucleons have isospin 1 and 1/2, respectively. 
Any matrix in the isospin  space of nucleons can be expressed as a superposition of
Pauli matrices $\tau^a$, $a=1,2,3$, and the 2$\times$2 identity matrix. 
Let us consider two pions with isospin Cartesian indices $a$ and $a'$, then the tensors with good properties
under isospin rotations that can be built having $I=0$ and 1 are $\delta_{aa'}$ and
$[\tau_a,\tau_{a'}]$, respectively.\footnote{No tensor of rank 2 is allowed  because its combination with  isospin an 1/2 cannot
	give rise again to an isospin 1/2.}
In this way, we can write
\begin{align}
\label{181103.2}
T_{aa'}&=\delta_{a'a} T^+ + \frac{1}{2}[\tau_a,\tau_{a'}]T^-~.
\end{align}
The two amplitudes $T^{\pm}$ are operators acting in the space of the Dirac
spinors and can expanded as a linear combination of the sixteen 
matrices $I$, $\gamma^\mu$, $\sigma^{\mu\nu}$, $\gamma_5$ and $\gamma_5\gamma^\mu$ with the Lorentz
indices contracted with four-momenta. The matrices $\gamma_5$ and $\gamma_5\gamma^\mu$
violate parity and they do not appear.  In addition to the identity and the gamma matrices $\gamma^\mu$, the matrices $\sigma^{\mu\nu}$ do not contribute with any new structure  because of the Dirac equations:  
$p\!\!\! / u(\vp,\sigma)=mu(\vp,\sigma)$ and $\bar{u}(\vp',\sigma')p'\!\!\!\! / =m\bar{u}(\vp',\sigma')$. For instance, let us consider $\sigma^{\mu\nu}p_\mu r_\nu$, where $r_\nu$ is a vector built from particle momenta. Then, we anticommute $\gamma^\mu p_\mu$ and $\gamma^\nu r_\nu$ in the first term of $\sigma^{\mu\nu}p_\mu r_\nu=(\gamma^\mu\gamma^\nu-\gamma^\nu\gamma^\mu)p_\mu r_\nu$ until $\gamma^\mu p_\mu$ always acts on $u(\vp,\sigma)$, giving then $m u(\vp,\sigma)$ because of the Dirac equation. The rest of terms resulting from the anticommutation relation of the gamma matrices can be reabsorbed in the ones involving the identity matrix and the $\gamma^\mu$. The same would apply to $\sigma^{\mu\nu}p'_\mu r_\nu$.  On the other hand,  $\sigma^{\mu\nu}q_\mu q'_\nu$ is the same as $\frac{1}{2}\sigma^{\mu\nu}(q+q')_\mu(q'-q)_\nu$ because $\sigma^{\mu\nu}$ is antisymmetric. However, since $q'_\nu-q_\nu=p_\nu-p'_\nu$ we again come back to the first case already discussed. All in all, we then arrive to
the standard form \cite{Alarcon:2012kn}
\begin{align}
\label{181103.4}
T^{\pm}=\bar{u}(p',\sigma')\left[A^\pm(s,t,u)+\frac{1}{2}(q\!\!\! /+q'\!\!\!\!\!/ \,\,)B^\pm(s,t,u)\right]u(p,\sigma)~.
\end{align}
The combination $q'-q$ does not appear in the previous equation once the Dirac equations  are taken into account because $q'-q=p-p'$. 
The analytical properties of the Lorentz invariant functions $A^\pm$ and $B^\pm$ are determined in the same way as explained
for the scattering amplitude of zero-spin mesons. The other factors in Eq.~\eqref{181103.4} are, of course, important to
finally do actual calculations to  establish relations between analyticity and experimental results.

\bigskip
$\bullet$ \underline{\it Cuts in the $s$ plane for PWAs due to poles and cuts in crossed channels:}
\bigskip

    The crossed-channel poles, stemming from poles in the $t$- and $u$-channels,  
produce  crossed cuts in the complex $s$ plane for PWAs.
 For example, for $\pi^- n\to \pi^- n$ scattering consider the  $u$-channel pole due to the exchange of a proton, $u=m_p^2$. 
 For pion-nucleon scattering in the CM the $u$ variable is given by
($m_p$ and $m_\pi$ are the nucleon and pion masses, respectively) 
\begin{align}
\label{181103.5}
u&=m_p^2+m_\pi^2-2\omega E-2\vp^2\cos\theta~,
\end{align}
where  $E$ and $\omega$ are the nucleon and pion CM energies, respectively. 
In the process of calculating a PWA, cf. Eq.~\eqref{240812.22},  the scattering angle is integrated $\theta\in[0,\pi]$.  
Thus, setting $u=m_p^2$ in Eq.~\eqref{181103.5} and giving $\omega$, $E$ and $\vp^2$ in terms of $s$,
\begin{align}
\label{181103.6}
\omega&=\frac{s+m_\pi^2-m_p^2}{2\sqrt{s}}~,\\
E&=\frac{s+m_p^2-m_\pi^2}{2\sqrt{s}}~,\nn\\
\vp^2&=\frac{\lambda(s,m_p^2,m_\pi^2)}{4s}~,\nn\\
\lambda(s,m_1^2,m_2^2)&=(s-(m_1+m_2)^2)(s-(m_1-m_2)^2)~,\nn
\end{align}
with  $\lambda(s,m_1^2,m_2^2)$   the K\"all\'en triangle function,
 the values of $s$ that are solution of Eq.~\eqref{181103.5}  as a function of $x=\cos\theta$ are:
\begin{align}
\label{181103.7}
s_1(x)=\frac{m^2 x + m_\pi^2(1+x)-\sqrt{m^4+2m_\pi^4(1+x)+2m^2m_\pi^2(-1+x+2x^2)}}{1+x}~,\\
s_2(x)=\frac{m^2 x + m_\pi^2(1+x)+\sqrt{m^4+2m_\pi^4(1+x)+2m^2m_\pi^2(-1+x+2x^2)}}{1+x}~.\nn
\end{align}

The first solution $s_1(x)$ produces a cut along the  negative real axis. We clarify that the radicand is larger than  the square of 
the terms in the numerator before the square root, with the difference between them being  $(1-x^2)(m^2-m_\pi^2)^2$. 
 This also shows that the radicand is positive for  $x\in[-1,1]$ and arbitrary values for the masses. This cut extends from $-\infty$ (for $x\to -1$) up to 0 ($x\to +1)$. This is a clear example of a left-hand cut (LHC).
 For $s_2(x)$ one has a finite cut which ranges along the positive real axis from  $(m^2-m_\pi^2)^2/m^2$ up to $m^2+2m_\pi^2$. 

Crossed-channel unitarity cuts give also rise to cuts in the complex $s$ plane for PWAs. 
For instance, let us take  $\pi\pi$ scattering. Then, the two-pion unitarity cuts along the $t$- and $u$-channels
happen for $t=-2(s/4-m_\pi^2)(1-x)\geq 4m_\pi^2$ and  
 $u=-2(s/4-m_\pi^2)(1+x)\geq 4m_\pi^2$. Solving $s$ in terms of $x$ we find that both cases  give
rise to LHCs with $s\in]-\infty,0]$ when $\cos\theta$ moves along $[-1,1]$.

\bigskip
$\bullet$ \underline{\it Crossed cuts in  PWAs for nonrelativistic scattering:}
\bigskip

In the case of a nonrelativistic theory  crossing does not apply. From the point of view of Quantum Field Theory, in the nonrelativistic case the quantum fields only involve annihilation operators (or creation ones
for the Hermitian conjugate field) \cite{thirring.181101.1}.
However,  a LHC stems in this case  due to the  exchanged particles producing the potential.   For instance, let us take a  Yukawa potential
\begin{align}
\label{181104.1}
V(r)&=\alpha\frac{ e^{-r m_\pi}}{r}~.  
\end{align}
Its Fourier transform is
\begin{align}
\label{181104.2}
V(\vq^2)&=\alpha\int d^3 r e^{-i\vq \vr}\frac{e^{-r m_\pi}}{r}=\frac{4\pi\alpha}{\vq^2+m_\pi^2}~,
\end{align}
with $\vq=\vp'-\vp$. Its partial-wave projection for particles
without spin, apply Eq.~\eqref{240812.22a}, is:
\begin{align}
\label{181104.3}
V_J(p,p')&=\frac{1}{2}\int_{-1}^{+1}d\cos\theta \,V(\vq^2)\,P_J(\cos\theta)\\
&=-\frac{\pi \alpha}{pp'}\int_{-1}^{+1}d\cos\theta \frac{P_J(\cos\theta)}{\cos\theta-(p^2+{p'}^2+m_\pi^2)/(2pp')}~. \nn
\end{align}
The LHC is produced when the denominator vanishes, and this can be isolated in the following way
\begin{align}
\label{181104.4}
V_J(p,p')&=-\frac{\pi \alpha}{pp'}\int_{-1}^{+1}d\cos\theta \frac{P_J(\cos\theta)}{\cos\theta-\xi}
=-\frac{\pi \alpha}{pp'}\int_{-1}^{+1}d\cos\theta \frac{P_J(\cos\theta)-P_J(\xi)}{\cos\theta-\xi}\\
&-\frac{\pi \alpha}{pp'}P_J(\xi)\int_{-1}^{+1} \frac{d\cos\theta}{\cos\theta-\xi}~,\nn
\end{align}
with
\begin{align}
\xi&=\frac{p^2+{p'}^2+m_\pi^2}{2pp'}~.
\end{align}
There is no LHC from the term before the last one in Eq.~\eqref{181104.4}  because when the denominator vanishes the numerator also does. The LHC then stems from the last term and the integration in
$\cos\theta$ can be done explicitly:
\begin{align}
\label{181104.5}
-\frac{\pi \alpha}{pp'}P_J(\xi)\int_{-1}^{+1} \frac{d\cos\theta}{\cos\theta-\xi}
&=-\frac{\pi \alpha}{pp'}P_J(\xi)\left[
\ln(1-\xi)-\ln(-1-\xi)
\right]~.
\end{align}
Now, for real and positive $p$ and $p'$ the difference of the two logarithms in the last term can be rewritten as
\begin{align}
\label{181104.6}
\frac{\pi \alpha}{pp'}P_J(\xi)\left[
\ln((p+p')^2+m_\pi^2)-\ln((p-p')^2+m_\pi^2) \right]~.
\end{align}
This form is specially suitable for the analytical continuation to the complex
 planes of $p$ and $p'$ and so determine the position of the cuts, as fully exploited in
Ref.~\cite{Oller:2018zts}. The  cuts in the $p$ variable for given $p'$ 
happen for the first logarithm when $(p+p')^2+m_\pi^2<0$, and for the second one when $(p-p')^2+m_\pi^2<0$. These cuts are vertical and are given by
\begin{align}
\label{181104.7}
p=(\pm)p'\pm i \sqrt{m_\pi^2+x^2}~,~x\in \mathbb{R}~,
\end{align}
where the first $\pm$ symbol is uncorrelated with the second one. Analogous reciprocal relations exist for the
cuts in the variable $p'$ for  given $p$.

In the case of on-shell scattering, $p=p'$, and particularizing Eq.~\eqref{181104.7},  the
only meaningful case happens  by selecting the minus sign in the $\pm$ sign between the first brackets, 
\begin{align}
\label{181104.8}
p=-p\pm i\sqrt{m_\pi^2+x^2}~.
\end{align}
Its solution is
\begin{align}
\label{181104.9}
p=\pm \frac{i}{2}\sqrt{m_\pi^2+x^2}~,
\end{align}
and  for the variable $p^2$ we have a cut for the values
\begin{align}
\label{181104.10}
p^2\leq -\frac{m_\pi^2}{4}~.
\end{align}
This is the LHC for nonrelativistic $NN$ scattering \cite{Oller:2018zts}.

\section{The $N/D$ method and CDD poles}
\def\theequation{\arabic{section}.\arabic{equation}}
\setcounter{equation}{0}   
\label{sec.181104.1}

\bigskip
$\bullet$ \underline{\it Set up for the $N/D$ method:}
\bigskip

The crossed cuts for PWAs in the complex $s$ plane are of two types. For 
processes with equal masses, like  $a+a\rightarrow a+a$, as  $\pi\pi$ scattering, there is only a 
LHC for $s<s_{\rm L}$ (for $\pi\pi$ scattering $s_{\rm L}=0$ since the lightest states exchanged in the crossed channels are two-pion states themselves). However, for scattering of the type 
$a+b\rightarrow a+b$ with $m_1=m_3=m_a$ and $m_2=m_4=m_b$, in addition to a LHC  
there is  a circular cut too in the complex $s$ plane at $|s|=m_2^2-m_1^2$ \cite{martin.200705.1}, 
where we have taken $m_2>m_1$. For simplicity 
in the formalism, we  refer to the LHC as if it comprises all the unphysical cuts. This is 
 enough for our introductory purposes here. It also worth mentioning that had we worked in the
complex $p^2$ plane all the cuts for elastic scattering would be LHC ones  for this variable. This is because the expression for $t$ and $u$ are linear in  the modulus squared of the momentum, as in the equal mass case for the $s$ variable.  
Then, the analysis would be analogous to the one developed here in the complex $s$ plane.

We first analyze the elastic case and then we generalize the results for  coupled
partial-wave amplitudes.
Let us indicate simply by $T(s)$ a PWA with total angular momentum $J$, together with other quantum numbers that would be needed to completely determine it. Above the threshold of the reaction $s_{\rm th}$ along the unitarity cut, according to Eq.~\eqref{240813.3b}, 
\begin{align}
\label{181105.2a}
\Ima T^{-1}(s)&=-\rho(s)=-\frac{|\hvp|}{8\pi\sqrt{s}}~,~s\geq s_{\rm th}~.
\end{align}
 Along the LHC for  $s<s_{\rm L}$ the discontinuity and Schwarz reflection theorem\footnote{This theorem states that an analytical function $f(z)$ that is real along an open domain in the real axis then it satisfies $f(z^*)=f(z)^*$. This is the case for the PWAs analyzed here with $s_{\rm L}<s_{\rm th}$.\label{foot.240817.1} } leads to
\begin{equation}
\label{lhc}
T(s+i\epsilon)-T(s-i\epsilon)=2 i \Ima T(s)~,~s<s_{\rm L}\,.
\end{equation}

The $N/D$ method \cite{Chew:1960iv} allows to obtain $T(s)$ fulfilling Eqs.~(\ref{181105.2a}) and (\ref{lhc}) by expressing it as the quotient of two functions separating the LHC and unitarity cuts as
\begin{equation}
\label{n/d}
T(s)=\frac{N_L(s)}{D_L(s)}~,
\end{equation}
such that  $D(s)$ only has unitarity cut for $s>s_{\rm th}$  and 
 $N(s)$ has only LHC for $s<s_{\rm L}$. From Eqs.~(\ref{181105.2a}) and \eqref{n/d} these functions satisfy:

\begin{align}
\label{eqs1}
\Ima  D(s)&=\Ima T^{-1}(s) N(s)=-\rho(s) N(s)~,  &s>s_{\rm th} \\
\Ima D(s)&=0~,   &s<s_{\rm th}\nn\\
\label{eqs2}
\Ima  N(s)&=\Ima T(s) \; D(s)~,  &s<s_{\rm L}  \\
\Ima N(s)&=0~.  &s>s_{\rm L}\nn
\end{align}

Now, we write a dispersion relation (DR) for $D(s)$. We assume that $D(s)/s^n$ vanishes  in the complex $s$ plane for $s\to\infty$. Therefore, we can a write an $n$-times subtracted DR. We notice that because along the unitarity cut $\Ima D(s)=-\rho(s)N(s)$, Eq.~\eqref{eqs1}, this asymptotic behavior also implies that $\lim_{s\to\infty}N(s)/s^n=0$ (note that, cf. Eq.~\eqref{181105.2a}, $\rho(s)\to \text{constant}$ for $s\to\infty$).\footnote{Because of the Kanasawa-Tomozawa theorem \cite{sugawara.180929.1,Oller:2019rej}, summarized in the Appendix \ref{sec.240322.2}, for a function with only a cut without poles on it the same degree of divergence occurs in the whole complex plane for any direction of $s\to\infty$.} This also allows to write an $n$-times subtracted DR for $N(s)$. For that, we apply the Cauchy integration theorem to an integration contour in the complex $s$ plane closed at infinity by a circle of infinite radius engulfing the unitarity cut along the real axis for $s>s_{\rm th}$. Therefore, 
\begin{align}
  \label{240814.7}
  \oint dz\frac{D(z)}{(z-s)(z-s_0)^n}&=
  2\pi i\frac{D(s)}{(s-s_0)^n}
  +2\pi i\frac{1}{(n-1)!}
  \left.\frac{d^{n-1}}{dz^{n-1}}\frac{D(z)}{z-s}\right|_{z=s_0}\,.
\end{align}
On the other hand, by dropping the contribution along the circle at infinity we are left only with the contribution to the integration along the unitarity cut,
\begin{align}
\label{240814.8}
\oint dz\frac{D(z)}{(z-s)(z-s_0)^n}&=\int_{s_{\rm th}}^{+\infty} ds'\frac{D(s'+i\ep)-D(s'-i\ep)}{(s'-s)(s'-s_0)^n}=2i\int_{s_{\rm th}}^{+\infty} ds'\frac{\Ima D(s')}{(s'-s)(s'-s_0)^n}\,.
\end{align}
Equating Eqs.~\eqref{240814.7} and \eqref{240814.8}, and taking into account Eq.~\eqref{eqs1}, we then have the following dispersive expression for $D(s)$,
\begin{align}
\label{240814.9}
D(s)&=-\frac{(s-s_0)^n}{\pi}\int_{s_{\rm th}}^{+\infty} ds'\frac{\rho(s')N(s')}{(s'-s)(s'-s_0)^n}-\frac{(s-s_0)^n}{(n-1)!}\left.\frac{d^{n-1}}{dz^{n-1}}\frac{D(z)}{z-s}\right|_{z=s_0}\,.
\end{align}
The last term is a polynomial of degree $n-1$,
\begin{align}
\label{240814.10}
-\frac{(s-s_0)^n}{(n-1)!}\left.\frac{d^{n-1}}{dz^{n-1}}\frac{D(z)}{z-s}\right|_{z=s_0}&=
 \sum_{m=0}^{n-1}a_ms^m\,.
\end{align}
Taking this to Eq.~\eqref{240814.9},
\begin{equation}
\label{d'}
D(s)=-\frac{(s-s_0)^n}{\pi}\int^{+\infty}_{s_{\rm th}} ds' 
\frac{\rho(s') {N}(s')}{(s'-s)(s'-s_0)^n}+\sum_{m=0}^{n-1}  a _m s^m\,.
\end{equation}

We can proceed analogously for the function $N(s)$ by closing the integration contour at infinity with a circle of infinite radius that engulfs the LHC for $s<s_{\rm L}$. Considering the discontinuity of $N(s)$ along this cut, Eq.~\eqref{eqs2}, and that $\lim_{s\to\infty}N(s)/s^n=0$, as discussed above,  we then have
\begin{equation}
\label{n2}
N(s)=\frac{(s-s_0)^{n}}{\pi}\int_{-\infty}^{s_{\rm L}} ds' 
\frac{{\Ima  T}(s') {D}(s')}{ (s'-s)(s'-s_0)^{n}}+
\sum_{m=0}^{n-1}  b_m s^m~.
\end{equation}                         

Equations~(\ref{d'}) and (\ref{n2}) are a system of coupled linear integral equations for the functions 
$N(s)$ and $D(s)$, whose input  is ${\Ima  T}(s)$ along the LHC (together with the coefficients $a_m$ and $b_m$ in the subtractive polynomials).

\bigskip
$\bullet$ \underline{\it CDD poles, and additional solutions:}
\bigskip

However, Eqs.~(\ref{d'}) and (\ref{n2}) are not the most general solutions 
to Eqs.~(\ref{eqs1}) and (\ref{eqs2}) because the inverse of a PWA along the unitarity cut is not defined when it crosses a zero. Therefore, Eq.~\eqref{eqs1} is actually ambiguous at such points. For other possible zeroes in the complex $s$ plane one can always imagine that these are accounted for by zeroes in $N(s)$ which would not pose any special problem for the DR of this function, only those along the unitarity cut have to be treated separately. These zeros are the Castillejo-Dalitz-Dyson (CDD) poles after Ref. \cite{Castillejo:1955ed}.

Let $\{s_i\}$ be the set of zeroes  along the unitarity cut at every of which ${T}(s_i)=0$.  We introduce a function $\lambda(s)$ such that  between 
two consecutive zeroes $s_i$ and $s_{i+1}$,
\begin{equation}
\label{dl1}
{\Ima D}(s)=-\rho(s)N(s)=\frac{d \lambda (s)}{ds}~.
\end{equation} 

Therefore, for $s\in(s_i,s_{i+1})$
\begin{equation}
\label{l1}
\lambda(s)=-\int_{s_i}^{s} \rho(s') {N}(s') ds'+\lambda(s_i)~,
\end{equation}
where $\lambda(s_i)$ is unknown. From this intermediate result we can then write along the $s$ real axis that \begin{equation}
\label{l2}
\lambda(s)=-\int_{s_{\rm th}}^s  \rho(s') {N}(s') ds'+
\sum_i \lambda(s_i)\theta(s-s_i)~,
\end{equation}
with $\theta(s)$ the Heaviside function. Then,  taking the derivative of $\lambda(s)$ according to Eq.~(\ref{dl1}), we have
\begin{align}
  \label{240815.1}
\Ima D(s)&=-\rho(s)N(s)+\sum_i \lam(s_i)\delta(s-s_i)\,.
\end{align}
Next, using this result for $\Ima D(s)$ in Eq.~\eqref{240814.8}, instead of Eq.~\eqref{d'}, it follows the more general result:
\begin{align}
\label{tower1}
&{D} = \frac{(s-s_0)^n}{\pi}\int_{s_{\rm th}}^{+\infty} \frac{
{\Ima  D}(s') ds'}{(s'-s)(s'-s_0)^n}+\sum_{m=0}^{n-1}  a_m
s^m =  \\ 
&-\frac{(s-s_0)^n}{\pi}\int_{s_{\rm th}}^{+\infty} \frac{\rho(s') 
{N}(s')}{(s'-s)(s'-s_0)^n}ds' +
\frac{(s-s_0)^n}{\pi} \int_{s_{\rm th}}^{+\infty} \frac{\sum_i \lambda(s_i) 
\delta(s'-s_i)}{(s'-s)(s'-s_0)^n}ds'+\sum_{m=0}^{n-1}  a_m s^m\nonumber\\
&=-\frac{(s-s_0)^n}{\pi}\int_{s_{\rm th}}^{+\infty} \frac{ \rho(s') 
{N}(s')}{(s'-s)(s'-s_0)^n}ds'
+\sum_i \frac{\lambda(s_i)}{\pi(s_i-s) }\frac{(s-s_0)^n}{(s_i-s_0)^n}+\sum_{m=0}^{n-1}  a_m s^m~.\nonumber
\end{align}

The term before the last one on the right-hand side of Eq.~(\ref{tower1}) can also be 
written more conveniently  without extra $s$ dependence affecting the pole terms:
\begin{align}
\label{cano}
\frac{(s-s_0)^n}{s-s_i}&=(s-s_0)^{n-1} \frac{s-s_i+s_i-s_0}{s-s_i}=
(s-s_0)^{n-1}\left(1+\frac{s_i-s_0}{s-s_i}\right) \\
&=(s-s_0)^{n-1}+(s_i-s_0)\frac
{(s-s_0)^{n-1}}{s-s_i}=\sum_{i=0}^{n-1}(s-s_0)^{n-1-i}(s_i-s_0)^i+\frac{(s_i-s_0)^n}{s-s_i}~.\nn
\end{align}

The terms 
\begin{align}
\sum_{i=0}^{n-1}(s-s_0)^{n-1-i}(s_i-s_0)^i 
\end{align}
can be reabsorbed in the polynomial
\begin{align}
\sum_{m=0}^{n-1}  a _m s^m\,,
\end{align}
which we still denote in the same manner after this reshuffling. 
As a result, we rewrite more simply Eq.~\eqref{tower1} as
\begin{align}
\label{d'2}
{D}(s)&=-\frac{(s-s_0)^n}{\pi}\int_{s_{\rm th}}^{+\infty} 
\frac{\rho(s') {N}(s')}{(s'-s)(s'-s_0)^n}ds' + 
\sum_{m=0}^{n-1}{a}_m s^m + 
\sum_i \frac{\gamma_i}{s-s_i}\,,\\
N(s)&=\frac{(s-s_0)^{n}}{\pi}\int_{-\infty}^{s_{\rm L}} ds' 
\frac{{\Ima  T}(s') {D}(s')}{ (s'-s)(s'-s_0)^{n}}+
\sum_{m=0}^{n-1}  b_m s^m\,.\nn
\end{align}                         
where $\{a_m,b_m,m=\text{min}(n-1,1),\ldots,n-1)$ and ${\gamma}_i$, $s_i$, are left undetermined by the DR.  Each of the poles terms in Eq.~\eqref{d'2} is referred to as a CDD pole \cite{Castillejo:1955ed}.
Every pole requires two free parameters: Its pole position $s_i$ and its residue $\gamma_i$, which is real because $\lambda(s)$ is real, cf. Eq.~\eqref{l1}.  

The scattering by an ordinary potential can be standardly solved by resolving the Lippmann-Schwinger equation, Eq.~\eqref{240810.11}, without including any free parameter at all. This has its correspondence in the $N/D$ method by requiring that $N(s)$ vanishes for $s\to\infty$. Then, fixing $D(0)=1$ by an including a subtraction in the DR for $D(s)$  we have:\footnote{Notice that $T(s)$ is left invariant if $N(s)$ and $D(s)$ are divided simultaneously by the same constant. Fixing the normalization of either $D(s)$ or $N(s)$ is required to solve the $N/D$ equations.}
\begin{align}
\label{240814.11}
D(s)&=1-\frac{s}{\pi}\int_{s_{\rm th}}^{+\infty} ds'\frac{\rho(s')N(s')}{s'(s'-s)}\,,\\
N(s)&=\frac{1}{\pi}\int_{-\infty}^{s_{\rm L}}ds'\frac{\Ima T(s')D(s')}{s'-s}\,.\nn
\end{align}
Any extra free parameters in Eq.~\eqref{d'2} compared to the solution in Eq.~\eqref{240814.11}, which is completely fixed by $\Ima T(s)$, $s<s_{\rm L}$, should be concerned with the microscopic dynamics in the scattering process reflecting the impact of underlying degrees of freedom. For example, in hadron physics these extra degrees of freedom would reflect the QCD dynamics.  This point is discussed in further details by Refs.~\cite{Entem:2016ipb,Oller:2018zts,Sanchez:2024xzl}. 

Already, in Ref.~\cite{Castillejo:1955ed} the CDD poles were related with the internal structure of the scatterer, by analogy with the Wigner-Eisenbud dispersion formula that expresses all the information that can be deduced about a scattering process, if one remains ignorant of the internal structure of the scatterer. The unknown internal structure is represented by an infinite number of adjustable free parameters. In this respect, Ref.~\cite{Chew:1961cer} linked the presence of CDD poles  to  elementary particles with the same 
quantum numbers as those of the PWA, that is, particles 
that are not originated from  given exchange forces between 
the scattering particles. A CDD pole can be added to a given ${D}(s)$  with its two parameters adjusted so as to get a zero of the real 
part of the new ${D}(s)$ with the sought position and residue, reproducing a resonance or a bound state.  
In this way, the parameters of a CDD pole can be 
related with the coupling constant and mass of an 'elementary' particle, not generated dynamically by the rescattering of the interacting particles due to the exchange forces.

The typical procedure to solve the $N/D$ equations consists of substituting the DR for $N(s)$ into that for $D(s)$, so that at the end one has a linear integral equation for $D(s)$ for $s<s_{\rm L}$. Let us exemplify this procedure by considering Eq.~\eqref{240814.11}:
\begin{align}
  \label{240905.1}
  D(s)&=1-\frac{s}{\pi^2}
  \int_{s_{\rm th}}^{+\infty}
  ds'\frac{\rho(s')}{s'(s'-s)}\int_{-\infty}^{s_{\rm L}}ds''\frac{\Ima T(s'')D(s'')}{s''-s'} 
\end{align}
Now, exchanging the order of the integrations
\begin{align}
  \label{240905.2}
  D(s)&=1+\frac{s}{\pi^2}
\int_{-\infty}^{s_{\rm L}}ds'' \Ima T(s'')D(s'')   \int_{s_{\rm th}}^{+\infty}
  ds'\frac{\rho(s')}{s'(s'-s)(s'-s'')}\,.
\end{align}
The last integral can be done algebraically. It is convenient to split the denominator and write
\begin{align}
  \label{240907.1}
-\frac{s}{\pi}\int_{s_{\rm th}}^{+\infty}
ds'\frac{\rho(s')}{s'(s'-s)(s'-s'')}=
-\frac{s}{\pi(s-s'')}\int_{s_{\rm th}}^{+\infty}ds'\rho(s')\left(\frac{1}{s'(s'-s)}-\frac{1}{s'(s'-s'')}\right)\,.
\end{align}
Each of these integrals is convergent and are discussed in the next section, cf. Eq.~\eqref{240815.3}.\footnote{One can easily show that every  integral in the right-hand side of Eq.~\eqref{240907.1} can be rewritten  in terms of $g(s)-g(0)$ evaluated at $s$ and $s''$. This is left as an exercise for the interested reader.} Then, Eq.~\eqref{240905.2} is a linear integral equation for $D(s)$ with $s<s_{\rm L}$. Once this is known, e.g. by solving it numerically after discretization, one can use the same DR and Eq.~\eqref{240814.11} for evaluating, respectively, $D(s)$ and $N(s)$ at any $s\in\mathbb{C}$. Then, $T(s)=N(s)/D(s)$ is also known.

\bigskip
$\bullet$ \underline{\it General formula for PWAs without LHC:}
\bigskip

Equation~(\ref{d'2}) gives the general integral equations for ${D}(s)$ and ${N}(s)$. Let us assume that the LHC can be neglected because its typically smooth contribution along the physical region ($s>s_{\rm th}$) can be accounted for by the polynomials and pole terms in these DRs or, simply, because it is weak enough (as estimated for scalar and isoscalar meson-meson scattering in Ref.~\cite{Oller:1998zr}).  Then, taking ${\Ima  T}(s)=0$ in Eq.~(\ref{n2})
\begin{equation}
\label{naprox}
{N}(s)=\sum_{m=0}^{n-1} b_m s^m~,\
\end{equation} 
and ${N}(s)$ becomes just a polynomial of degree $n-1$. Of course, this polynomial can be written in a factorized form in terms of its zeroes $\tilde{s}_j$ as
\begin{equation}
\label{ojo}
N(s)={\mathcal{C}} \prod_{j=1}^{n-1}(s-\tilde{s}_j)~.
\end{equation}
Here,  it should be understood that if $n-1=0$ then ${N}$ is  constant. In this way, the only effect of ${N}(s)$, apart from the normalization constant ${\mathcal{C}}$, is to produce $n-1$ zeros in ${T}(s)$. At this point, we notice that one can always divide ${N}(s)$ and ${D}(s)$ by Eq.~(\ref{ojo}). Then, once the LHC is neglected, we can always take ${N}(s)=1$ 
and all the zeros of ${T}(s)$ manifests as poles terms of $D(s)$, in addition to the CDD poles already included in Eq.~\eqref{d'2}. These extra poles in $D(s)$, because of the Cauchy integration theorem, produce extra terms in the DR for $D(s)$, and one has to add them on the right-hand side of Eq.~\eqref{240814.7}. After straightforward manipulations they give the same type of extra pole contributions as the CDD poles in Eq.~\eqref{d'2} to which they add:
\begin{eqnarray}
\label{fin/d}
{T}(s)&=&\frac{1}{{D}'_L(s)}~,\\
{N}(s)&=&1~,\nn\\
{D}(s)&=&-\frac{(s-s_0)^{n}}{\pi}\int_{s_{\rm th}}^{+\infty} ds' \frac
{ \rho (s')}{(s'-s)(s'-s_0)^{n}}+\sum_{m=0}^{n-1} a_m s^m+
\sum_i \frac{\gamma_i}{s-s_i}~.\nn
\end{eqnarray}
The DR for $D(s)$ can be further simplified because $\rho(s')/s'$ vanishes for $s'\to \infty$. Therefore, for evaluating the integral we can get rid of  $n-1$ factors of $s'-s_0$ in the denominator, by repeatedly applying
\begin{align}
  \label{240815.2}
  \frac{(s-s_0)^n}{(s'-s)(s'-s_0)^n}=(s-s_0)^{n-1}\frac{s-s'+s'-s_0}{(s'-s)(s'-s_0)^n}=\frac{(s-s_0)^{n-1}}{(s'-s)(s'-s_0)^{n-1}}-\frac{(s-s_0)^{n-1}}{(s'-s_0)^n}\,.
\end{align}
The term before the last one gives the same kind of integration but with $n\to n-1$, while the last term is a polynomial of degree $n-1$ that can be reabsorbed in $\sum_m a_m s^m$ of Eq.~\eqref{fin/d}. Implementing these steps $n-1$ times more, we can write
\begin{eqnarray}
\label{fin/dd}
{T}(s)&=&\frac{1}{{D}'_L(s)}~,\\
{N}(s)&=&1~,\nn\\
{D}(s)&=&-\frac{s-s_0}{\pi}\int_{s_{\rm th}}^{+\infty} ds' \frac
{ \rho (s')}{(s'-s_0)(s'-s)}+\sum_{m=0}^{n-1} a_m s^m+
\sum_i \frac{\gamma_i}{s-s_i}\,.\nn
\end{eqnarray}
   If a pole position $s_i$, with residue $\gamma_i$, is complex (corresponding to a complex zero in the original $N(s)$ function) then, because of the Schwarz reflection principle, its complex conjugate $s_i^*$ is also a pole with residue $\gamma_i^*$. For this case of a pair of complex conjugated poles four free parameters are added instead of the two of a CDD pole at real $s_i$. Notice that in Eq.~\eqref{fin/dd} the function $N(s)$ is the one that has been normalized to 1, instead of $D(s)$ (as we did above in Eq.~\eqref{240814.11} when requiring $D(0)=1$).

It is worth noticing that Eq.~(\ref{fin/dd}), first deduced in Ref.~\cite{Oller:1998zr}, is the most general structure of  an 
 elastic PWA, with arbitrary $J$,  when the LHC is neglected. The free parameters there could be fitted to 
the experiment or calculated by matching with  the basic underlying theory. In Ref.~\cite{Oller:1998zr}  the 
basic dynamics is QCD, but Eq.~(\ref{fin/dd}) could  be 
applied to other interactions, like in Refs.~\cite{Blas:2020dyg,Blas:2020och,Oller:2024neq} for graviton-graviton interactions [which also have the symmetries  needed to derive Eq.~(\ref{fin/dd})].

\begin{figure}
  \begin{center}
  \includegraphics[width=0.25\textwidth]{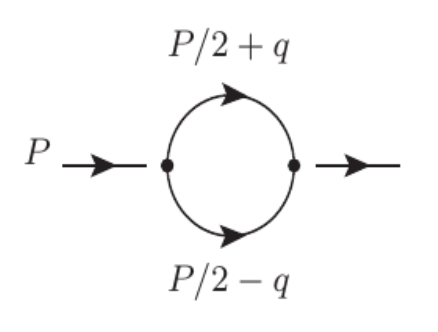}
  \caption{{\small The function $g(s)$ as the loop function in Eq.~\eqref{240815.4},  through which a total momentum $P$ flows.}\label{fig.240815.1}}
  \end{center}
  
  \end{figure}
We call the integral along the unitarity cut in Eq.~\eqref{fin/dd} as $g(s)$:
\begin{align}
  \label{240815.3}
g(s)=g(s_{\rm th})-\frac{s-s_{\rm th}}{\pi}\int_{s_{\rm th}}^{+\infty}ds'\frac{\rho(s')}{(s'-s_{\rm th})(s'-s)}\,.
\end{align}
This integral actually corresponds to the loop integral indicated in Fig.~\ref{fig.240815.1},
\begin{align}
  \label{240815.4}
  g(s)&=i\int\frac{d^4q}{(2\pi)^4}\frac{1}{[(P/2-q)^2-m_1^2+i\ep][[(P/2+q)^2-m_2^2+i\ep]}\\\
    &=\frac{1}{(4\pi)^2}\left\{
    a(\mu)+\ln\frac{m_1^2}{\mu^2}-x_+\ln\frac{x_+-1}{x_+}-x_-\ln\frac{x_--1}{x_-}    \right\}\,,\nn\\
    x_\pm&=\frac{E_2\pm |\vp|}{\sqrt{s}}\,.\nn
  \end{align}
\footnote{The algebraic expression for $g(s)$ in Eq.~\eqref{240815.4} is actually obtained by integrating in dimensional regularization the loop in Eq.~\eqref{240815.3} and removing the divergent term $1/(d-4)$ ($d$ is the number of dimensions), plus some constant terms,   by the subtraction constant $a(\mu)$.} 
 By calculating $g(s_{\rm th})$  from Eq.~\eqref{240815.4}, we have the relation: 
\begin{align}
  \label{240815.5}
g(s_{\rm th})&=\frac{a(\mu)}{(4\pi)^2}+\frac{m_1\ln(m_1/\mu)+m_2\ln(m_2/\mu)}{8\pi^2(m_1+m_2)}\,.
\end{align}
In terms of $g(s)$ the DR for $D(s)$ in Eq.~\eqref{fin/dd} becomes (the polynomial term is reduced to just a constant since this is what is strictly necessary to end with a meaningful DR once $N(s)=1$)
\begin{align}
  \label{240815.6}
  D(s)&=\alpha+\sum_i\frac{\gamma_i}{s-s_i}+g(s)\,,\\
  \label{240815.6b}
  T(s)&=\frac{1}{\alpha+\sum_i\frac{\gamma_i}{s-s_i}+g(s)}\\
  &=\frac{\left(\alpha+\sum_i\frac{\gamma_i}{s-s_i}\right)^{-1}}{1+\left(\alpha+\sum_i\frac{\gamma_i}{s-s_i}\right)^{-1}g(s)}\,.\nn
\end{align}
Let us note that $(\alpha+\sum_i\frac{\gamma_i}{s-s_i})^{-1}$ is a rational function of polynomials in $s$, both of them with degree the number of CDD poles. The constant $\alpha$ can also be interpreted as a CDD pole at infinity with the ratio  $-\gamma_k/s_k\to \alpha$\,. 

\bigskip
$\bullet$ \underline{\it Generalization to coupled channels:}
\bigskip

We now employ a matrix formalism to  generalize Eq.~\eqref{240815.6}
to coupled channels, as we did already when discussing unitarity in coupled channels in Sec.~\ref{sec.240811.1}. 
From the beginning we neglect the unphysical cuts and then $\hat{T}$ has only unitarity cut. We write $\hat{T}$ as: 
\begin{equation}
\label{A.4}
\hat{T}(s)=\hat{D}(s)^{ -1} \hat{N}(s)\,,
\end{equation}
where $\hat{N}$ has only LHC (which is neglected by now) and $\hat{D}(s)$  only  unitarity cut. It is clear that one can  always take $\hat{N}(s)$ free of poles, which could be reabsorbed as zeroes in $\hat{D}(s)$ to account for the corresponding bound states. In this way,  $\hat{N}(s)$ is just a matrix of polynomials in $s$ of degree  $n-1$, namely, 
\begin{equation}
\label{A.5}
\hat{N}(s)=\hat{Q}_{n-1}(s)~.
\end{equation}

Next, from Eqs.~(\ref{240813.3e}) and (\ref{A.4}) one has
\begin{equation}
\label{A.6}
\Ima \hat{D}(s)=-\hat{N}(s) \hat{\rho}(s)~,
\end{equation}
Given the fact that $\hat{N}(s)$ is a matrix of polynomials, it can be reabsorbed in $\hat{D}$, such that $\hat{D}^{-1}\hat{N}=(\hat{N}^{-1}\hat{D})^{-1}$. Therefore, in the case of neglecting the LHC we can take $\hat{N}(s)$ and a $\hat{D}(s)$ that satisfies Eq.~\eqref{A.6} but with $\hat{N}(s)=I$. Let us notice that because
\begin{align}
  \label{240815.7}
\hat{N}^{-1}&=\frac{1}{\text{det}\hat{N}}\text{adj}\hat{N}\,,
\end{align}
where $\text{adj}\hat{N}$ is the adjoint matrix of $\hat{N}$ and $\text{det}\hat{N}$ is its determinant,  $\hat{N}^{-1}\hat{D}=\text{adj}\hat{N}\,\hat{D}/\text{det}\hat{N}$ has poles at the position of the zeroes of $\text{det}\hat{N}(s)$. Furthermore, we introduce the diagonal matrix $\hat{G}(s)$ whose diagonal elements correspond to the function $g(s)$ in Eq.~\eqref{240815.4} but for each channels separately (using the corresponding masses and subtraction constants $a_i(\mu)$).

Proceeding analogously as for the one-channel case in order to get Eq.~\eqref{fin/dd}, we can then write\footnote{We restrict ourselves to possible poles of first degree.}
\begin{align}
\label{A.8}
\hat{T}(s)&=\hat{D}^{-1}\,, \\
\hat{N}(s)&=I~, \nonumber\\
\hat{D}(s)&=\underbrace{\hat{G}(\hat{s}_{\rm th})-\frac{s-\hat{s}_{\rm th}}{\pi}\int_{\hat{s}_{\rm th}}^{+\infty} ds'
\frac{\hat{\rho}(s')}{(s'-\hat{s}_{\rm th})(s'-s)}}_{\hat{G}(s)}+\sum_{m=0}^n \hat{a}_m s^m+\sum_i\frac{\hat{\gamma}_i}{s-s_i}\,,
\end{align}
where now the subtractive coefficients $\hat{a}_m$ and the CDD-pole residues $\hat{\gamma}_i$  are symmetric matrices. In addition, $\hat{s}_{\rm th}$ is the diagonal matrix with the threshold values of the $s$ variable for each channel. 
We can also write $\hat{T}(s)$ in Eq.~\eqref{A.8} as
\begin{align}
  \label{240815.8}
  \hat{T}(s)&=\left(I+\left(\sum_{m=0}^n \hat{a}_m s^m+\sum_i\frac{\hat{\gamma}_i}{s-s_i}\right)^{-1}\hat{G}(s)\right)^{-1}\left(\sum_{m=0}^n \hat{a}_m s^m+\sum_i\frac{\hat{\gamma}_i}{s-s_i}\right)^{-1}\,.
\end{align}

\section{General parameterization for a PWA and reaching  unphysical Riemann sheets}
\def\theequation{\arabic{section}.\arabic{equation}}
\setcounter{equation}{0} 

\begin{figure}
  \begin{center}
    \includegraphics[width=0.35\textwidth]{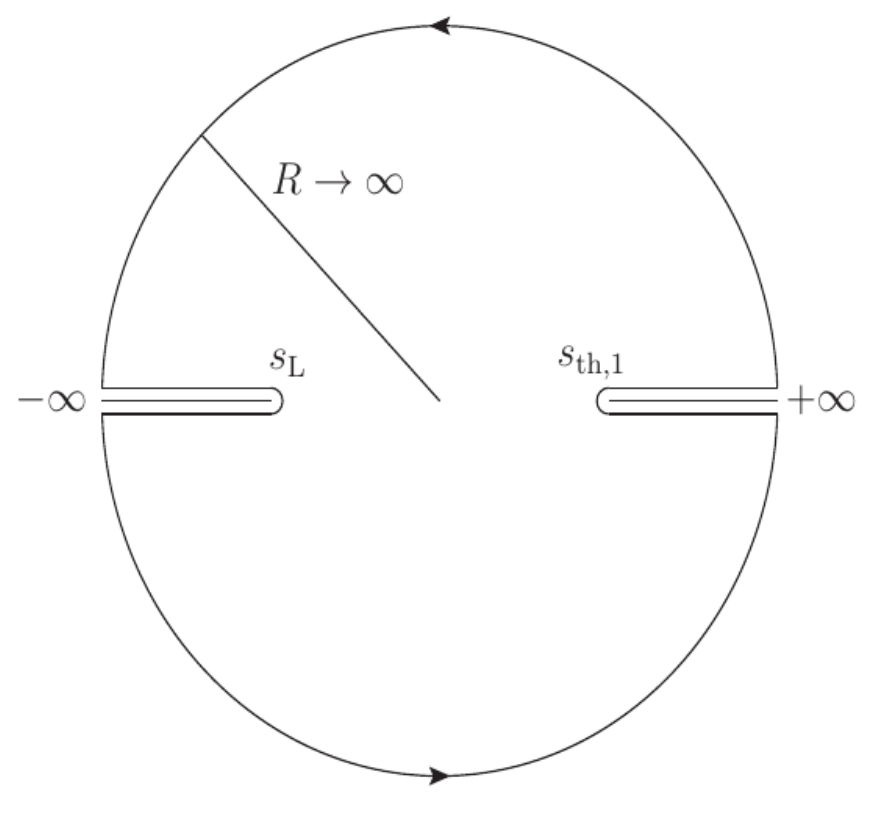}
\caption{{\small Contour of integration ${\cal C}$ used to settle the parameterization in Eq.~\eqref{181109.6} for $T(s)$.}\label{fig.240815.3}}
  \end{center}
  \end{figure}

\bigskip
$\bullet$ \underline{\it General formula for a PWA, the matrices  $\hat{\cal N}(s)$ and $\hat{G}(s)$:}
\bigskip

Let us start by deriving a  generic parameterization for a
$T$ matrix by explicitly isolating the unitarity cut and respecting the analytic properties associated to this cut.
For that, we perform a  DR of the inverse of the $T$ matrix by employing Eq.~\eqref{240813.3e}, which provide us with the discontinuity of the $T$ matrix of PWAs along the unitarity cut, the right segment indicated in Fig.~\ref{fig.240815.3}. This contribution to the DR gives rise to the diagonal $\hat{G}(s)$ matrix of functions $g_i(s)$, introduced in the previous section. In addition, there are also the crossed cuts, which are generically represented  in Fig.~\ref{fig.240815.3} by a LHC. However, the same type of construction with the appropriate contour associated with these crossed-channel cuts can be realized for other more involved ones, like e.g. the circular cut in $K\pi$ elastic scattering. The point is that the contribution from the DR along these extra cuts give rise to the matrix $\hat{\cal N}^{-1}$. Then, we write
\begin{align}
\label{181109.6}
\hat{T}(s)&=\left(\hat{\cal N}(s)^{-1}+\hat{G}(s)\right)^{-1}\,,\\
\label{181109.6b}
\hat{T}(s)&=\left[I+\hat{\cal N}(s)\hat{G}(s)\right]^{-1}\hat{\cal N}(s)\,,
\end{align}
By construction $\hat{\cal N}(s)^{-1}$ is a matrix that only has crossed-channel cuts (although it could  include CDD poles too).
In the limit in which crossed cuts are neglected we can compare with Eq.~\eqref{240815.8} and then
\begin{align}
\label{240815.9}
\hat{\cal N}(s)^{-1}&=\sum_{m=0}^n \hat{a}_m s^m+\sum_i\frac{\hat{\gamma}_i}{s-s_i}\,.
\end{align}

\bigskip
$\bullet$ \underline{\it Change to the second Riemann sheet of $g(s)$, and different Riemann sheets of $T(s)$:} 
\bigskip

\begin{figure}
  \begin{center}
    \includegraphics[width=0.8\textwidth]{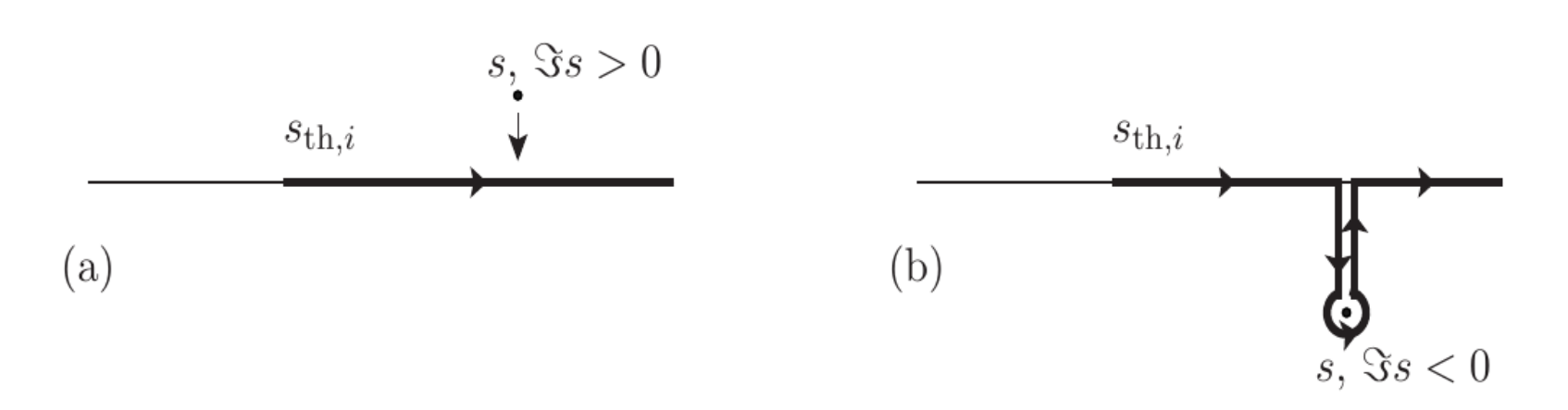}
\caption{{\small Contour deformation (thick solid line) for passing to the second Riemann sheet (RS) of  $g_i(s)$ when  crossing the RHC from top (a) to bottom (b).
The deformation of the integration contour is required so as to  avoid the pole singularity of the integrand in Eq.~\eqref{240815.3} at $s'=s$, 
for real  $s>s_{{\rm th},i}$.  Of course, the
RHC could also be crossed from bottom to top, and the  deformed contour is then  the mirror image of the one pictured in (b).}
\label{fig.240815.2}}
\end{center}
\end{figure}

Equation~\eqref{181109.6b} gives $\hat{T}(s)$ in the first Riemann sheet. We are also interested in reaching other Riemann sheets, generically called unphysical ones. For example, this is necessary in order to search for poles of the $T$ matrix corresponding to resonances and antibound states.\footnote{Let us recall that bound states correspond to poles in the first or physical Riemann sheet.} 
This is accomplished by performing the analytical continuation of the functions $g_i(s)$ in $\hat{G}(s)$.  
The function $g_i(s)$ has a branch-point singularity at the threshold $s_{{\rm th},i}$, and a unitarity cut starting from this
point along the positive real $s$ axis up to $+\infty$, also called right-hand cut (RHC). To move smoothly to the second Riemann sheet (RS) of $g_i(s)$ one should go through the  RHC and proceed by analytical continuation to the second RS. This process requires to deform the integration contour
 in the integral representation of $g_i(s)$, cf. Eq.~\eqref{240815.3}, so that the singularity at the integrand of this equation for $s'=s$ is avoided by the deformation of the integration contour  \cite{olive.181102.1}, as shown  in Fig.~\ref{fig.240815.2}. 

Thus, if we denote by $g_{II,i}(s)$ the function $g_i(s)$ in its second RS, once we add to $g_i(s)$ the result of the counterclockwise  integration  of the small circle around $s$, shown in panel (b)  of Fig.~\ref{fig.240815.2}, we have
\begin{align}
  \label{181110.1}
g_{II,i}(s)&=g_i(s)-2i \rho_{II,i}(s)=g_i(s)+2i\rho_{I,i}(s)\,.
\end{align}
Here care is needed in the extrapolation of the integrand down to $s$ to guarantee that the process is smooth. This is why we have changed to the second RS of $\rho_i(s)$ in Eq.~\eqref{181110.1}.  To be clear, we have denoted by $\rho_{I,i}(s)$ the function $\rho_i(s)$ in its first RS, 
\begin{align}
\label{181110.2}
\rho_{I,i}(s)=\frac{1}{16\pi}\sqrt{\frac{\lambda(s,m_1^2,m_2^2)}{s^2}}~,
\end{align}
with the square root $\sqrt{z}$ taken in its first RS, i.e. ${\rm arg}z\in[0,2\pi)$. 
Notice, that the minus sign between the second and third terms in the right-hand side of Eq.~\eqref{181110.1}  is because $\rho_{II,i}(s)=-\rho_{I,i}(s)$

Equation ~\eqref{181110.1} also shows that this is a two-sheet cut, because by turning around threshold in a complete circle and then  crossing again the RHC above the threshold we would have to add once more $-2i\rho_{II,i}(s)$, but the piece previously added becomes $-2i\rho_{II,i}(s)\to -2i\rho_{I,i}(s)$. Then, they both cancel and we come back to $g_i(s)$ in the first Riemann sheet. 
This discussion  shows that the branch-point singularity is of square-root type,  like the momentum itself. This is why it is customary to designate the different RSs as the ones of the linear momentum   in the CM,
\begin{align}
\label{181110.3}
|\vp(s)|&=\pm \sqrt{\frac{\lambda(s,m_1^2,m_2^2)}{4s}}~.
\end{align}
A possible convention to nominate all the possible $2^n$ RSs is by indicating the sign in front of the square root in the previous equation: 
The physical or first RS corresponds to  $(+,+,\ldots)$, 
the second RS to $(-,+,\ldots)$, the third RS to $(+,-,+,\ldots)$,
the fourth RS to $(-,-,+,\ldots)$, etc. Thus, before the sign of the momentum for the $m_{\rm th}$ channel is flip we have
$2^{m-1}$ RSs.

\begin{figure}[H]
  \begin{center}
  \begin{tabular}{lr}
    \hspace{-1cm}\includegraphics[width=0.4\textwidth]{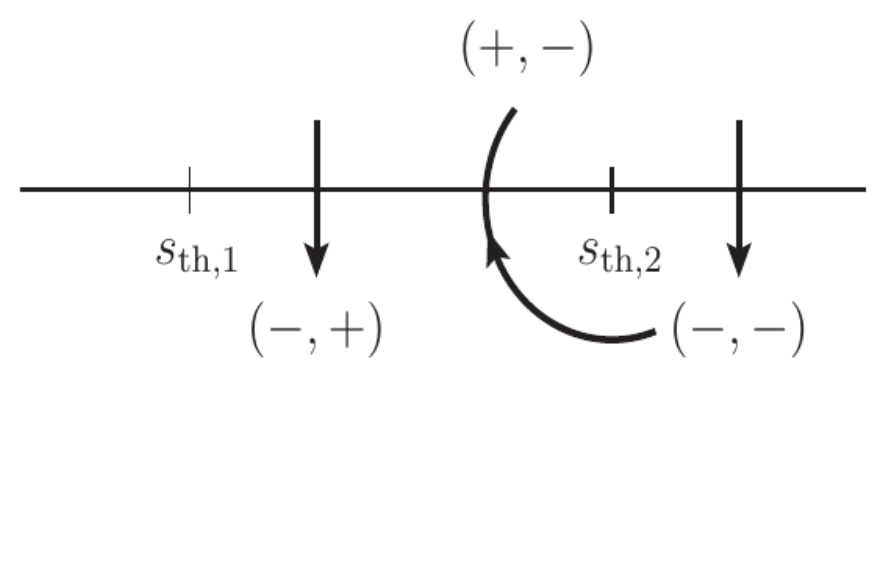} &
\hspace{1cm}\includegraphics[width=0.5\textwidth]{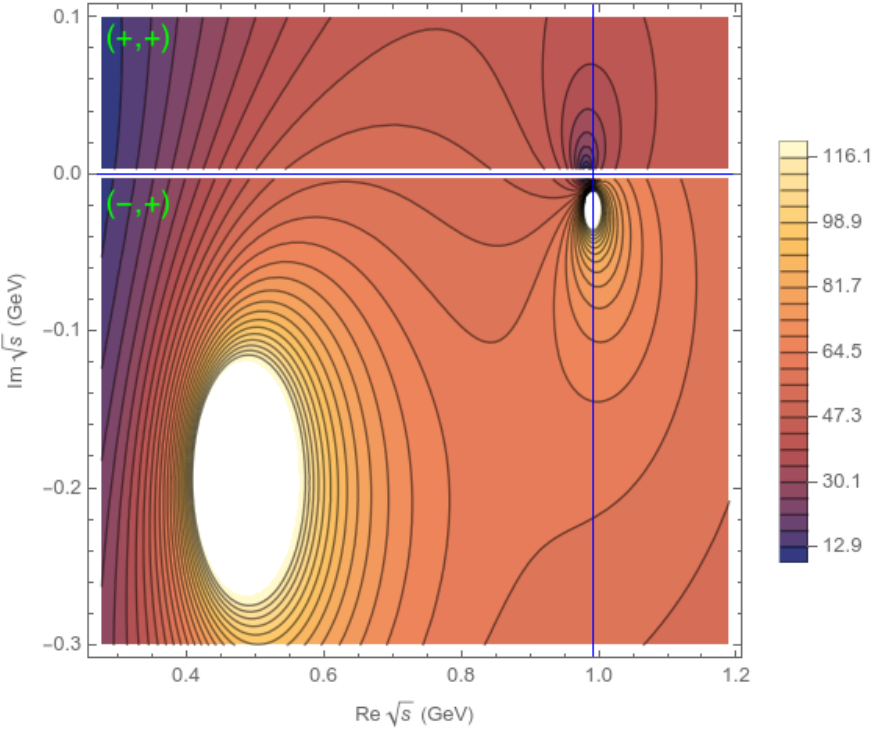}     
  \end{tabular}
  \caption{{\small Left panel: RSs for the two-channel case. Right panel: Signal stemming from the second RS ($\Ima s<0$) with two poles affecting the first RS ($\Ima s>0$). Notice how above the second-channel threshold the contour lines do not match above and below the real axis. This plot corresponds to $S$-wave isoscalar $\pi\pi$ and $K\bar{K}$ coupled-channel scattering, with the $\sigma/f_0(500)$ pole to the left and the $f_0(980)$ pole to the right. In the figure we have plotted $|T_{00}(s)|$ for $\pi\pi\to \pi\pi$ scattering. }\label{fig.240816.1}}
  \end{center}
\end{figure}
  
In the left panel of Fig.~\ref{fig.240816.1} we show the case of two channels with four RSs. Notice that in order to reach the third RS one has to cross again the real axis from bottom to top. This makes that there is no region around the real axis that smoothly connects in a direct way the first and third RSs. As a result, the features from the latter are ``hidden'' to the former. In the right panel of the same figure we show a prominent resonance pole lying in the second RS ($\Ima s<0$) in between the thresholds of the first and second channels (ordered by increasing threshold, from lighter to heavier one), and how its signal affects the physical RS ($\Ima s>0$)  above the real axis.  The resonance signal stemming from the Laurent series around the resonance pole in the second RS continuously imprints the physical  RS.  There is another  pole near the threshold of the heavier channel, whose contour lines do not match between the first and second RSs above this threshold. The reason is because the RS that smoothly connects with the physical RS is the fourth  and not the second RS. This example actually corresponds to $S$-wave $\pi\pi$ and $K\bar{K}$ scattering in coupled channels with null isospin. The resonance poles depicted in the right panel of Fig.~\ref{fig.240816.1}  are the $\sigma/f_0(500)$ (left) and the $f_0(980)$ (right) \cite{ParticleDataGroup:2024cfk}.  

\subsection{Dynamically generated resonances \& pre-existing ones}

\bigskip
$\bullet$ \underline{\it Low lying $\sigma$ pole, radius of convergence, and calculation of the $I=\ell=0$ $\pi\pi$ PWA:}
\bigskip

The lightest QCD resonance is the $\sigma$ or $f_0(500)$, while the most prominent one for energies below 1~GeV is by far the $\rho(770)$ \cite{ParticleDataGroup:2024cfk}. The  low-energy effective field theory of QCD for $s\ll 1$~GeV$^2$ is Chiral Perturbation Theory (ChPT), with an expected expansion scale in $s$ of around $1$~GeV$^2$.  However, the pole position of the $\sigma$ resonance,
\begin{align}
  \label{240816.10}
  \sqrt{s}_\sigma\approx 450-i\,250~\text{MeV},
\end{align}
obtained from sophisticated dispersive studies \cite{Garcia-Martin:2011nna} and also from unitarizing higher-order ChPT scattering amplitudes, by applying Eqs.~\eqref{240815.6b} \cite{Oller:1997ti} and \eqref{181109.6b}  \cite{Albaladejo:2012te}, implies that $|s_\sig|\approx 0.25$~GeV$^2\ll 1$~GeV$^2$. Therefore, such a low pole gives a radius of convergence for the perturbative ChPT series that is actually much smaller than the expected one around $1$~GeV$^2$. This was a strong caveat to accept the existence for this resonance.

The  $\sigma$ pole  can be understood without any free parameter and just in terms of $m_\pi$, $f_\pi\approx 92$~MeV (this is the pion-weak decay constant for $\pi^-\to\mu\bar{\nu}_\mu$), together with chiral symmetry and unitarity. 
As dynamical input we  take the $\pi^0\pi^0\to\pi^+\pi^-$ scattering amplitude, $A(s,t,u)$,  at leading order in ChPT. Then, 
\begin{align}
  \label{240816.1}
A(s,t,u)&=\frac{s-m_\pi^2}{f_\pi^2}\left(1+{\cal O}(\frac{s}{\Lambda^2})\right)\,.
\end{align}
Here,  the expansion scale $\Lambda$ is typically written as $4\pi f_\pi\approx 1.2$~GeV, and other times is taken around the $\rho$ mass squared.   The isoscalar $\pi\pi$ amplitude by plugging Eq.~\eqref{240816.1} into Eq.~\eqref{240814.6} is
\begin{align}
  \label{240816.2}
A_0(s,t,u)&=2\frac{s-m_\pi^2/2}{f_\pi^2}\,.
\end{align}
Its partial-wave projection in $S$-wave, $T_{00}(s)$, is obtained by applying Eq.~\eqref{240812.22} with $\chi_1=\chi_2=1$:
\begin{align}
  \label{240816.3}
A_{00}(s)&=\frac{1}{4}\int_{-1}^{+1}d\cos\theta T_0(s,t,u)=\frac{s-m_\pi^2/2}{f_\pi^2}\,. 
\end{align}
At this order this PWA has no LHC and we then apply Eq.~\eqref{240815.6b} to obtain an expression that fulfills unitarity to all orders. Since $T_{00}(s)$ has a zero at $s=m_\pi^2/2$ we then include only one CDD pole at this location with residue $\gamma_1=f_\pi^2$, with no CDD at infinite either ($\alpha=0$).\footnote{The position of a zero corresponds to the location of the CDD pole and the derivative of the PWA  at this zero is the inverse of the residue of the CDD pole.} Then, we write
\begin{align}
  \label{240816.4}
T_{00}(s)&=\left(\frac{f_\pi^2}{s-m_\pi^2/2}+g(s)\right)^{-1}\,.
\end{align}
For the equal mass case the expression for $g(s)$ in Eq.~\eqref{240815.4} simplifies as
\begin{align}
  \label{240816.5}
  g(s)&=\frac{1}{16\pi^2}\left(a(\mu)+\ln\frac{m^2}{\mu^2}-\sig(s)\ln\frac{\sig(s)-1}{\sig(s)+1}\right)\,,\\
  \sigma(s)&=\sqrt{1-\frac{4m_\pi^2}{s}}\,.\nn
\end{align}

\bigskip
$\bullet$ \underline{\it Natural-value estimate for $a(\mu)$, pole of the $\sigma$, and dynamically generated resonances:}
\bigskip

We now fix the subtraction constant $a(\mu)$ by an estimate of natural size. Here the idea is that within the low-energy effective field theory it does not really make sense to use momenta larger than 1~GeV to evaluate $g(s)$, cf. Eq.~\eqref{240815.4}. Then, we also consider $g(s)$ in Eq.~\eqref{240816.5} with its version calculated with a momentum cutoff $\Lambda$~GeV,
\begin{align}
  \label{240816.6}
g_\Lambda(s)&=i\int^\Lambda\frac{d^4 q}{(2\pi)^4}\frac{1}{[(P/2-q)^2-m_\pi^2+i\ep][(P/2+q)^2-m_\pi^2+i\ep]}=\int_0^\Lambda\frac{p^2 dp}{2\pi^2 w(p)}\frac{1}{s-4w(p)^2+i\ep}\,.
\end{align}
An explicit expression for this integral can be found in Ref.~\cite{Oller:1998hw}. We then match $g(s)$ and $g_\Lambda(s)$ for $s=s_{\rm th}$, which is the point with null momentum. Using that
\begin{align}
  \label{240816.7}
  g_\Lambda(s_{\rm th})&=-\frac{1}{8\pi^2}\left(\ln\bigg(1+\sqrt{1+\frac{m_\pi^2}{\Lambda^2}}\bigg)-\ln\frac{m_\pi}{\Lambda}\right)
\end{align}
and equating with $g(s_{\rm th})$ given in Eq.~\eqref{240815.5}, we then find ($m_1=m_2=m_\pi$)
\begin{align}
  \label{240816.8a}
  a(\mu)&=-2\ln(1+\sqrt{1+\frac{m_\pi^2}{\Lambda^2}})-\ln\frac{\Lambda^2}{\mu^2}\,.
\end{align}
Taking $\mu=\Lambda=1$~GeV we have
\begin{align}
  \label{240816.8}
 a(\Lambda) \approx -2 \ln 2=-1.40\,.
\end{align}
 This is called the natural value for the subtraction constant after Ref.~\cite{Oller:2000fj}.

\bigskip
{\bf Exercise 11:} Check that the same result for $a(\mu)$ is obtained by matching  $g(s)$ with the leading term of $g_\Lambda(s)$ in an expansion in powers of $m_\pi^2/\Lambda^2\ll 1$. 
\bigskip

We then have all the needed ingredients to use Eq.~\eqref{240816.4} for studying $I=J=0$ $\pi\pi$ scattering. By extrapolating it to the second RS
\begin{align}
  \label{240816.9}
T_{00}^{\rm II}(s)&=\left(\frac{f_\pi^2}{s-m_\pi^2/2}+g_{II}(s)\right)^{-1}\,,
\end{align}
with $g_{II}(s)$ given in Eq.~\eqref{181110.1}, we can look for the $\sigma$ pole. Within this approach we find
\begin{align}
  \label{181106.9b}
  s_\sigma&=(0.47-i\,0.20)^2~\text{GeV}^2,
\end{align}
without any free parameter, see the right panel of Fig.~\ref{fig.240816.1}. This value is  remarkably close to the value given in Eq.~\eqref{240816.10} and it is also  compatible with the values given in Particle Data Group (PDG) \cite{ParticleDataGroup:2024cfk}
\begin{align}
  \label{181106.9c}
s_\sigma&=(0.4-0.5-i\,(0.20-0.35))^2~{\rm GeV}^2. 
\end{align}
Then, it is clear that the $\sigma$ resonance has a dominant dynamical origin from $\pi\pi$ self-interactions dictated by chiral symmetry, unitarity and analyticity (RHC).

\bigskip
$\bullet$ \underline{\it Calculation of the $I=\ell=1$ $\pi\pi$ PWA, and fine tuning of $a(\mu)$ to a large negative value:}
\bigskip

We now proceed for the $P$-wave $\pi\pi$ interactions which are isovector, according to the rule $\ell+S+I-2i=$\,even, demonstrated in Eq.~\eqref{240812.28} applied with $S=0$, $\ell=1$ and $i=1$. Then, we take the $I=1$ amplitude from the second line of Eq.~\eqref{240814.6}, with $A(s,t,u)$ given in Eq.~\eqref{240816.1} above. The projection formula used is Eq.~\eqref{240812.22}  as before but with $\ell=1$. Then, we find
\begin{align}
  \label{240816.11}
A_{11}(s)&=\frac{s-4m_\pi^2}{6f_\pi^2}\left(1+{\cal O}(\frac{s}{\Lambda^2})\right)\,.
\end{align}
In comparison with Eq.~\eqref{240816.3} for the $I=0$ $S$-wave scattering amplitude we notice that this PWA is suppressed by a factor 6 for $|s|\gg m_\pi^2$. Again we need a CDD pole, which is located at threshold $4m_\pi^2$ because of the $P$-wave nature of the interaction, with its residue given by $6f_\pi^2$. Then, similarly as before, cf. Eq.~\eqref{240816.4}, we have now
\begin{align}
  \label{240816.12}
T_{11}(s)&=\left(\frac{6f_\pi^2}{s-4m_\pi^2}+g(s)\right)^{-1}\,,\\
T^{II}_{11}(s)&=\left(\frac{6f_\pi^2}{s-4m_\pi^2}+g_{II}(s)\right)^{-1}\,. \nn
\end{align}
However, with this construction the subtraction constant $a(\Lambda)$ in Eq.~\eqref{240816.8} (having natural value) is too small in absolute value to obtain a good $\rho$ pole. This is related to the mentioned suppression by a factor $6$ of $A_{11}(s)$, so that its inverse is six times larger. As a result, one needs to try with a much larger subtraction constant in absolute value to find a good $\rho$ pole. For instance, for $a(\mu)=-14$, $\mu=1$~GeV,
 $T^{II}_{11}(s)$ has a pole at 
\begin{align}
  \label{181106.10}
s_\rho&=(0.777-i\,0.072)^2~\text{GeV}^2~,~a(1~\text{GeV})=-14~,
\end{align}
in good agreement with the $\rho(770)$ pole $s_\rho=(0.761-0.762-i\,(0.071-0.074))^2~\text{GeV}^2$ from the Review of particle properties \cite{ParticleDataGroup:2024cfk}. To obtain $a(\mu)=-14$  by employing Eq.~\eqref{240816.8a} with $\mu=1$~GeV,  requires an extremely large cutoff $\Lambda=600$~GeV (as $a(\mu)$ depends logarithmically on it), and this value makes no physical sense for strong interactions (for which $\Lambda$ should be around 1~GeV).  Thus, the presence of the $\rho(770)$ cannot be explained as a $\pi\pi$ dynamically generated resonance, which indicates that it has a very different nature as compared with  the $\sigma$.

\bigskip
$\bullet$ \underline{\it Second CDD pole, and the $\rho(770)$ as pre-existing or elementary resonance:}
\bigskip

The tree-level leading-order ChPT amplitude, Eq.~\eqref{240816.11}, plus the bare exchange of a $\rho$ resonance  can be easily calculated  \cite{Oller:1998zr,Bernard:1991zc} 
\begin{equation}
\label{t1n2}
A_{11}(s)=\frac{2}{3}\frac{s-4m_\pi^2}{f_\pi^2}\left[1
 +g_v^2 \frac{s}{M_\rho^2-s}\right]~. 
\end{equation} 
The KSFR \cite{Kawarabayashi:1966kd,Riazuddin:1966sw} relation\footnote{It can be deduced from vector-meson dominance and imposing the appropriate QCD high-energy behavior \cite{Ecker:1988te}.} requires the coupling $g_v$ to be equal 1.
We can match this tree-level amplitude by adding two CDD poles in Eq.~\eqref{240815.6b}. 
The new CDD pole location and its residue in $T_{11}(s)$ are
\begin{align}
\label{181108.1}
s_2&=\frac{M_\rho^2}{1-g_v^2}~,\\
\gamma_2&=\frac{6f_\pi^2}{1-g_v^2}\frac{g_v^2M_\rho^2}{M_\rho^2-4(1-g_v^2)m_\pi^2}~.\nn
\end{align}
In the limit $g_v^2\to 1$,  $s_2\to \infty$ in such a way that
\begin{align}
\label{181108.2}
\lim_{s_2\to \infty} \frac{\gamma_2}{s-s_2}=-\frac{6f_\pi^2}{M_\rho^2}~
 .
\end{align}
This is why before we could  generate a good $\rho(770)$ pole  by just adjusting the  subtraction $a(\mu)$ in $g(s)$, Eq.~\eqref{240816.12}. Indeed, Eq.~\eqref{181108.2} 
times $16\pi^2$ gives 
\begin{align}
\label{181108.3}
-\frac{96\pi^2f_\pi^2}{M_\rho^2}=-13.6~,
\end{align}
which is the value above  for $a(1~{\rm GeV})$ in order to reproduce the $\rho(770)$ in $T_{11}(s)$. Therefore, this number reflects the elementary nature of the $\rho(770)$ resonance from the
point of view of pionic degrees of freedom. The final expression for $T_{11}(s)$ is 
\begin{align}
  \label{181108.4}
  T_{11}(s)&=\left[\frac{6f_\pi^2}{s-4m_\pi^2}-\frac{6 f_\pi^2}{M_\rho^2}
 +\frac{1}{16\pi^2}\left(-2\ln 2+\ln\frac{m_\pi^2}{\mu^2}-\sig(s)\ln\frac{\sig(s)-1}{\sig(s)+1}\right)\right]^{-1}\,,
\end{align}
with $\mu= 1$~GeV. Analogous discussions to the $f_0(500)$ and $\rho(770)$ for $\pi\pi$ can also be made for the $ \kappa$ and $K^*(890)$ in  $K\pi$ scattering  \cite{Oller:1998zr}. 

\bigskip
    {\bf Exercise 12:} Reproduce $|{T}_{00}(s)|$ depicted in the right panel of Fig.~\ref{fig.240816.1} by using the formalism of this chapter based on Eq.~\eqref{181109.6b}, and on the natural size estimate for $a(\mu)$, Eq.~\eqref{240816.8a} (replace there $m_\pi\to m_K$ for kaons, with $m_K$ the kaon mass in the isospin limit). In addition to $A_{00}(s)$ in Eq.~\eqref{240816.3} use the tree-level ChPT scattering amplitudes for $\pi\pi\to K\bar{K}$, $A_{00;12}(s)$, and $K\bar{K}\to K\bar{K}$, $A_{00;22}(s)$, given by:
    \begin{align}
      \label{240819.1}
      A_{00;12}(s)&=\frac{\sqrt{3}s}{4f_\pi^2}~,~A_{00;22}=\frac{3s}{4f_\pi^2}\,.
      \end{align}
\bigskip

\section{Final(Initial)-state interactions. Watson theorem.}
\def\theequation{\arabic{section}.\arabic{equation}}
\setcounter{equation}{0}

\bigskip
$\bullet$ \underline{\it Feeble probes, unitarity, PWAs, and Watson's theorem: }
\bigskip

The typical situation to have in mind is a process initiated from the imprint of a probe that interacts with the system weakly, as opposed to the typical stronger self-interactions within the particles in the system. It could also be that the probe generates the system, like for example the two photon fusion into mesons. In general the weaker process can give rise to different final states (channels) that
interact strongly among them. These interactions are called final-state interactions (FSI). 
From the previous discussion it is clear by crossing that  strong interaction processes could drive  
to feeble probes in the final state, in this case we would have initial-state interactions. Of course, there are also reactions having both initial- and final-state interactions.  
Though, we refer in the following to  FSI, the formalism developed would be also applicable to initial-state interactions.

The total $T$ and $S$ matrices comprise the weaker and strong interacting processes. 
Keeping only terms linear in the feeble interactions (with $F_i$ its  matrix element for producing the channel $i$, generally called form factor), we can write from unitarity, Eq.~\eqref{240809.17}, that
\begin{align}
  \label{181118.1}
F_i-F_i^\dagger&=i\sum_j \int dQ_j \theta(s-s_{\rm th,j}) T^\dagger_{ij}F_j ~,
\end{align}
where only states that are open contribute to the sum.  This relation is valid  even if some of the
final states $|i\rangle $ are closed (as long as the unitarity cut does not overlap a crossed one).

As usual, the unitarity relation becomes simpler  in
PWAs. For example, for the pion form factor, 
$\langle \pi^+\pi^-|T|\gamma\rangle$
\begin{align}
\label{181118.2}
\langle \pi^+(p)\pi^-(p')|T|\gamma(q)\rangle&=e 
 \ve(q)_\mu(p-p')^\mu F_{\pi\pi}(s)\,,
\end{align}
only the $P$-wave  $\pi\pi$ scattering contributes. We also restrict our discussion to two-body  final states, and assume time-reversal symmetry, so
that  PWAs are symmetric. As a result, we deduce from the unitarity relation Eq.~\eqref{240813.3} the expression\footnote{Depending on  the probe, it might be necessary to decompose it as well in PWAs. As an example, the reader can consult \cite{Abarbanel:1967wk,Morgan:1987gv,Oller:2007sh}  for $\gamma\gamma$ to meson-meson .}
\begin{align}
\label{181118.3}
\Ima F_i(s)&=\sum_j F_j(s)\rho_j(s) T_{ij}(s)^*=\sum_j F_j(s)^*\rho_j(s)  T_{ij}(s)~,
\end{align}
where the subscripts $i$, $j$ denote the different partial-wave projected states.

Particularizing Eq.~\eqref{181118.3} to the uncoupled case, we then have
\begin{align}
\label{181119.1}
\Ima F_1(s)&=F_1(s)\rho_1(s)T_{11}(s)^* ~. 
\end{align}
Since the left-hand side is real so must be the right-hand side and from this observation it follows that the phase of $F(s)$
above threshold and up to the next higher threshold is the same as the phase of $T(s)$ modulo $\pi$. This is known as
the Watson's theorem  in FSI. 

\bigskip
$\bullet$ \underline{\it Generalization to coupled channels, using the $N/D$ method, and an Omn\'es function:}
\bigskip

We proceed to its  generalization to coupled channels by employing 
Eq.~\eqref{181109.6b}, which expresses  the $T$ matrix of PWAs in terms of $\hat{\cN}(s)$ and $\hat{G}(s)$.
We rewrite the left-hand side of Eq.~\eqref{181118.3} as $(F_i(s)-F_i(s)^*)/2i$ and group together the $F_i(s)^*$
on the right-hand side:
\begin{align}
\label{181119.2}
F_i(s)&=\sum_j\left[\delta_{ij}+2i\rho_j(s)  T_{ij}(s)\right]F_j(s)^*~.
\end{align}
Next, we write this equation in matrix notation,
\begin{align}
\label{240910.1}
\hat{F}(s)&=\left[I+2i  \hat{T}(s)\hat{\rho}(s)\right]\hat{F}(s)^*\,,
\end{align}
and factor to the left $\hat{T}(s)=(\hat{\cN}^{-1}+\hat{G})^{-1}$,
\begin{align}
\label{181119.2b}
\hat{F}&=\left(\hat{\cN}^{-1}+\hat{G}\right)^{-1}\left(\hat{\cN}^{-1}+\hat{G}+2i\hat{\rho} \right)\hat{F}^*~.
\end{align}
Let us recall that $\hat{\rho}$ was included in connection with Eq.~\eqref{240813.3c} and it already includes the Heaviside functions $\theta(s-s_{\rm th,j})$.  Because  $\hat{G}(s)+2i\hat{\rho}(s)=\hat{G}(s)^*$, from Eq.~\eqref{181119.2b} we deduce that
\begin{align}
\label{181119.3}
\left(\hat{\cN}^{-1}+\hat{G}\right)\hat{F}&=\left(\hat{\cN}^{-1}+\hat{G}^* \right)\hat{F}^*~.
\end{align}
Multiplying  both sides by $\hat{\cN}$,
\begin{align}
\label{181119.4}
\left[I+\hat{\cN}(s) \hat{G}(s)\right]\hat{F}(s)&=\left[I+\hat{\cN}(s) \hat{G}(s)^* \right]\hat{F}(s)^*~,
\end{align}
and this is the generalization of the Watson's theorem to FSI in coupled channels. In particular it requires that  $(I+\hat{\cN} \hat{G})\hat{F}$ is real along the unitarity cut. It follows also then that 
\begin{align}
\label{181119.5}
\left[I+\hat{\cN}(s) \hat{G}(s)\right]\hat{F}(s)
\end{align}
is free of RHC, as  it is the same as its complex conjugate for $s>s_{\rm th,1}$.\footnote{$\hat{\cN}$ has no RHC and we could have taken its complex conjugate in Eq.~\eqref{181119.4}.} Therefore, we can write for  $\hat{F}(s)$ the expression
\begin{align}
\label{181119.6}
\hat{F}(s)&= \left[I+\hat{\cN}(s) \hat{G}(s)\right]^{-1}\hat{L}(s)~,
\end{align}
where $\hat{L}(s)$ is a column vector having only LHC, if any.

\bigskip
{\bf Exercise 13:} Calculate the pion vector form factor $F_{\pi\pi}(s)$ by applying Eq.~\eqref{181119.6} and using $A_{11}(s)$ given in Eq.~\eqref{t1n2}, with $g_v=1$. Take $L(s)=\alpha/(M_\rho^2-s)$ and fix $\alpha$ by guaranteeing that  $F_{\pi\pi}(0)=1$ because of charge conservation. Depict the resulting $|F_{\pi\pi}(s)|$ and compare it with experimental data for the pion-vector form factor. See e.g. Refs.~\cite{Oller:2000ug,Gounaris:1968mw}. 
\bigskip

These results can equally be deduced  if instead the $N/D$ method were used, with $\hat{T}=\hat{D}^{-1}\hat{N}$ in Eq.~\eqref{181119.2}.
Then,  instead of Eq.~\eqref{181119.2b} we have the relation
\begin{align}
\label{181120.1}
\hat{F}(s)=&\hat{D}^{-1}(s)\left[\hat{D}(s)+ \hat{N}(s) 2i\hat{\rho} \right]\hat{F}(s)^*=\hat{D}^{-1}(s)\hat{D}(s)^*\hat{F}(s)^*~,
\end{align}
 since $\Ima \hat{D}(s)=- \hat{N}(s)\hat{\rho}(s)$. Then,  
\begin{align}
\label{181120.1b}
\hat{D}(s)\hat{F}(s)&=\hat{D}(s)^*\hat{F}(s)^*~.   
\end{align}
By the same token as used to get Eq.~\eqref{181119.6} we can write $\hat{F}(s)$ as
\begin{align}
\label{181120.2}
 \hat{F}(s)&=\hat{D}(s)^{-1}\hat{L}(s)~,
\end{align}
with $\hat{L}(s)$ having at most LHC (if any).
 This expression separates better between the RHC and LHC than Eq.~\eqref{181119.6} because $\hat{D}(s)$ has  only RHC, while $I+\hat{\cN}(s) \hat{G}(s)$ has both RHC and LHC (as $\hat{\cN}(s)$ has LHC and $\hat{G}(s)$ has RHC). For the one-channel case $D(s)^{-1}$ is an Omn\'es function: Its phase is equal to the strong phase shift modulo $\pi$, and it has only a unitarity cut. For more details see \cite{Oller:2019rej}. 

As an example, if Eq.~\eqref{181120.2} is applied to  the pion vector form factor in Eq.~\eqref{181118.2}, the
function $L(s)$ is free of any cut because the pion vector form factor $F_{\pi\pi}(s)$ has only RHC and no LHC.  To understand this point notice that the Mandelstam variable $s$
is the only Lorentz invariant  that can be constructed out of the four-momenta of the two pions, since $p^2={p'}^2=m_\pi^2$ and $p_1 p_2=s/2-m_\pi^2$.

\section*{Acknowledgements}
I am grateful to Marcela Pel\'aez and Miguel Campiglia who invited me to lecture  at the Instituto de Física de la UdelaR, to which I acknowledge  its warm hospitality during my stay there.  Partial financial support to the Grant PID2022-136510NB-C32 funded
by MCIN/AEI/10.13039/501100011033/ and FEDER, UE, and to the EU Horizon 2020 research and innovation program, STRONG-2020 project, under grant agreement no. 824093 is acknowledged as well. The author thanks financial support from the ANII-FCE-175902 project.

\appendix

\section{The Sugawara-Kanazawa theorem and number of subtractions in dispersion relations}
\label{sec.240322.2}
\def\theequation{\Alph{section}.\arabic{equation}}
\setcounter{equation}{0}\label{app.240815.1}

\begin{figure}
  \begin{center}
    \includegraphics[width=0.35\textwidth]{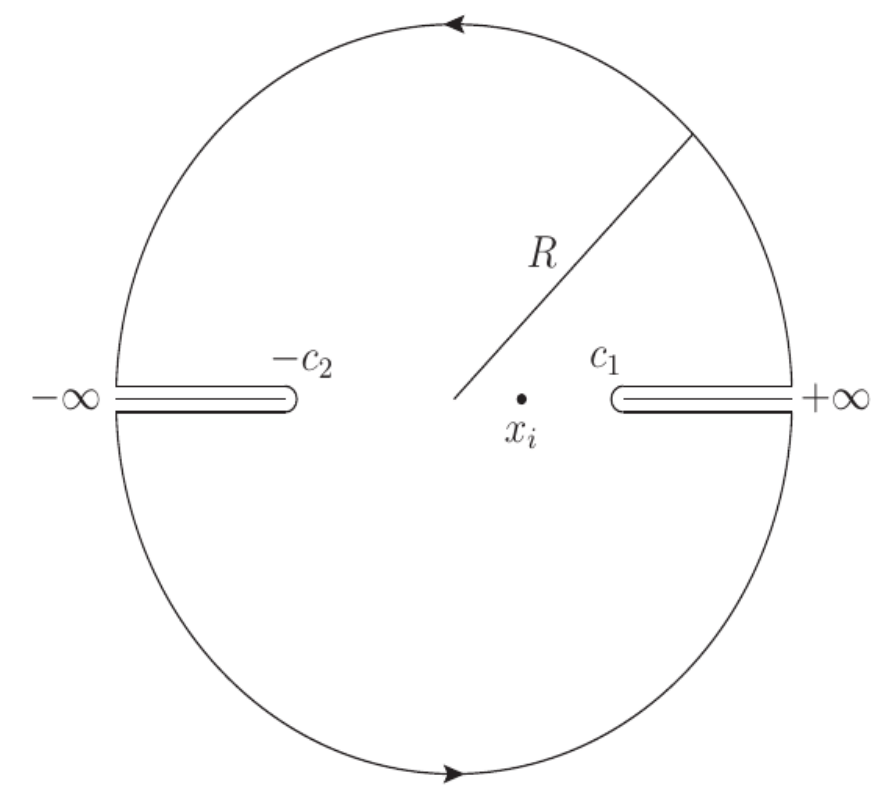}
    \caption{{\small  The integration contour ${\cal C}$ for  the DR in Eq.~\eqref{181011.2}. }\label{fig.181202.1}}
    \end{center}
\end{figure}

\bigskip
$\bullet$ \underline{\it The Sugawara-Kanazawa theorem \cite{sugawara.180929.1}:}
\bigskip

Let  $f(z)$
be an analytic function  everywhere in the complex $z$ plane except for two cuts along the real axis and
poles between them, see Fig.~\ref{fig.181202.1}.
 It is also assumed that $f(z)$ fulfills: 
\begin{itemize}
\item[i)]~ $f(z)$  has finite limits $f(+\infty\pm i\ve)$ as  $z\to +\infty \pm i\ve$ along the $c_1$-cut or RHC 
  ($\ve\to 0^+$).

\item[ii)]~ There exists a finite natural number $N\geq 1$ such that $\lim_{z\to \infty}f(z)/z^N=0$.

\item[iii)]~ $f(z)$ has definite (not necessarily finite) 
limits for  $z\to -\infty\pm i\ve$ along the $c_2$-cut or LHC.\footnote{The notation of 
Ref.~\cite{sugawara.180929.1} for a infinite definite limit implies that 
 the function should diverge steadily without oscillating.}
\end{itemize}

 The theorem then probes that $f(z)$ has  the limits
\begin{align}
  \label{181011.1}
  \lim_{z\to \infty} f(z)&= f(+\infty+i\ve)~,~{\Ima z>0}~, \\
  \lim_{z\to \infty} f(z)&= f(+\infty-i\ve)~,~{\Ima z<0}~, \nn
\end{align}
and that it can be represented as
\begin{align}
  \label{181011.2}
  f(z)=\sum_i\frac{R_i}{z-x_i}+\frac{1}{\pi}\left(\int_{c_1}^{+\infty}+\int_{-\infty}^{-c_2}\right)
  \frac{\Delta f(x)}{x-z}dx+\bar f(\infty)~,
  \end{align}
    where
    \begin{align}
  \label{181011.3}
     \Delta f(x)&=\frac{1}{2i}\left[ f(x+i\ve) - f(x-i\ve) \right]~,\\
     \bar{f}(x)&=\frac{1}{2}\left[ f(x+i\ve) + f(x-i\ve) \right] ~.\nn
    \end{align}

    The  Sugawara-Kanazawa  theorem applies then to a function $f(z)$ that can be found  in many physical applications. This is e.g. the case for a meson-meson PWA with equal-mass mesons  because
    \begin{align}
      T_{ij}=\frac{S_{ij}-\delta_{ij}}{2i \rho_i^{1/2}{\rho_j^{1/2}}}\,,
    \end{align}
    and $|S_{ij}|\leq 1$, cf. Eq.~\eqref{240817.1}. In addition, the Schwarz reflection theorem is satisfied so that the complex conjugate of these finite limits would apply for $s\to +\infty-i\ve$.  For QCD, which is an asymptotically free theory, one would expect that the $S$ matrix for $s\to\infty$ to be bounded in modulus by some finite power $s^N$. In addition, there are no bound states in pure $\pi\pi$ strong interaction physics, and those from QED lie very close to threshold. 

\bigskip
$\bullet$ \underline{\it Corollaries:}
\bigskip


\begin{enumerate}
\item If the RHC and LHC have finite extent,  the results of the theorem are trivial (the infinite point
  is isolated).
\item If the function only has one infinite cut, then  
$\De f(\infty)=0$, since $f(z)$ is continuous at the  opposite side of cut in the real axis. Then, 
  $\bar{f}(\infty)=f(\infty\pm i\ve)$.
This is the case for form factors.\footnote{At first sight it could be strange for  a form factor that $f(+\infty+i\ep)=f(+\infty-i\ep)$ because  the Schwarz reflection theorem implies that $f(+\infty+i\ep)=f(+\infty-i\ep)^*$. But the convergence of the integral in Eq.~\eqref{181011.2} with only one integral requires $f(+\infty\pm i\ep)=0$.}

\item If the Schwarz theorem is fulfilled, footnote \ref{foot.240817.1}, then $ \Delta f(x)=\Ima f(x)$ and $\bar{f}=\Rea f(\infty)$.

\item In i) it is assumed that {\it both} limits $f(+\infty\pm i\ve)$ are finite.
  This is necessarily the case when only one of them is assumed to be finite and the Schwarz reflection principle is satisfied.
  Even  with only  one being finite, let us say $f(+\infty + i\ve)[f(+\infty - i\ve)]$, and the other $f(+\infty - i\ve)[f(+\infty + i\ve)]$
  infinite, the first[second] line of Eq.~\eqref{181011.1} still applies.

\item If $f(z)$ is known to approach zero as $z\to \infty\pm i\ve$, the theorem states that $f(z)$ approaches zero in any
  other direction and  
 Eq.~\eqref{181011.2} with $\bar{f}(\infty)=0$ holds.

\item If $f(z)$  diverges as $z$ goes to either one or both of the limits $+\infty\pm i\ve$, we can introduce an auxiliary 
  function $F(z)$ which at least diverges  as strong as $f(z)$, and then apply the theorem to  $f(z)/F(z)$. For that,  
  this new function must have finite limits for $+\infty\pm i\ve$ and cannot have branch cuts
out of the real axis, with the cuts lying separately along the latter, and with the possible poles in between the two cuts. 

As a result of applying the theorem to $f(z)/F(z)$ we conclude that $f(z)$ diverges at infinity as $F(z)$ times constants for positive/negative $\Ima z$
that are the limits of $f(z)/F(z)$ for $z\to +\infty\pm i\ve$, respectively.

\end{enumerate}

\bibliographystyle{unsrt}
\bibliography{references2}

\end{document}